\UseRawInputEncoding
\documentclass[aps,pre,twocolumn,superscriptaddress]{revtex4-2}
\usepackage{amsfonts,amsmath}
\usepackage{hyperref}
\usepackage{csquotes}
\usepackage{graphicx}
\usepackage{xcolor}
\usepackage{dcolumn}
\usepackage{bm}
\usepackage[final]{pdfpages}
\makeatletter
\AtBeginDocument{\let\LS@rot\@undefined}
\makeatother

\begin{document}

\title{Permutation Jensen-Shannon distance: a versatile and fast symbolic tool\\for complex time series analysis}

\author{Luciano Zunino}
\email[Corresponding author: ]{lucianoz@ciop.unlp.edu.ar}
\affiliation{Centro de Investigaciones \'Opticas (CONICET La Plata - CIC - UNLP), C.C. 3, 1897 Gonnet, La Plata, Argentina}
\affiliation{Departamento de Ciencias B\'asicas, Facultad de Ingenier\'ia, Universidad Nacional de La Plata (UNLP), 1900 La Plata, Argentina}

\author{Felipe Olivares}
\email[]{olivaresfe@gmail.com}
\affiliation{Instituto de F\'isica Interdisciplinar y Sistemas Complejos CSIC-UIB, Campus Universitat de les Illes Balears, E-07122 Palma de Mallorca, Spain}

\author{Haroldo V. Ribeiro}
\email[]{hvr@dfi.uem.br}
\affiliation{Departamento de F\'isica, Universidade Estadual de Maring\'a, Maring\'a, PR 87020-900, Brazil}

\author{Osvaldo A. Rosso}
\email[]{oarosso@gmail.com}
\affiliation{Instituto de F\'isica, Universidade Federal de Alagoas, Macei\'o, Alagoas 57072-970, Brazil}

\date{\today}

\begin{abstract}
The main motivation of this paper is to introduce the \textit{permutation Jensen-Shannon distance}, a symbolic tool able to quantify the degree of similarity between two arbitrary time series. This quantifier results from the fusion of two concepts, the Jensen-Shannon divergence and the encoding scheme based on the sequential ordering of the elements in the data series. The versatility and robustness of this ordinal symbolic distance for characterizing and discriminating different dynamics are illustrated through several numerical and experimental applications. Results obtained allow us to be optimistic about its usefulness in the field of complex time series analysis. Moreover, thanks to its simplicity, low computational cost, wide applicability and less susceptibility to outliers and artifacts, this ordinal measure can efficiently handle large amounts of data and help to tackle the current big data challenges.
\end{abstract}

\maketitle

\section{Introduction}
\label{sec-int}
Nowadays, investigations related to complex systems are spread along all scientific fields and, typically, time series of measured variables are employed to analyze the dynamical behavior of such systems. Consequently, the identification of possible hidden dynamical structures from time series is crucial to achieve a more reliable comprehension of the mechanisms that govern the system that generates these complex temporal fluctuations. Unveiling the nature of the underlying process from observed data is indispensable for classification, modeling, and forecasting purposes. Different methods have been developed to deal efficiently with this non-trivial task. Without being exhaustive, the interested readers are directed to Refs.~\cite{costa2002,gao2006,marwan2007,lacasa2008,podobnik2008,domenico2010,donges2013,li2015,gao2016,bernaola2017} for a brief survey of some of these methods. It has also been found that symbolic techniques are especially efficient for recognizing distinct features of the phenomenon represented by the sequence of observations of a suitable observable~\cite{daw2003,porta2015,miyano2017}. Moreover, this fact is true not only for low-dimensional dynamical systems since symbolic approaches can be successfully applied to time series generated by systems of much higher dimensionality~\cite{xu2017}. In particular, during the last years, the ordinal symbolization introduced by Bandt and Pompe (BP)~\cite{bandt2002} has proven great potential for extracting useful information about the intrinsic dynamics of complex systems~\cite{amigo2007,rosso2007,parlitz2012,zunino2012,zhao2013,li2014,zunino2015,aragoneses2016,politi2017,ribeiro2017}. This success can be mainly attributed to the fact that the causal information that stems from the system's dynamical evolution is naturally taken into account in the BP recipe. As stated by Amig\'o~\textit{et~al.}~\cite{amigo2015}, ``ordinal patterns are not symbols \textit{ad hoc} but they actually encapsulate qualitative information about the temporal structure of the underlying data.'' Furthermore, this approach does not rely on generating partitions~\cite{garland2014,bradley2015}, resulting in a practical option to symbolize datasets generated by unknown dynamic processes with unknown levels of noise. Due to these useful properties, permutation entropy, the most representative and widely used descriptor from the BP encoding scheme, that estimates the average rate at which new information appears in a time series per observation~\cite{neuder2021}, has been successfully applied in countless applications since its introduction almost twenty years ago (please see Refs.~\cite{zanin2012,zanin2021} for a review of some of these applications).

On the other hand, different methodologies have been previously introduced to measure the degree of similarity between the symbol sequence statistics of two signals. To the best of our knowledge, up to now, there is not an optimal algorithm for quantifying it in practice. Particularly, Yang~\textit{et al.}~\cite{yang2003} have proposed a method based on rank order statistics of symbolic sequences and it has been successfully applied to detect temporal structures in human heartbeat time series and to categorize biological signals~\cite{peng2007}. However, there exist some discrepancies regarding its properties and practical utility~\cite{kraskov2004,yang2004}. Ouyang~\textit{et al.}~\cite{ouyang2010} have defined a dissimilarity measure based on the rank-frequency distribution of ordinal patterns to investigate the dynamical characteristics of epileptic EEG data, while Parlitz~\textit{et al.}~\cite{parlitz2013} have identified equivalent dynamics quantifying the similarity of their relative frequencies of occurrence of ordinal patterns through the Hellinger distance. Much more recently, an approach based on the differences between R\'enyi entropy spectra has been implemented by Xu and Beck~\cite{xu2017}. Finally, the \textit{alphabetic Jensen-Shannon divergence}, a relative distance (in a distributional sense) between time series, has been introduced by Mateos~\textit{et al.}~\cite{mateos2017} for detecting dynamical changes. A binary encoding is undertaken as a coarse-grained representation of the signals, and then used to estimate the Jensen-Shannon divergence. Discrimination between stochastic and chaotic dynamics and the determination of change points in time series are two successful applications achieved by implementing this scheme.

Following a similar philosophy to that of Refs.~\cite{ouyang2010,parlitz2013,mateos2017}, we shed some light on this issue by introducing the \textit{permutation Jensen-Shannon distance}. The Jensen-Shannon divergence is implemented as a discriminant quantifier between two signals but using the ordinal symbolic representation introduced by BP for estimating it. It is plausible to assume that the distribution of ordinal patterns is a fingerprint of the underlying dynamics and that, in practical contexts, different dynamics are associated with different ordinal patterns probability distributions. Consequently, the Jensen-Shannon divergence between the ordinal mapping of two time series provides an alternative to quantify the degree of similarity between the temporal structures of arbitrary complex systems. Moreover, this tool straightforwardly inherits all the important advantages associated with the BP scheme, namely, simplicity, low computational cost, noise robustness, and invariance under scaling of the data. These features make our proposed ordinal distance especially well-suited for the analysis of experimental data. Besides, another advantage to highlight is the versatility of the proposed approach since hypothesis tests related to the nature of an arbitrary time series can be easily carried out by estimating its permutation Jensen-Shannon distance to reference time series appropriately generated according to the null model. In the following, we introduce the permutation Jensen-Shannon distance, present numerical analysis and empirical applications, and finally, summarize the main conclusions achieved in this work. A performance comparison with other approaches has been addressed in the Appendix.

\section{Permutation Jensen-Shannon distance}
\label{sec-pjsd}
The Jensen-Shannon divergence~\cite{lin1991} between the probability distributions $P=\{p_1,\dots,p_n\}$ and $Q=\{q_1,\dots,q_n\}$ is a symmetrized version of the Kullback-Leibler divergence $D_{KL}(P,Q)=\sum_{i=1}^{n} p_i \ln{(p_i/q_i)}$ given by
\begin{equation}
\begin{split}
D_{JS}(P,Q)&=\frac{1}{2} \left[ D_{KL}\left(P,\frac{P+Q}{2}\right) + D_{KL}\left(Q,\frac{P+Q}{2}\right) \right]\\
&=S\left(\frac{P+Q}{2}\right) - \frac{1}{2} S(P) - \frac{1}{2} S(Q)\,,
\end{split}
\label{JSD}
\end{equation}
where $S$ is the Shannon entropy function, \textit{i.e.} $S(P)=-\sum_{i=1}^{n} p_i \ln p_i$ (as usual, we assume the convention that $0 \ln 0=0$ and $\ln{(0/0)}=0$). This is a dissimilarity measure, bounded between 0 and $\ln 2$, that can also be interpreted as the entropy of the average distribution minus the average of the entropies~\cite{nielsen2020}. The minimum value is achieved if and only if the two distributions under comparison are identical, while the maximum value is obtained whenever their supports are disjoints (that is, $p_iq_i=0$ for $i=1,\dots,n$). The entropic definition of the Jensen-Shannon divergence leads to a natural generalization that allows to weigh differently the compared distributions. Denoting $\pi_P$ and $\pi_Q$ ($\pi_P \geq 0$, $\pi_Q \geq 0$ and $\pi_P + \pi_Q=1$) the weights associated with the probability distributions $P$ and $Q$, respectively, this generalization is defined as 
\begin{equation}
D_{JS}^{\pi_P,\pi_Q}(P,Q)=S\left(\pi_P P + \pi_Q Q\right) - \pi_P S(P) - \pi_Q S(Q).
\label{WJSD}
\end{equation}
Furthermore, this measure of discernability can also be generalized to quantify the difference between more than two distributions~\cite{lin1991}:
\begin{equation}
\begin{split}
D_{JS}^{\pi_1,\pi_2,\dots,\pi_n}(P_1,P_2,\dots,P_n)&=S\left(\sum_{i=1}^{n} \pi_i P_i \right) - \\ &- \sum_{i=1}^{n} \pi_i S(P_i)\,,
\end{split}
\label{GJSD}
\end{equation}
with $P_i$ the probability distributions and $\pi_i$ positive weights such that $\sum_{i=1}^{n} \pi_i=1$. But what is more important for our present purposes is the fact that the square root of the Jensen-Shannon divergence, $[D_{JS}(P,Q)]^{1/2}$, is a true metric for probability distributions. For further mathematical details please see Ref.~\cite{endres2003}. In fact, an entire monoparametric family of metrics is obtained for $[D_{JS}(P,Q)]^{\gamma}$ with $\gamma \in (0,1/2]$~\cite{osan2018}. The Jensen-Shannon divergence has previously allowed to shed some light on some relevant issues: the segmentation of nonstationary symbolic sequences, such as DNA sequences, into stationary subsequences~\cite{bernaola1996,grosse2002}, the definition of a statistical measure of complexity~\cite{lamberti2004}, the quantification of time symmetry breaking~\cite{feng2008}, the characterization of network evolution processes~\cite{carpi2011}, the study of proteins dynamics~\cite{sparacino2011} and the statistical analysis of language~\cite[and references therein]{gerlach2016}.

The estimation of the Jensen-Shannon distance, \textit{i.e.} $[D_{JS}(P,Q)]^{1/2}$, as any measure from Information Theory, requires first to know the probability distribution of the time series under analysis. We propose to implement this quantifier over the symbolic sequences obtained via the ordinal mapping procedure introduced by BP. This encoding scheme considers the order relations between some equidistant successive values of time series instead of the values themselves. Consequently, it avoids amplitude threshold dependencies that affect other more conventional symbolization recipes based on range partitioning~\cite{bollt2001}. The probability distribution is then approximated by the histogram of the ordinal patterns and it characterizes the shape of the time series waveforms in short temporal windows quantitatively. The ordinal approach offers the possibility to find an appropriate symbolic representation from a time series based on the temporal dynamics in a simple and natural way, with only a weak stationary assumption on the underlying process~\cite{bandt2019}.

This procedure can be better illustrated through a simple numerical example. Given the short time series $X=\{4,1,6,5,10,7,2,8,9,3\}$, two parameters, the order of the permutation symbols $D$ ($D \geq 2$ with $D \in \mathbb{N}$, length of the ordinal pattern) and the lag $\tau$ ($\tau \in \mathbb{N}$, time separation between elements) should be chosen. Next, the time series is partitioned into subsets of length $D$ with lag $\tau$ akin to phase space reconstruction by means of time-delay-embedding. The elements in each new partition (of length $D$) are replaced by their rank in the subset. This symbolic ordering reflects the visual pattern in this subset. For example, if $D=3$ and $\tau=1$, there are eight three-dimensional vectors of three consecutive data points associated with $X$. The first one $(x_0,x_1,x_2)=(4,1,6)$ is mapped to the ordinal pattern $(102)$. The second three-dimensional vector is $(x_0,x_1,x_2)=(1,6,5)$, and $(021)$ will be its related permutation. The procedure continues so on until the last sequence, $(8,9,3)$, is mapped to its corresponding motif, $(120)$. There are a couple of technical issues worth mentioning. First, note that we implement an overlapped scheme for successive ordinal patterns. A comparison between the overlapping versus the nonoverlapping ordinal pattern selection has confirmed that the former approach reduces the variance of motifs frequencies thanks to the larger number of ordinal patterns to count~\cite{little2017}. Second, if two elements in the vector have the same value, they are ranked according to their temporal order or, alternatively, a small random perturbation is added to break ties. The effect of a significant number of equalities on the BP encoding scheme, and more specifically on the permutation entropy estimated values, has been carefully analyzed in Refs.~\cite{zunino2017,cuesta-frau2018}.

The probability of each ordinal pattern can then be estimated by simply computing the relative frequencies of the $D!$ possible permutations $\pi_i$:
\begin{equation}
p(\pi_i)=\frac{C(\pi_i)}{N-(D-1)\tau}, i=1,\dots,D!\,,
\label{OP}
\end{equation}
with $C(\pi_i)$ the number of occurrences of the ordinal pattern $\pi_i$ in an arbitrary scalar time series $\{x_t\}_{t=1}^{N}$ of length $N$. In such a way, an ordinal pattern probability distribution, $P=\{p(\pi_i),i=1,\dots,D!\}$, is obtained. Returning to our numerical example: $p(\pi_1)=p(012)=1/8$, $p(\pi_2)=p(021)=1/4$, $p(\pi_3)=p(102)=3/8$, ${p(\pi_4)=p(120)=1/8}$, $p(\pi_5)=p(201)=0$, and ${p(\pi_6)=p(210)=1/8}$. When there is not any temporal dependence between the values in the time series, all possible ordinal patterns appear with the same probability, \textit{i.e.}, $P=\{p(\pi_i)=1/D!,i=1,\dots,D!\}$. Nontrivial dynamics manifest themselves in a non-uniform distribution of the ordinal patterns: some motifs are more often than others and this can be interpreted as an underlying dynamical signature. The distribution of ordinal patterns is invariant with respect to order-preserving changes, such as translations (adding a constant), scalings (multiplying by a positive constant) and/or any monotone increasing nonlinear transformations of the original data~\cite{unakafov2018}. Technically speaking, the ordinal pattern probability distribution $P$ is obtained once the order $D$ and the lag $\tau$ are fixed. This encoding scheme does not require the optimal reconstruction of the phase space that is necessary for estimating other quantifiers of chaotic signals. Consequently, $D$ and $\tau$ are not usually selected following the methodologies often employed in a conventional phase space reconstruction (\textit{e.g.}, the first zero of the autocorrelation function, the first minimum of the average mutual information, and the false nearest neighbor algorithm)~\cite{riedl2013}. Taking into account that there are $D!$ potential permutations for a $D-$dimensional vector, the condition $N \gg D!$, with $N$ the length of the time series, must be satisfied in order to obtain a reliable estimation of $P$~\cite{staniek2007}. It is clear that more temporal information is incorporated into the ordinal patterns as the order $D$ increases. For practical purposes, BP suggest in their seminal paper to estimate the frequency of ordinal patterns with $3 \leq D \leq 7$ and lag $\tau=1$ (consecutive data points). Since the lag $\tau$ physically corresponds to multiples of the sampling time of the signal under analysis, a multiscale analysis can be easily accomplished by analyzing the behavior of any statistic of $P$ as a function of this parameter. Particularly, the estimation with lagged data points, \textit{i.e.} $\tau \geq 2$, offers a more complete comprehension of the underlying dynamics in the case of continuous and/or scale-dependent systems~\cite{parlitz2012,zunino2012,bandt2019,bandt2020,olivares2020}.

The permutation Jensen-Shannon distance (PJSD) is hence obtained by calculating the square root of Eq.~(\ref{JSD}), with $P$ and $Q$ the ordinal probability distributions associated with the two signals under analysis. It is worth remarking here that the metric property of $[D_{JS}(P,Q)]^{1/2}$ and all the practical advantages of the ordinal coarse-graining are merged into the PJSD. Henceforth, normalized values of the PJSD with respect to its upper bound, \textit{i.e.} $(\ln 2)^{1/2}$, are estimated. Since it constitutes a measure of distinguishability between two probability distributions, the symbol composition between different sequences can be quantitatively compared through this metric. Obviously, larger values of this quantifier indicate less similarity between the symbolic mapping of the signals and vice versa. We hypothesized that signals coming from the same underlying dynamics would have small estimated values of this ordinal distance (close to but not exactly zero as a consequence of finite-size effects) whereas larger values (significantly different from zero) will be found when the signals have different origins. Actually, in the former instance, PJSD converges asymptotically to zero with the series size, as it will be shown shortly.

\section{Numerical analysis}
\label{sec-num-app}
Next, several numerical datasets are analyzed for characterizing the PJSD behavior and, also, for illustrating the versatility and robustness of the proposed symbolic quantifier within controlled frameworks. It is worth advancing that the main purpose in each of these examples is totally different, and that these applications seek to demonstrate the potentiality and flexibility of the PJSD for heterogeneous applications.

\subsection{Convergence to zero for signals with the same dynamics}
\label{subsec-zero-convergence}
\begin{figure}[!ht]
\centering
\includegraphics[width=\linewidth,trim={.8cm .15cm 1.4cm .3cm},clip=true]{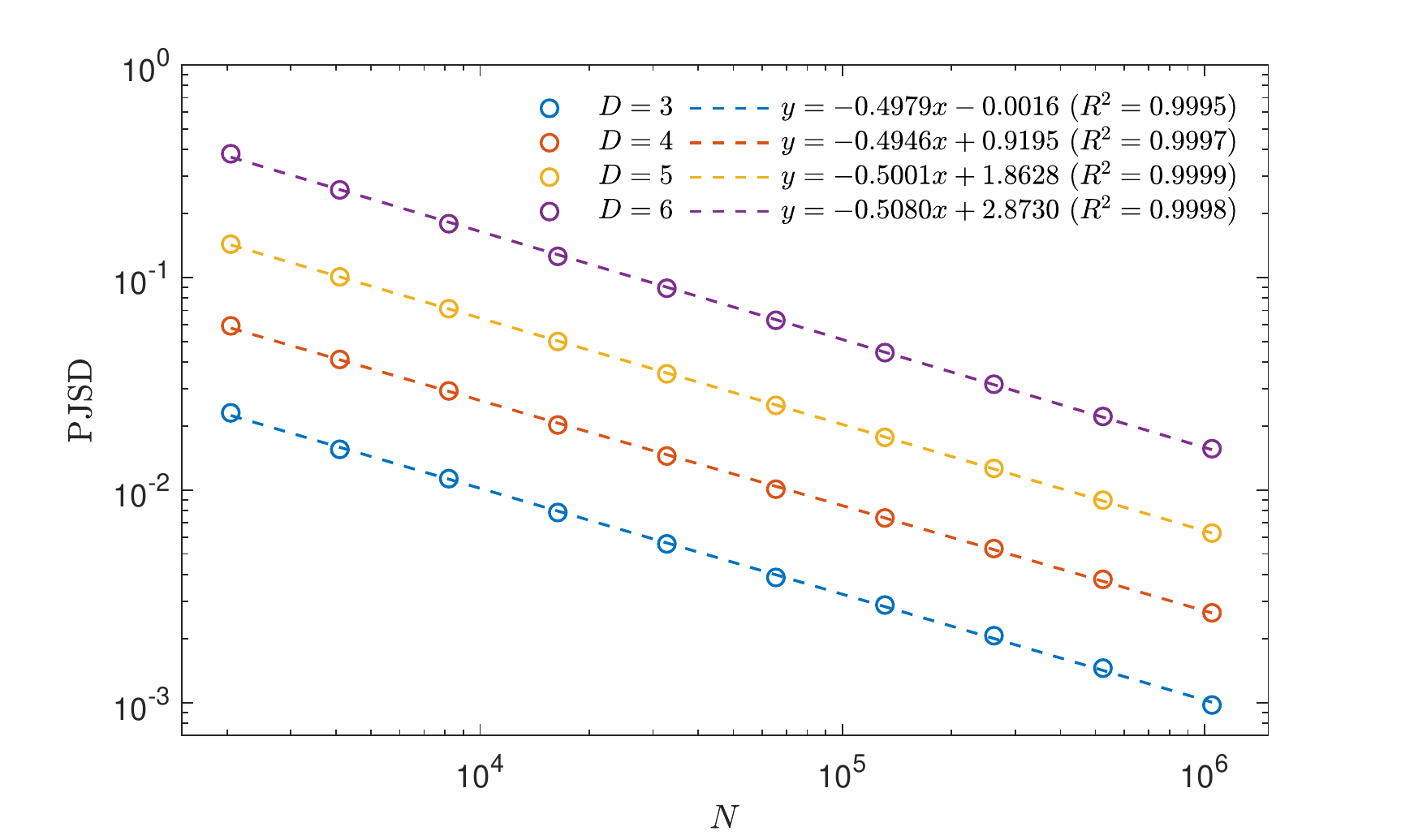}
\caption{Log-log plot of the PJSD estimations as a function of the time series size $N$ for two independent numerical realizations of a Gaussian white noise. Circles indicate the average value from an ensemble of one hundred independent estimations with several orders $D$ and lag $\tau=1$. Dashed lines correspond to the best linear fits to the data. Fitting parameters and $R^2$ values of these linear regressions are detailed in the legend.}
\label{fig:WN_error}
\end{figure}

Trying to characterize the behavior of the PJSD when two signals from the same dynamical source are analyzed, we have estimated the ordinal metric as a function of the time series size $N$ for two independent numerical realizations of a Gaussian white noise. Figure~\ref{fig:WN_error} shows the log-log plot of the average PJSD with $D \in \{3,4,5,6\}$ and lag $\tau=1$ from an ensemble of one hundred independent estimations of sizes $N \in \{2^{11},2^{12},\dots,2^{20}\}$. From the observed linear decrease in the log-log plot, it can be concluded that the PJSD tends asymptotically to zero following a power-law behavior with $N$. To be more precise, $\text{PJSD} \propto N^{-1/2}$, with a proportionality factor that depends on the order $D$. Consequently, the deviation from zero for the PJSD estimation in the case of two signals with the same dynamics is due to finite-size effects. We have confirmed qualitatively similar behaviors for other stochastic and deterministic dynamics~\cite{SM}.

It has been analytically shown by Grosse~\textit{et al.}~\cite{grosse2002} that, up to a first order approximation, the expected value of the Jensen-Shannon divergence $D_{JS}(P,Q)$ for two sequences of independent and identically distributed symbols decays inversely proportional to $N$, \textit{i.e.} $D_{JS}(P,Q)~=~C_{n}/N$, with $C_{n}$ a constant that depends on $n$, the number of elements in the probability distributions. Considering that the square root is taken when estimating the PJSD and the fact that $n=D!$ (number of possible permutations) in the BP symbolization scheme, our numerical results are consistent with this analytical approximation. It should also be highlighted here that $D_{JS}(P,Q)$ follows a $\chi^2$ distribution for asymptotically large values of $N$ under the null hypothesis that $P$ and $Q$ are generated from the same probability distribution~\cite{grosse2002,endres2003}. This result could, in principle, be used for defining a criterion or threshold in order to quantify the statistical significance of estimated values of PJSD.

\subsection{Characterizing long-range correlated time series}
\label{subsec-colored-noises}
It is well-known that a wide range of complex systems from diverse fields, including physiology, economy, geophysics and music among many others, generates output signals that display different degrees of long-range power-law correlations. Discriminating between these long-memory stochastic dynamics is crucial for several purposes. For example, scaling properties observed for the heart-rate variability are different in healthy and pathological conditions~\cite{goldberger2002}. Fractional Brownian motions (fBms) and their increments, the fractional Gaussian noises (fGns), are ubiquitous models for nonstationary and stationary long-range correlated time series, respectively. These stochastic processes are usually characterized by the Hurst exponent $H \in (0,1)$ that controls the power-law exponent of their associated spectra. In the stationary case, \textit{i.e.} for fGns, the power spectrum is proportional to $1/f^\beta$ with $\beta=2H-1$ ($-1<\beta<1$), while an analogous generalized power spectrum but with $\beta=2H+1$ ($1<\beta<3$) can be defined for the fBm nonstationary processes. Antipersistence, absence of correlations and persistence are obtained for fGns with $H \in (0,1/2)$, $H=1/2$ and $H \in (1/2,1)$, respectively. In fGns with antipersistent behaviors, the series are anticorrelated. That means that positive values are followed by negative values (or vice versa) more frequently than by chance and time series with higher degrees of roughness are observed. On the other hand, long-range correlated series are obtained for persistent fGns where positive values are followed by positive values and negative values are followed by negative values more likely than by chance. Memoryless dynamics are generated if $H=1/2$: Gaussian white noise for the fGn model and ordinary Brownian motion for the fBm one. Smoother time series with stronger trends are obtained for fBms with larger values of $H$. Further details about these fractal stochastic models can be found in the seminal paper by Mandelbrot and Van Ness~\cite{mandelbrot1968}.

\begin{figure}[!ht]
\centering
\includegraphics[width=\linewidth,trim={1.25cm .25cm 1.25cm .35cm},clip=true]{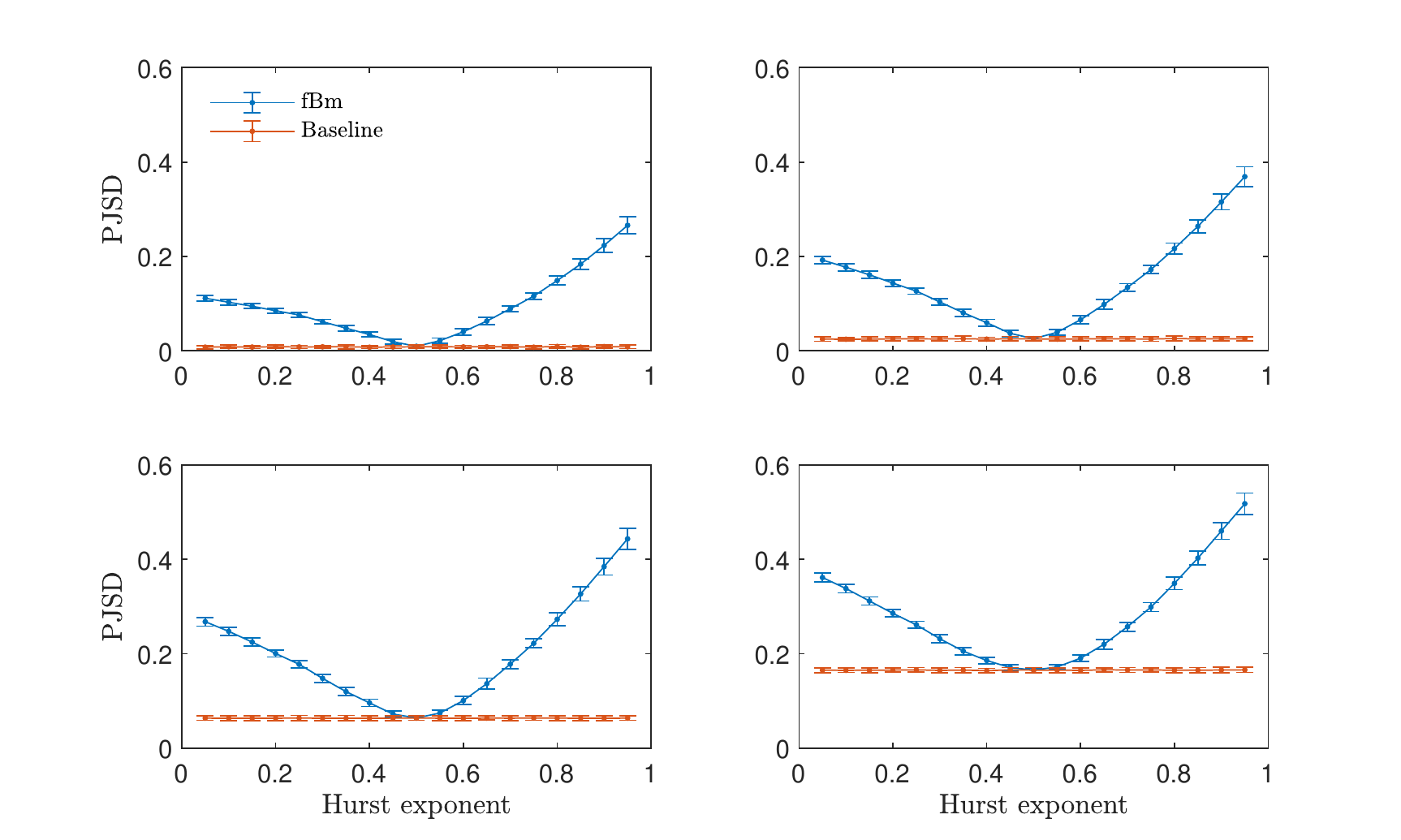}
\caption{PJSD estimations with several orders $D \in \{3,4,5,6\}$ (increasing from top left to bottom right) and lag $\tau=1$ are plotted as a function of the Hurst exponent for fBms. Mean and standard deviation (as error bar) from estimations of an ensemble of one hundred independent realizations of length $N=10^4$ data are depicted. PJSD baseline references resulting from the analysis of a pair of shuffled surrogate realizations from each simulation are also included.}
\label{fig:PJSD_fBm}
\end{figure}

\begin{figure}[!ht]
\centering
\includegraphics[width=\linewidth,trim={1.25cm .25cm 1.25cm .35cm},clip=true]{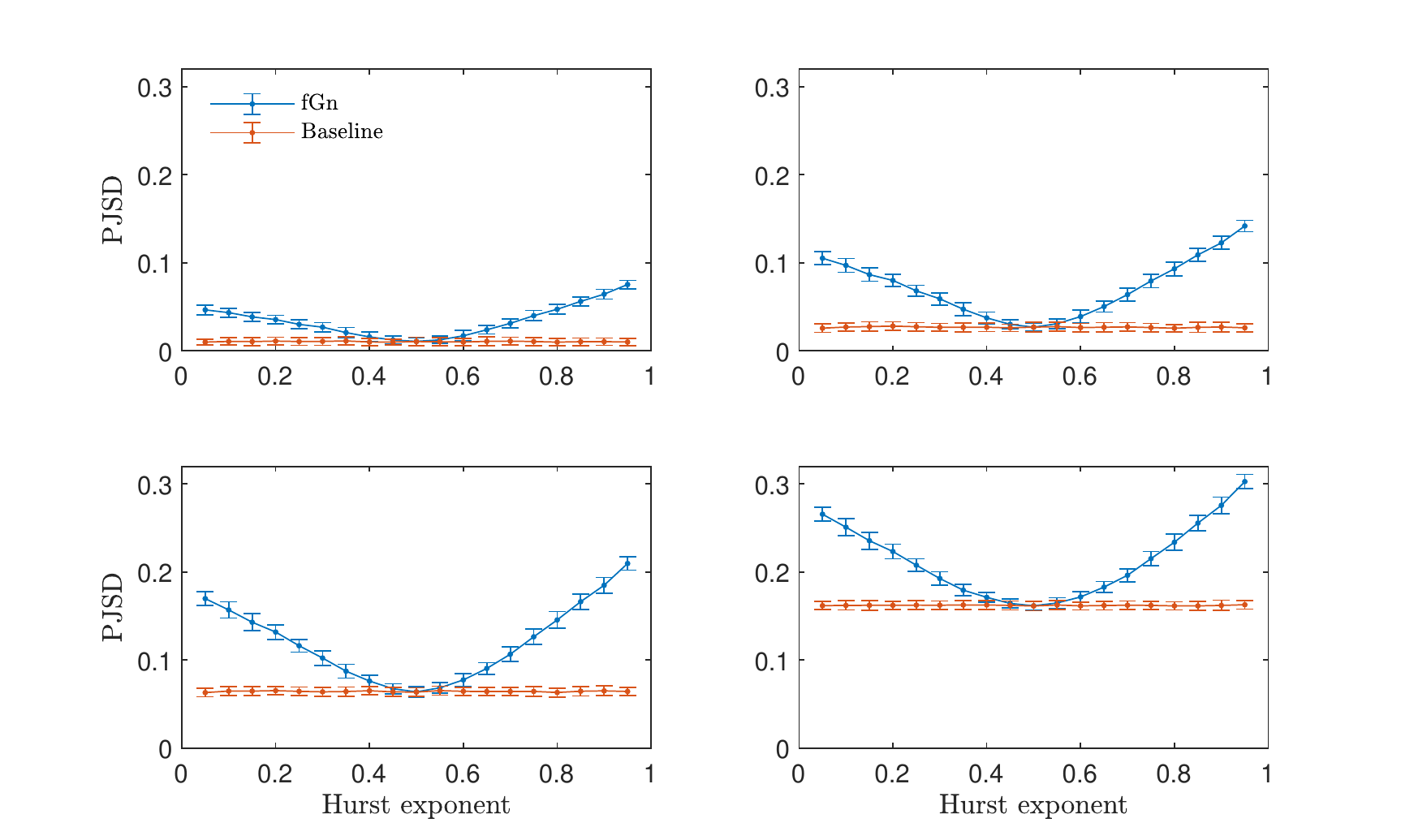}
\caption{Same as Fig.~\ref{fig:PJSD_fBm} but for fGns.}
\label{fig:PJSD_fGn}
\end{figure}

We have analyzed numerical simulations of fBm and fGn with Hurst exponents $H \in \{0.05,0.1,\dots,0.95\}$ generated by implementing the function \textit{wfbm} of MATLAB. One hundred independent realizations of length $N=10^4$ data were simulated for each Hurst exponent. Mean and standard deviation (included as error bar) of the PJSD, estimated for several orders $D \in \{3,4,5,6\}$ and lag $\tau=1$, between the simulated records and their shuffled counterparts (white noise and random walk null models for fGn and fBm, respectively~\cite{RWnullmodel}) are displayed in Fig.~\ref{fig:PJSD_fBm} for fBms and in Fig.~\ref{fig:PJSD_fGn} for fGns. Shuffled or scrambled surrogates are considered as time series of reference since we are trying to quantify the degree of temporal correlations, and these randomized resampled sequences are constrained realizations that satisfy the null hypothesis; \textit{i.e.} fully random data having exactly the same amplitude distribution as the original time series are generated~\cite{theiler1992,lancaster2018}. Moreover, a baseline is established by estimating the PJSD between two shuffled realizations of each simulation. In such a way, finite-size and amplitude distribution effects are taken into account. The presence of any temporal structure can be robustly concluded only when a significant deviation from this baseline is confirmed. Actually, a clear discrimination of the temporal correlations is observed for both models: the PJSD reaches values larger than the baseline for $H \neq 1/2$. This discrimination is improved for longer simulations with $N=10^5$ data~\cite{SM1}. The deviations are asymmetric, with larger distances for highly persistent dynamics than those reached for highly antipersistent ones. A similar asymmetric behavior is found when the permutation entropy is implemented for the fGns characterization~\cite{zunino2008,SM2}. In relation to the selection of the lag $\tau$, we have confirmed that the correlations discrimination is worse for fGns (especially for $H<1/2$) and remains qualitatively similar for fBms when larger values ($1<\tau \leq 20$) are used~\cite{SM1}. This latter finding for fBms will be further explored in Sec.~\ref{subsec-osd}. Finally, by comparing the results obtained for the two models, it can be concluded that PJSD analysis on the process (fBm) allows to unveil the presence of the underlying temporal correlations much more precisely. This is in agreement with previous findings~\cite{DeFord2017} and should be taken as a rule of thumb when possible. For a comparison with the results obtained by implementing other dissimilarity measures, please see Appendix.

\subsection{Distinguishing persistent from antipersistent stochastic dynamics}
\label{subsec-fGn-distance-matrix}
\begin{figure}[!ht]
\centering
\includegraphics[width=\linewidth,trim={.8cm .1cm 1.4cm .3cm},clip=true]{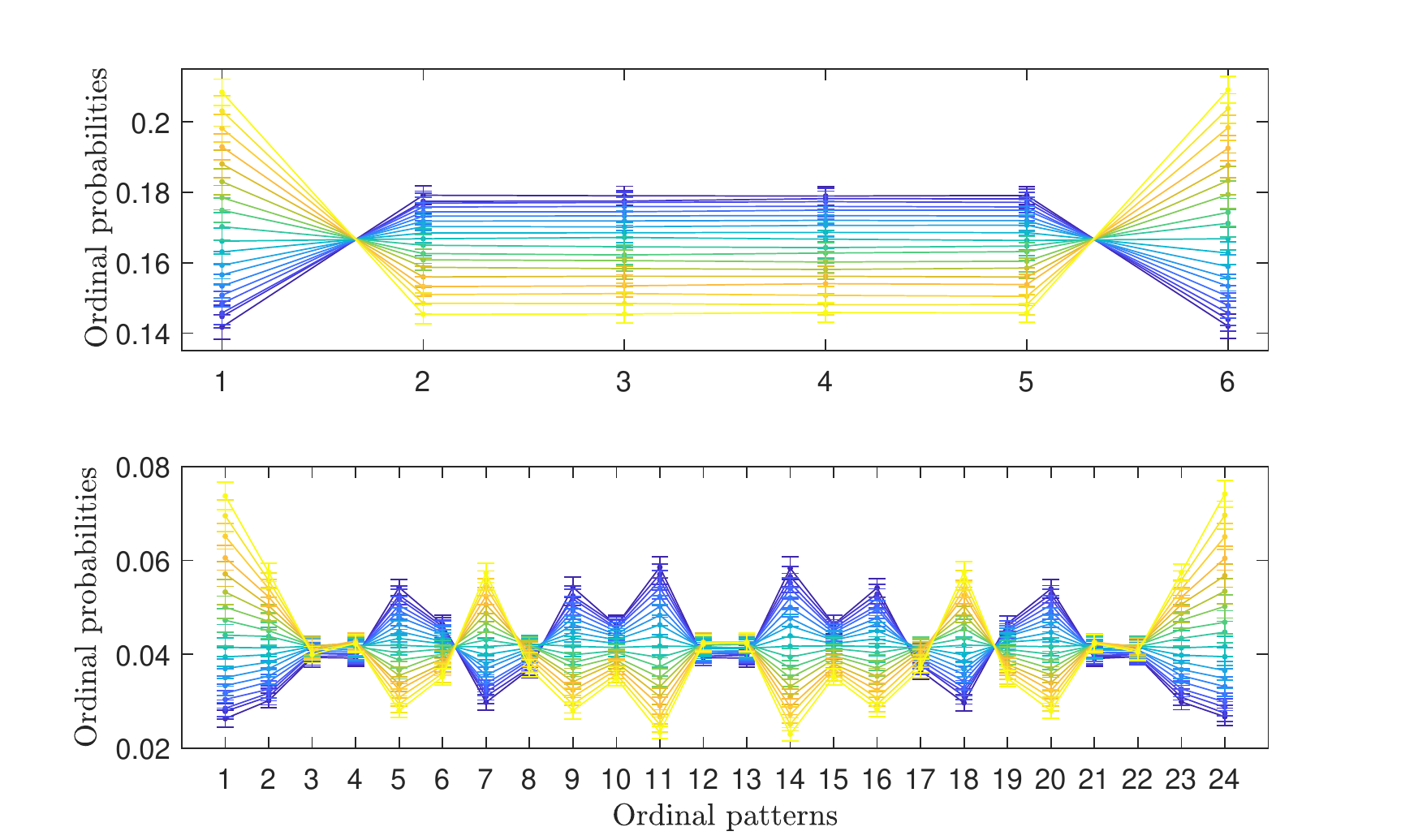}
\caption{Estimated ordinal pattern probabilities for fGns with Hurst exponents $H \in \{0.05,0.1,\dots,0.95\}$ (increasing from blue to yellow). For the sake of a better visualization, only results for $D=3$ (top plot) and $D=4$ (bottom plot), and lag $\tau=1$, are displayed. Mean and standard deviation (as error bar) from estimations of an ensemble of one hundred independent realizations of length $N=10^4$ data are depicted. Ordinal patterns are indexed following the lexicographic order (please see Fig.~2 of Ref.~\cite{parlitz2012}).}
\label{fig:OPP_fGn}
\end{figure}

\begin{figure}[!ht]
\centering
\includegraphics[width=\linewidth,trim={1.25cm .25cm 1.4cm .35cm},clip=true]{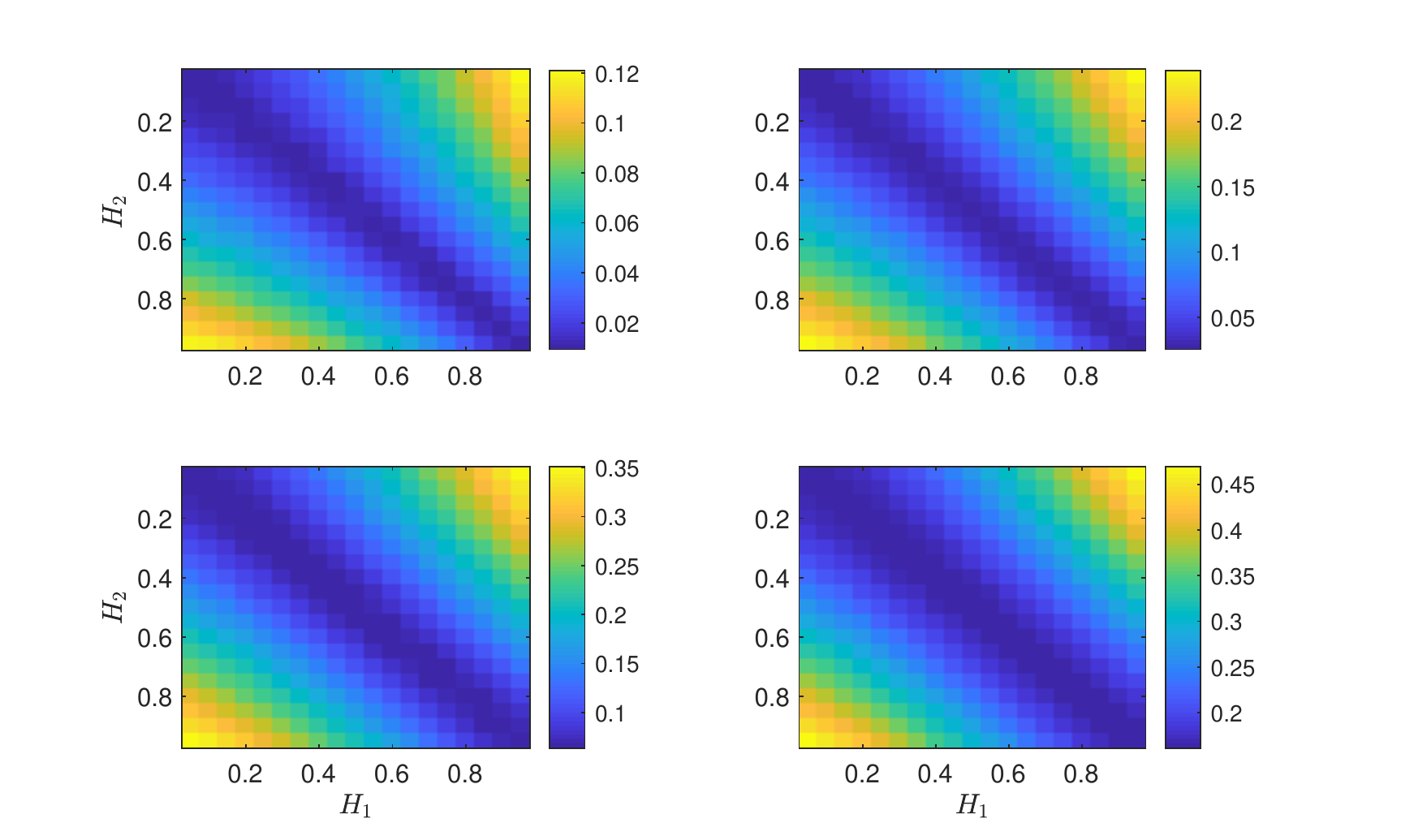}
\caption{PJSD matrices for two fGns with Hurst exponents $H_1$ and $H_2$ in the set $\{0.05,0.1,\dots,0.95\}$. Mean from estimations of an ensemble of one hundred independent realizations of length $N=10^4$ data for several orders $D \in \{3,4,5,6\}$ (increasing from top left to bottom right) and lag $\tau=1$ are depicted.}
\label{fig:Matrix_PJSD_fGn}
\end{figure}

Results previously obtained for the fGns show that persistent and antipersistent dynamics are indistinguishable when the PJSD to the shuffled surrogate realizations is implemented to characterize the underlying temporal correlations. This ambiguous behavior, also observed when the permutation entropy is used as a discriminative measure~\cite{SM2}, could be considered an important weakness, especially for classification tasks. Ordinal probability distributions for fGns with $H>1/2$ and $H<1/2$ are clearly different as you can conclude from Fig.~\ref{fig:OPP_fGn}. However, the problem lies in the fact that their distances to the equiprobable distributions (associated with the uncorrelated dynamics of the shuffled realizations) achieve similar values. Beyond this limitation, it is important to emphasize that PJSD can discriminate persistence from antipersistence. To demonstrate this, we have estimated the PJSD between two fGns with Hurst exponents $H_1$ and $H_2$ in the set $\{0.05,0.1,\dots,0.95\}$, generated by following the same conditions described in the previous section. Figure~\ref{fig:Matrix_PJSD_fGn} shows the associated distance matrices with several orders $D$ and $\tau=1$ for these synthetic fGns of length $N=10^4$ data. More precisely, mean distance from one hundred independent estimations for each pair of Hurst exponents is plotted. As expected, the minimal distances are obtained along the diagonal with values near (but not) zero due to finite-size effects. Furthermore, the distance between two fGns increases as long as their Hurst exponents are more separated from each other. Improved discrimination is achieved for longer time series sizes~\cite{SM3}. Consequently, we can confirm that the PJSD is actually an efficient tool for distinguishing persistent from antipersistent fGns dynamics and potentially useful for classifying them.

\subsection{Ordinal self-dissimilarity}
\label{subsec-osd}
\begin{figure}[!ht]
\centering
\includegraphics[width=\linewidth,trim={.8cm .1cm 1.4cm .3cm},clip=true]{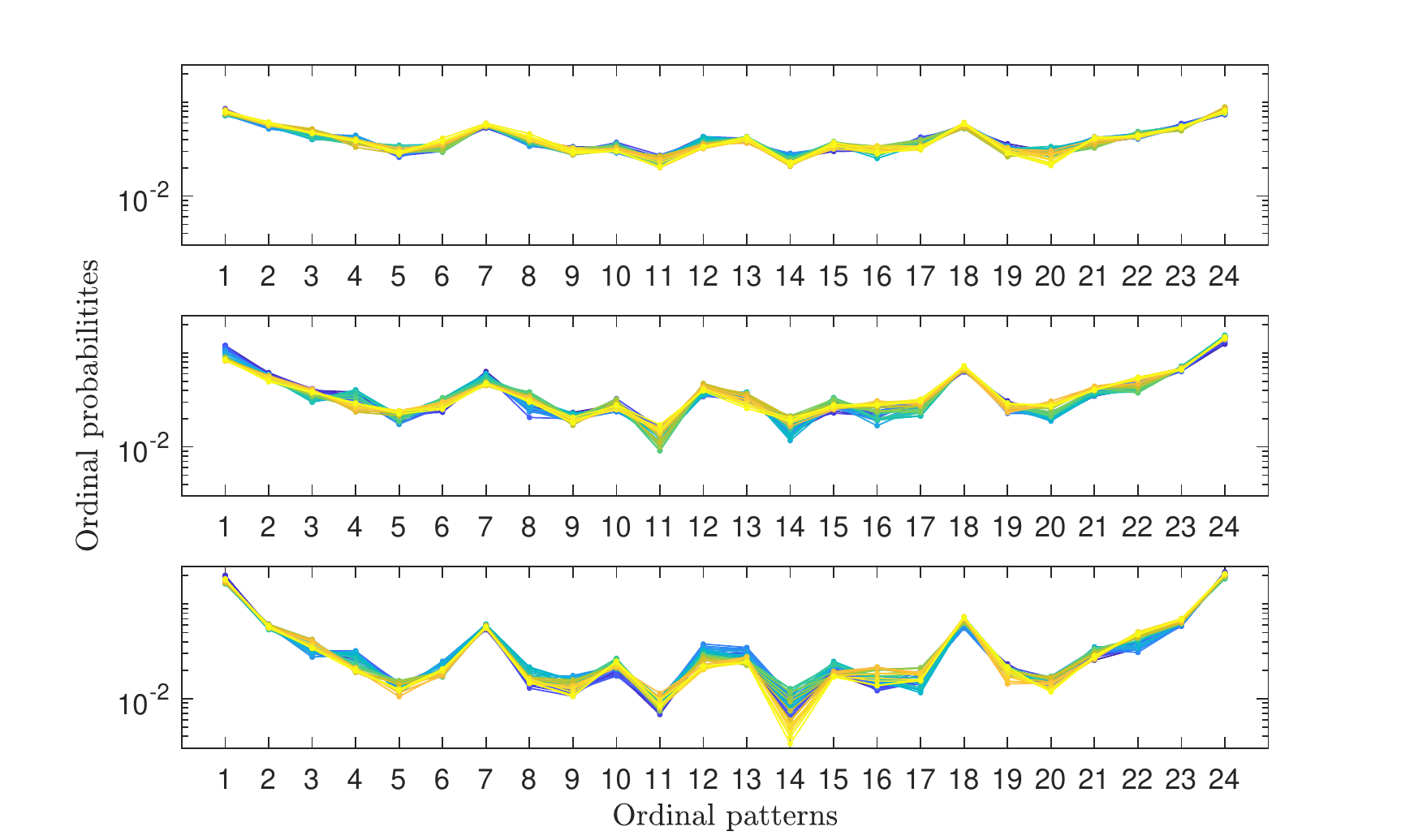}
\caption{Estimated ordinal pattern probabilities with $D=4$ and lags $1 \leq \tau \leq 40$ (increasing from blue to yellow) for one representative fBm numerical simulation of length $N=10^4$ data with Hurst exponent $H=0.25$ (top plot), $H=0.5$ (center plot) and $H=0.75$ (bottom plot). Ordinal patterns are indexed following the lexicographic order.}
\label{fig:SS_OP_fBm_semilog}
\end{figure}

\begin{figure}[!ht]
\centering
\includegraphics[width=\linewidth,trim={.8cm .15cm 1.4cm .3cm},clip=true]{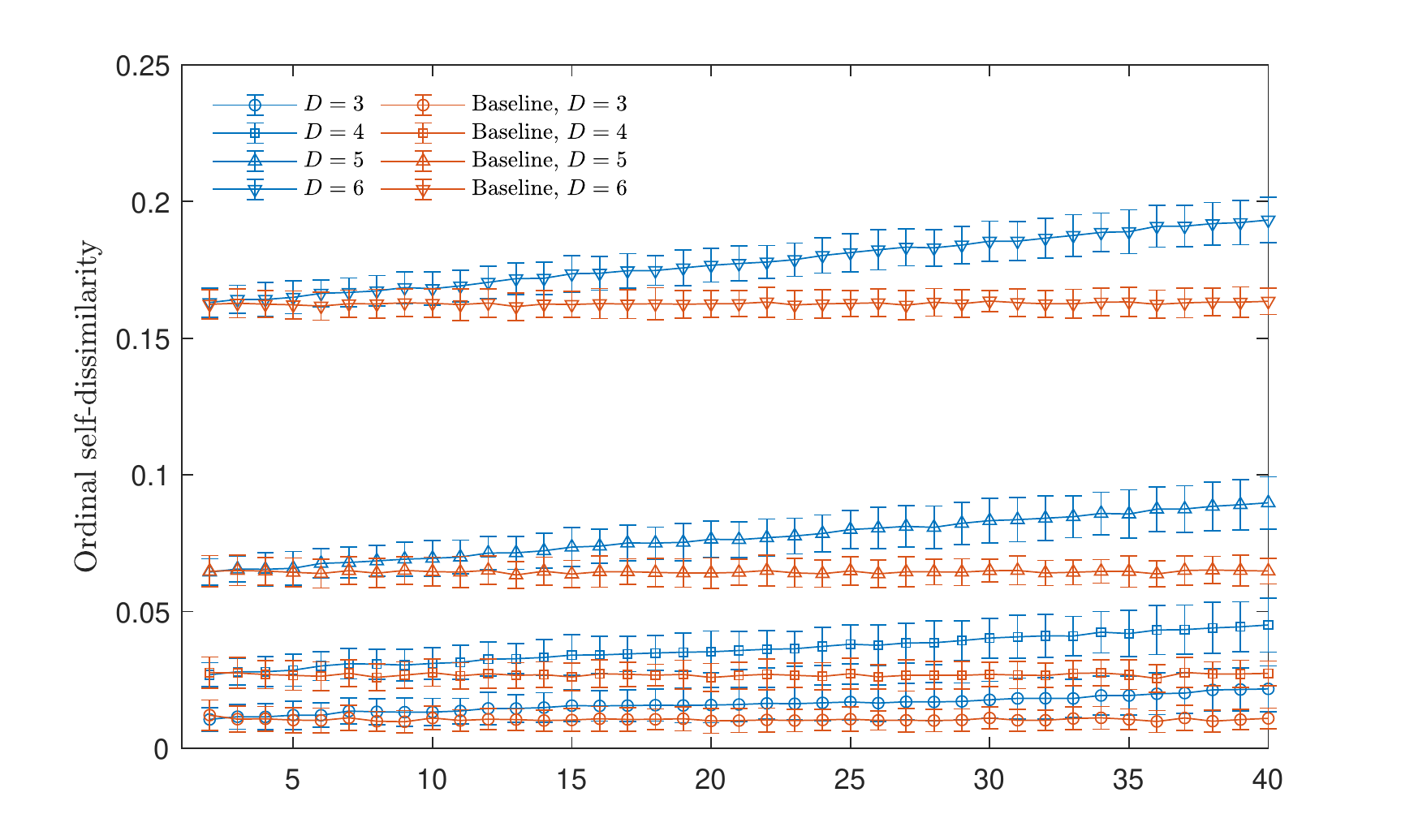}
\includegraphics[width=\linewidth,trim={.8cm .15cm 1.4cm .3cm},clip=true]{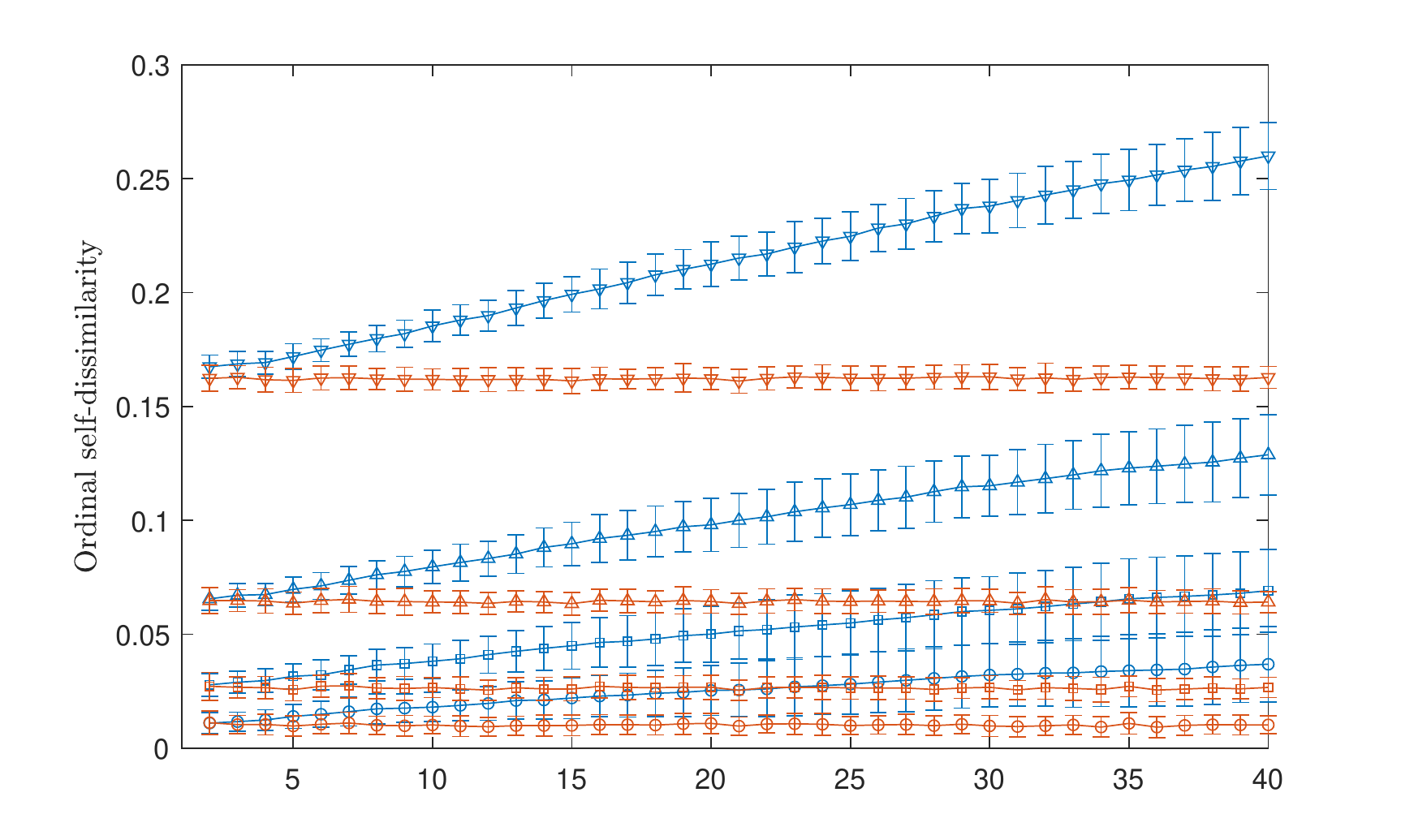}
\includegraphics[width=\linewidth,trim={.8cm .15cm 1.4cm .3cm},clip=true]{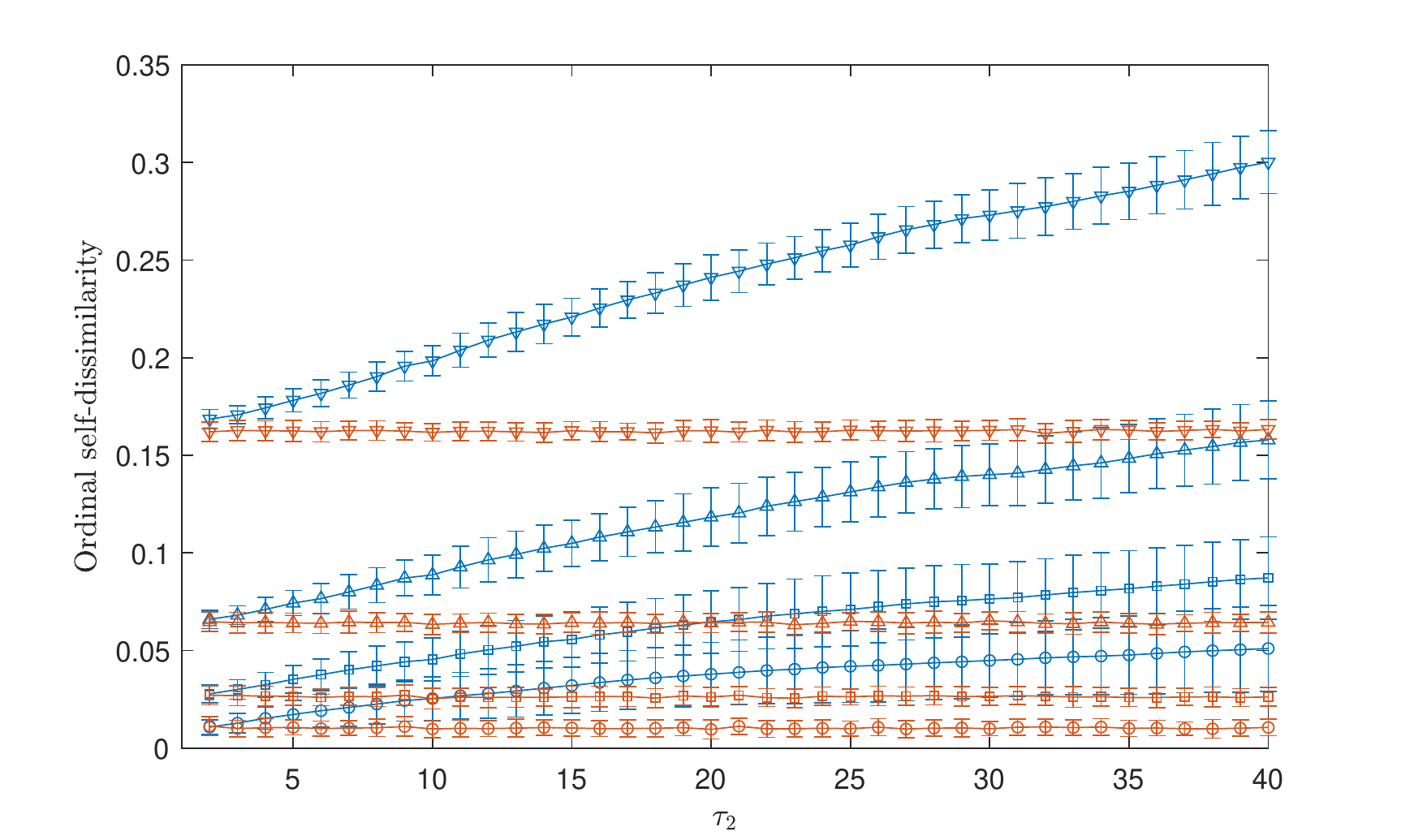}
\caption{Ordinal self-dissimilarity as a function of the observation scale $\tau_2$ with $\tau_2^{max}=40$ for fBms with Hurst exponents $H=0.25$ (top plot), $H=0.5$ (center plot) and $H=0.75$ (bottom plot). Mean and standard deviation (as error bar) from estimations of an ensemble of one hundred independent realizations of length $N=10^4$ data for several orders $D \in \{3,4,5,6\}$ are displayed. Baseline references are calculated from shuffled surrogates of the original simulations.}
\label{fig:SS_fBm}
\end{figure}

Our analysis for the fBms developed in Sec.~\ref{subsec-colored-noises} shows that results are invariant under changes of the lag $\tau$ (at least for small values of this parameter). The self-similar property of these processes analytically justifies this finding: fBms look qualitatively similar independently of the distance from which we look at them~\cite[see particularly page 48 for a more formal definition of self-similarity]{beran1994}. Indeed, it has been theoretically shown that the ordinal patterns probabilities do not depend on the lag $\tau$ for fBms~\cite{bandt2007}. Following the definition introduced by Bandt~\cite{bandt2020}, fBms fulfill the \textit{order self-similarity} property. For illustrative purpose, we show in Fig.~\ref{fig:SS_OP_fBm_semilog} the estimated ordinal pattern probabilities with $D=4$ and lags $\tau$ between 1 and 40 for fBms with three Hurst exponents $H \in \{0.25,0.5,0.75\}$. Since numerical realizations of length $N=10^4$ data are analyzed, the theoretical order self-similarity is only approximately verified and some discrepancies are visually evident when the lag $\tau$ increases from 1 (blue curve) to 40 (yellow curve). Keeping this in mind, we propose to estimate the PJSD of an arbitrary time series symbolized with an order $D$ and two different lags $\tau_1$ and $\tau_2$ to quantify the degree of self-dissimilarity of the time series itself at these two temporal scales. By fixing $\tau_1=1$ and varying $\tau_2$ from $2$ to $\tau_2^{max}$, this \textit{ordinal self-dissimilarity} seeks to characterize how different, from an ordinal perspective, the time series looks when the scale is increased with respect to the initial scale used to examine it. Smaller estimated values of the ordinal self-dissimilarities, close to the baseline reference, are expected in the case of self-similar dynamics, such as that associated with fBms processes, while larger values will be obtained for systems whose dynamics depend on the observation scale. Moreover, it is reasonable to conjecture that intrinsic characteristic scales of the dynamics that generates the time series, such as temporal delays and/or periodicities, could be revealed through this analysis. With the aim of illustrating this approach, we have tested fBms numerical realizations. Figure~\ref{fig:SS_fBm} shows the ordinal self-dissimilarity with different orders $D$ as a function of the observation scale $\tau_2$ with $\tau_2^{max}=40$ for fBm simulations of length $N=10^4$ data with the same three Hurst exponents previously considered. Once again, average and standard deviation from an ensemble of one hundred independent realizations have been plotted. Baseline references are calculated repeating the procedure for shuffled surrogates of the original simulations. The expected invariance is only verified for small scales. Moreover, the ordinal self-dissimilarity moves further away from the baseline reference for larger Hurst exponent and larger orders $D$. It is also confirmed that the behavior is qualitatively similar but with smaller values of the distances if longer numerical realizations are considered~\cite{SM4}. Even though ordinal pattern probabilities are a priori similar (on average) for different lags $\tau$, the ordinal self-dissimilarity seems to be a sensitive tool able to quantify and magnify their subtle differences.

\subsection{Distances between different complex deterministic dynamics}
\label{subsec-different-dynamics}
\begin{figure*}[!ht]
\centering
\includegraphics[width=.4\textwidth,trim={0cm .3cm 1.6cm .3cm},clip=true]{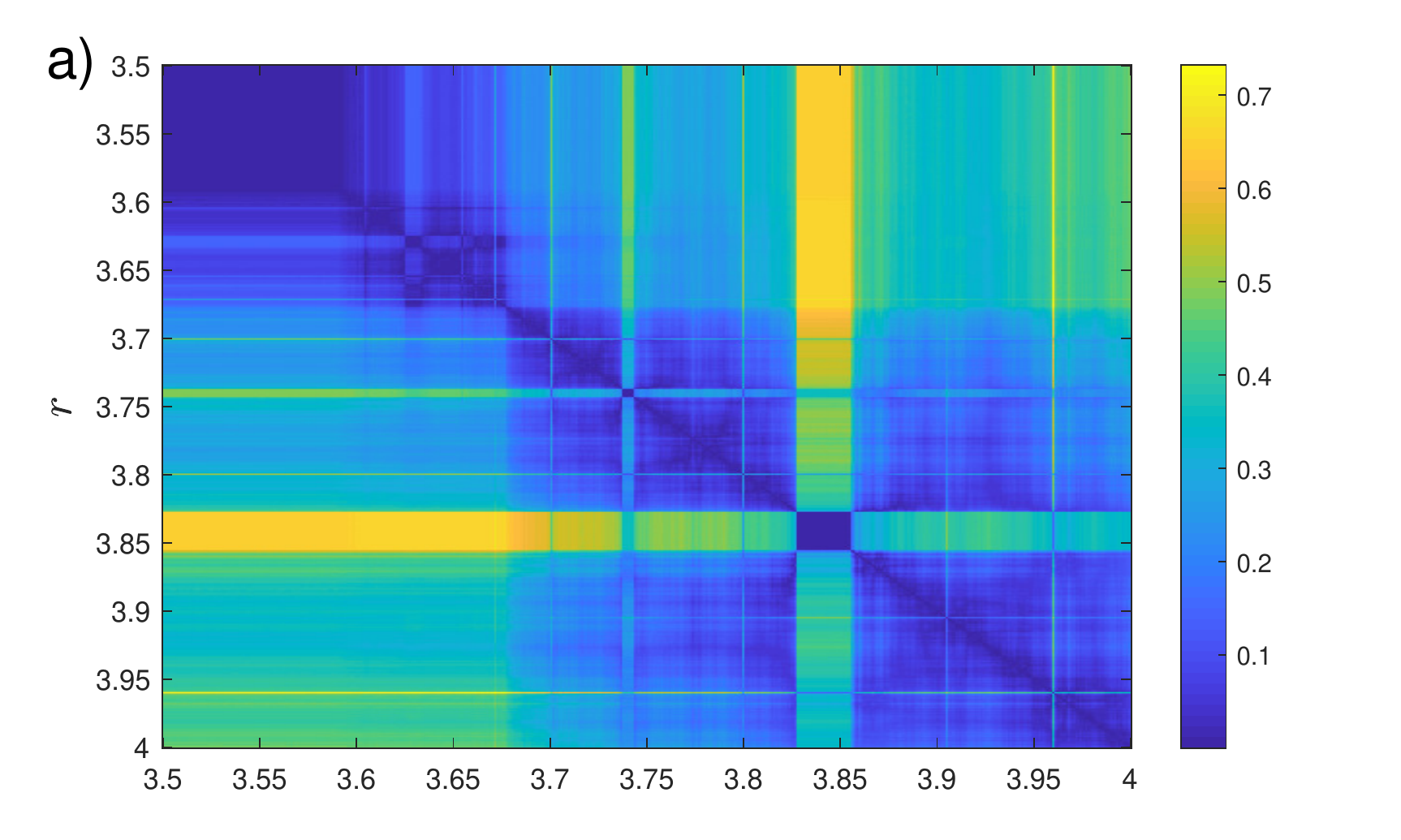}
\includegraphics[width=.4\textwidth,trim={0cm .3cm 1.6cm .3cm},clip=true]{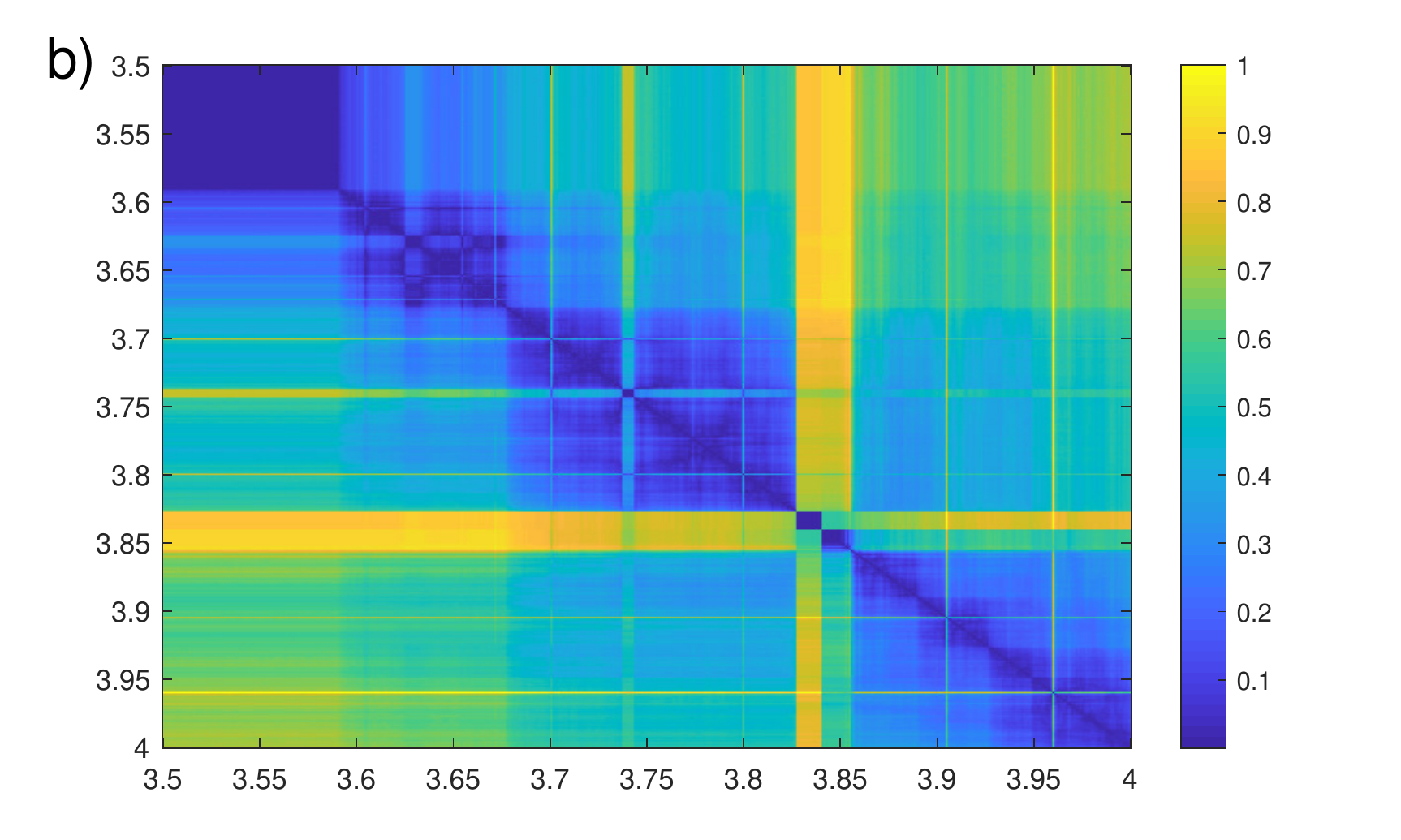}
\\
\includegraphics[width=.4\textwidth,trim={0cm .3cm 1.6cm .3cm},clip=true]{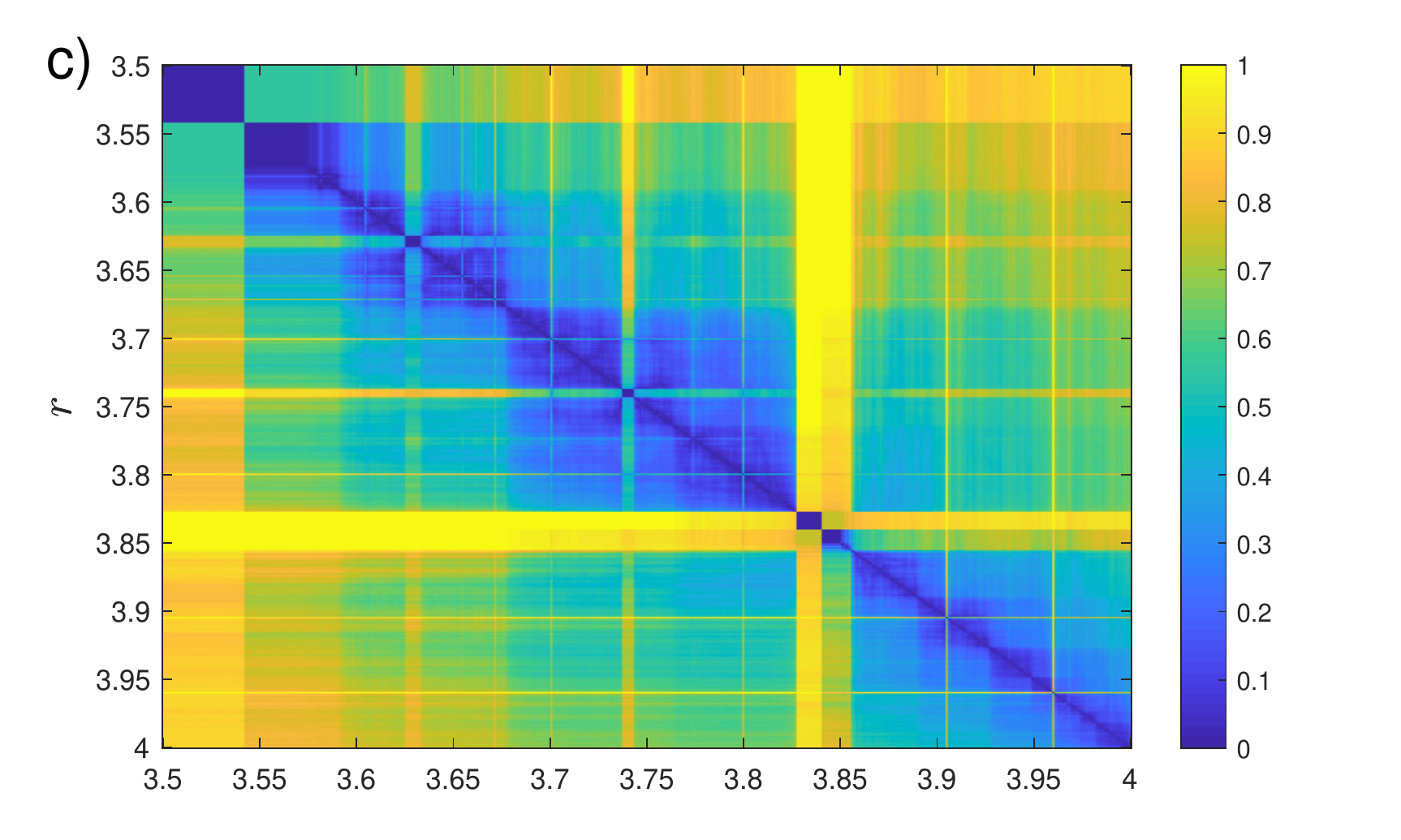}
\includegraphics[width=.4\textwidth,trim={0cm .3cm 1.6cm .3cm},clip=true]{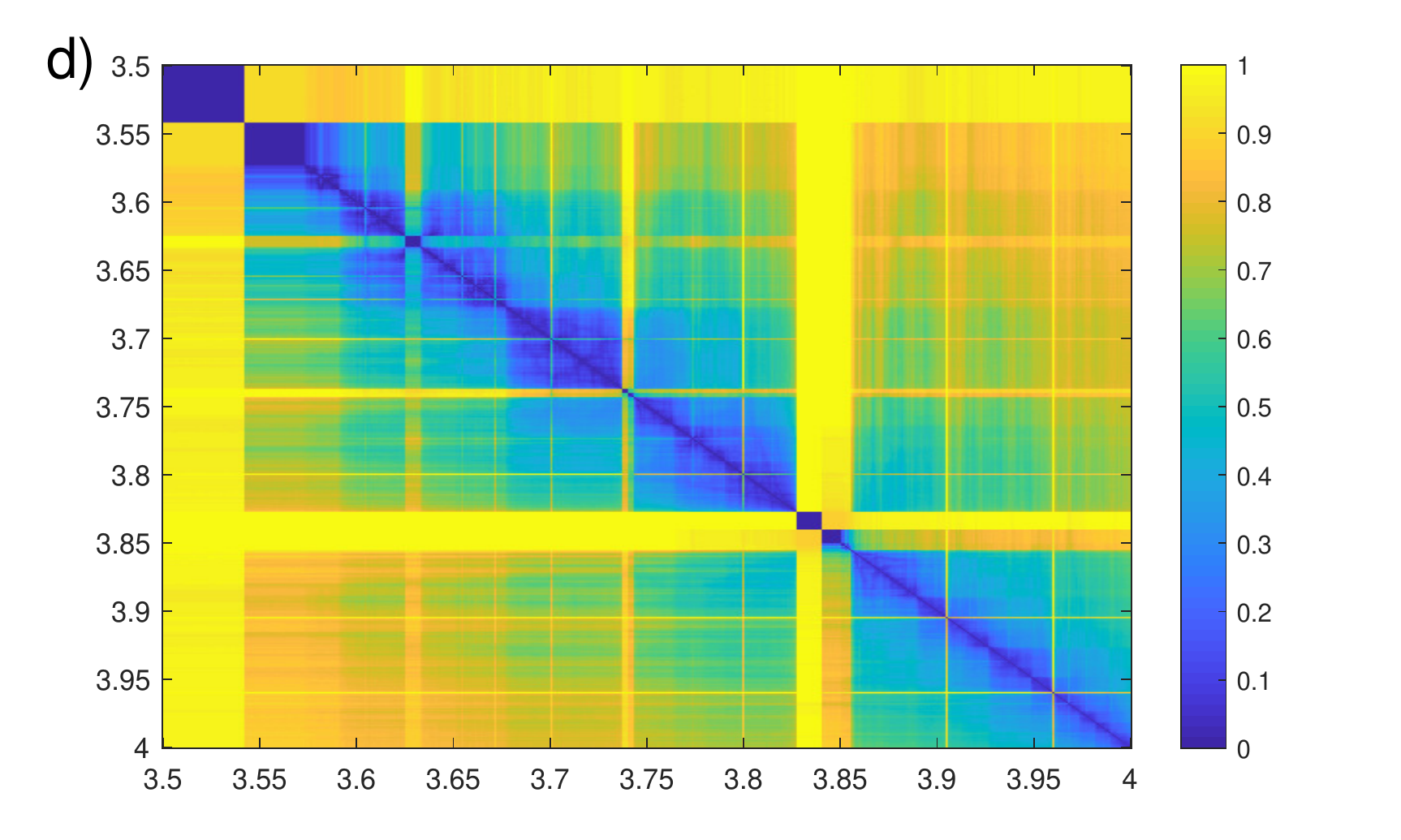}
\\
\includegraphics[width=.4\textwidth,trim={0cm .2cm 1.6cm .3cm},clip=true]{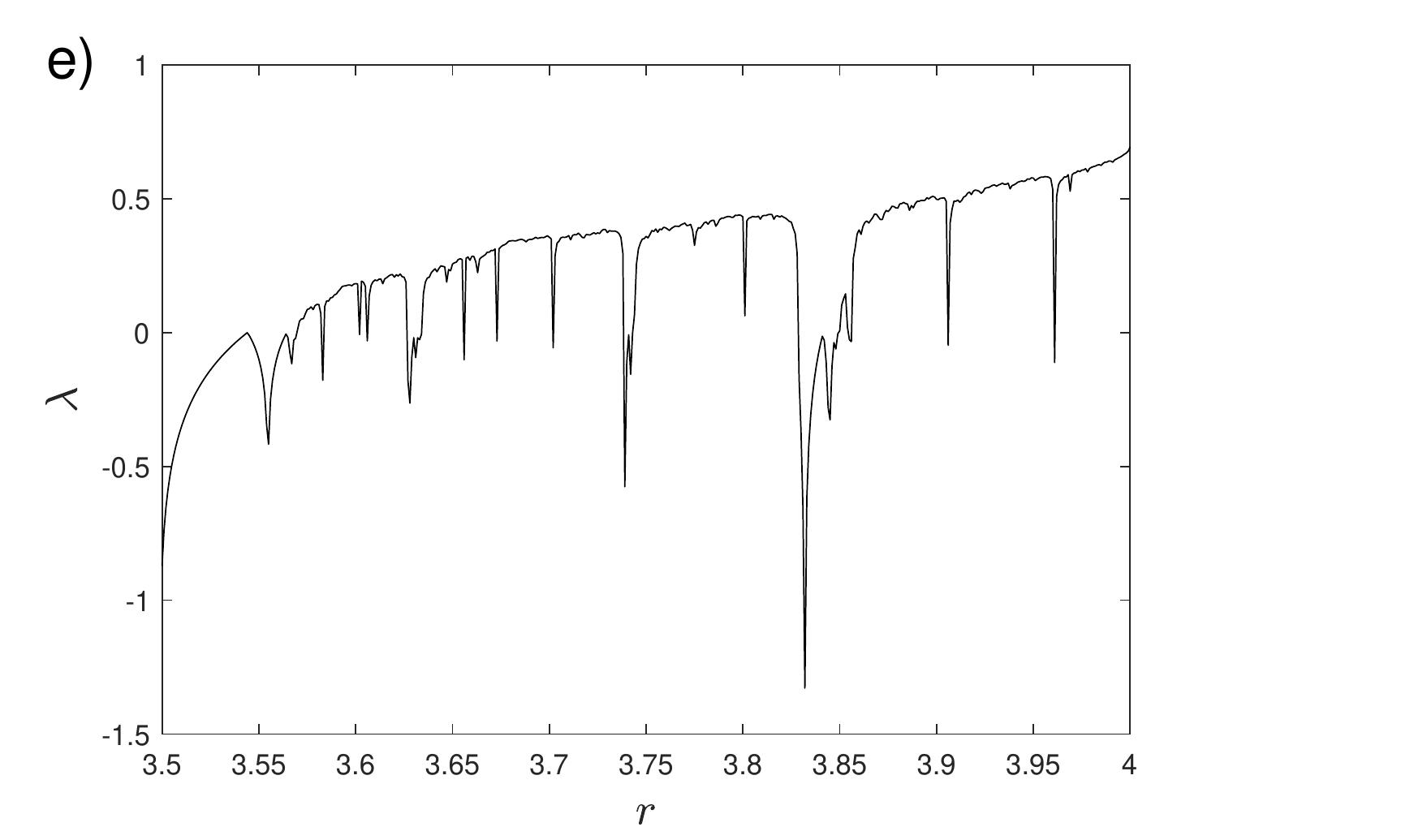}
\includegraphics[width=.4\textwidth,trim={0cm .2cm 1.6cm .3cm},clip=true]{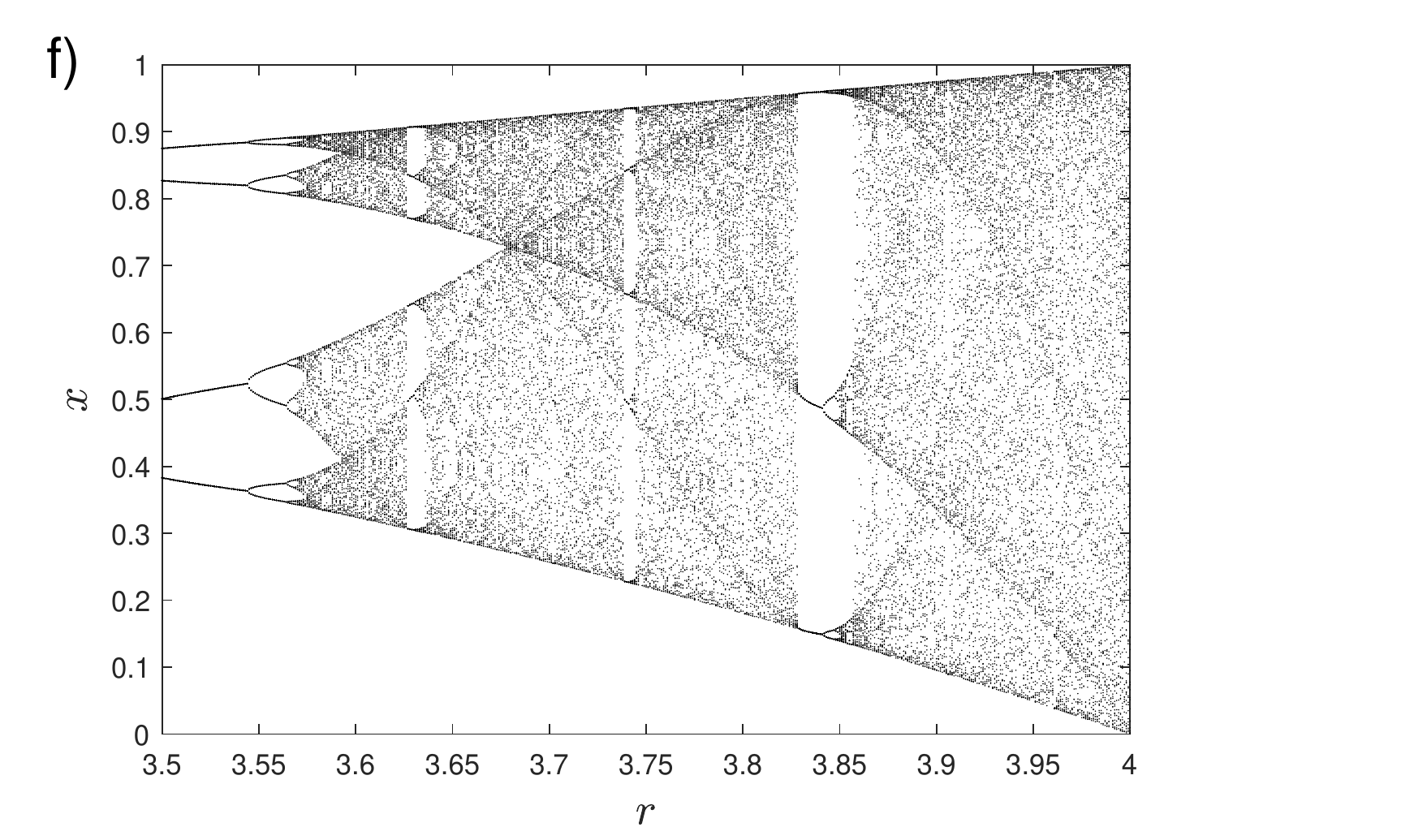}
\caption{a-d) PJSD matrices estimated with $D \in \{3,4,5,6\}$ textcolor{blue}{(increasing from top left to bottom right)} and lag $\tau=1$ for the logistic map with parameter $r$ between 3.5 and 4 and step $\Delta r=0.001$. Average values from estimations of an ensemble of twenty independent realizations of length $N=10^4$ data for each pair of values of the parameter $r$ are shown. e) Lyapunov exponents and f) bifurcation diagram of the logistic map for this parameter range are also included for comparison purposes.}
\label{fig:Logistic_analysis}
\end{figure*}

Within the proposed approach, we conjecture that the PJSD estimated between signals coming from different dynamical origins will achieve values significantly larger than the baseline or reference values obtained when two signals from the same system are compared. The rich dynamical behavior of the logistic map is considered as a testbed to illustrate this fact. Realizations of this one-dimensional quadratic map $x_{t+1}=rx_t(1-x_t)$ for values of the parameter $r$ between 3.5 and 4 with step $\Delta r=0.001$ (giving a total of 501 values of $r$) are mutually compared through the PJSD. It is well-known that different deterministic dynamics are generated within this range of the parameter value, from simple periodic to fully chaotic regimes. Average values of the PJSD matrices estimated with several orders $D$ and lag $\tau=1$ from an ensemble of twenty independent realizations of length $N=10^4$ data for each pair of values of the parameter $r$ are displayed in Fig.~\ref{fig:Logistic_analysis}. As was expected, minima of the PJSD near zero are observed along the diagonal. It is also concluded that the discrimination between different dynamics improves for larger orders $D$. This is especially evident in the periodic windows, in agreement with results previously obtained by Parlitz~\textit{et al.}~\cite{parlitz2013}. Thus, as a rule of thumb, the largest possible order $D$ that satisfies the condition $N \gg D!$ should be chosen to maximize the accuracy for discriminating between different dynamics. Please see Appendix to compare with results obtained by implementing other dissimilarity measures.

\subsection{Unveiling noise-induced effects}
\label{subsec-noise-effects}
\begin{figure}[!ht]
\centering
\includegraphics[width=\linewidth,trim={1.25cm .3cm 1.5cm .35cm},clip=true]{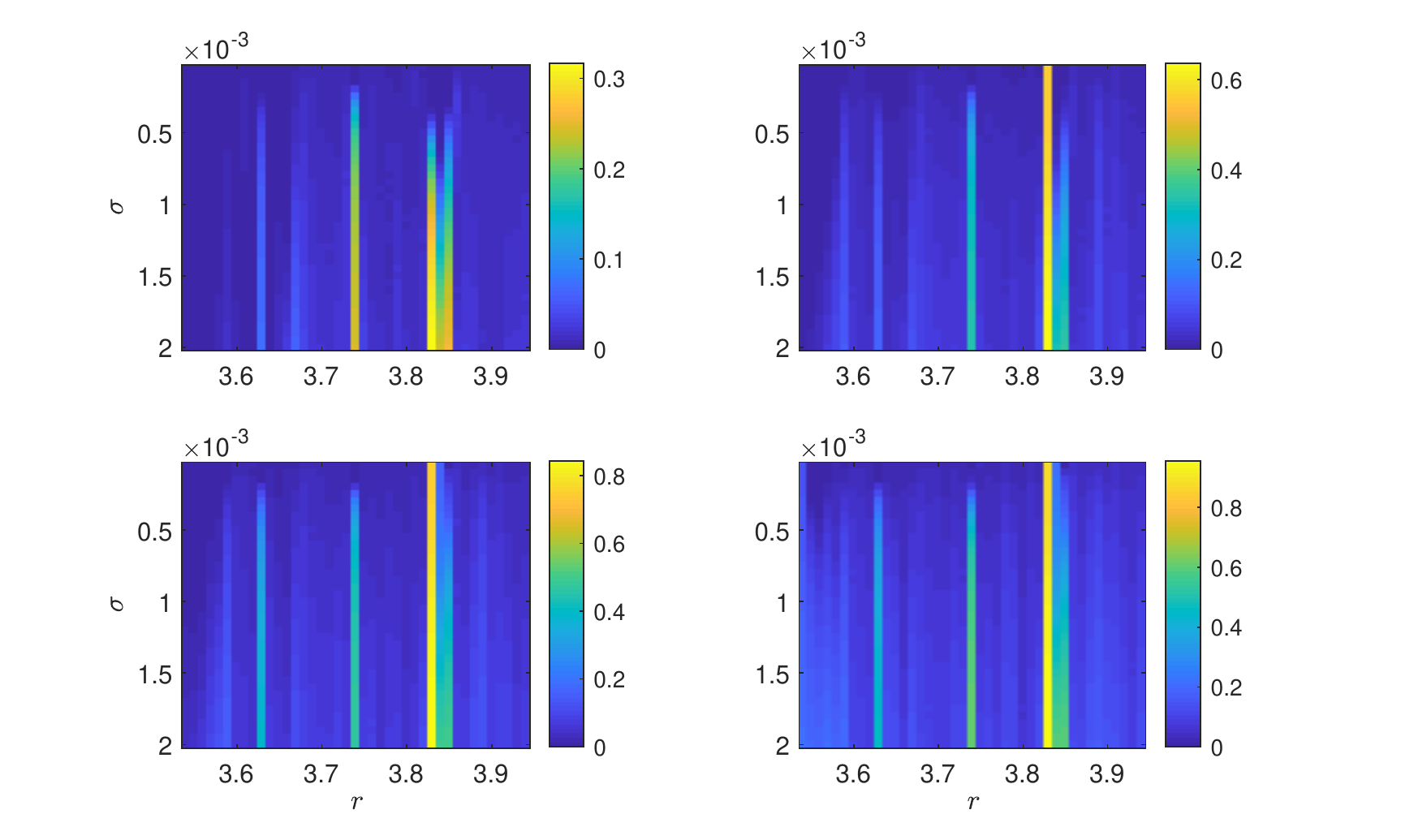}
\caption{PJSD matrices estimated with $D \in \{3,4,5,6\}$ (increasing from top left to bottom right) and lag $\tau=1$ for the noisy logistic map with parameter $r$ between 3.54 and 3.94 with step $\Delta r=0.01$ and noise levels $\sigma$ between 0.0005 and 0.002 with step $\Delta \sigma=0.0005$. Average values of PJSD estimations between an ensemble of one hundred independent noisy realizations of length $N=10^4$ data (for each value of $r$ and $\sigma$) and their noise-free counterparts are shown. PJSD reaches larger values for all orders $D$ when the bifurcation parameter $r \in \{3.63,3.74,3.83,3.84,3.85\}$ and the noise level $\sigma$ increases.} 
\label{fig:Noisy_logistic_analysis}
\end{figure}

It is well-known that the presence of dynamical noise in nonlinear dynamical systems can lead to several interesting phenomena such as stochastic resonance, noise-induced order and noise-induced chaos~\cite{gao1999}. Within this context, it is critical to appropriately identify the parameters of the nonlinear systems and the noise levels for which these phenomena are realized. To illustrate that the PJSD is useful for this purpose, we have analyzed the paradigmatic noisy logistic map defined by
\begin{equation}
x_{t+1}=rx_t(1-x_t)+\mu_t\,,
\end{equation}
with $\mu_t$ a Gaussian random variable with zero mean and standard deviation $\sigma$. As it was done in Ref.~\cite{gao1999}, the $\sigma$ value is interpreted as the noise level. Realizations of this noisy map for values of the bifurcation parameter $r$ between 3.54 and 3.94 with step $\Delta r=0.01$ (giving a total of 41 values of $r$) and noise levels $\sigma$ between 0.0005 and 0.002 with step $\Delta \sigma=0.0005$ (giving a total of 40 values of $\sigma$) are \textit{ordinally} contrasted against their noise-free counterparts via the PJSD. More precisely, average values of the PJSD with different orders $D$ and lag $\tau=1$ between one hundred independent noisy realizations of length $N=10^4$ data (for each value of $r$ and $\sigma$) and the noise-free realization with the same value of the bifurcation parameter are calculated. Results obtained are displayed in Fig.~\ref{fig:Noisy_logistic_analysis}. 

It is observed that the PJSD increases abruptly for some particular pair of $r$ and $\sigma$ values independently of the order D. In other words, noisy and original (noise-free) dynamics are significantly different from an ordinal perspective, and, consequently, noise-induced relevant effects on the original clean dynamics are concluded for such instances. Actually, noise-induced chaos has already been confirmed for some of the bifurcation parameter values our analysis highlights, namely $r=3.63$, $r=3.74$ and $r=3.83$~\cite{gao1999}. Although further research is needed to confirm the chaotic nature of the noisy realizations, we conjecture that transition from periodic to chaotic states due to the presence of dynamical noise is the main source of the observed behavior.

\subsection{Discriminating reversible from irreversible time series}
\label{subsec-reversibility}
The possibility to discriminate between reversible and irreversible dynamics from the analysis of time series generated by the underlying generating systems is relevant for modelling purposes. The irreversible nature of the time series confirms that Gaussian linear processes and nonlinear static transformations of them should be excluded as potential models, and this allows to shed some light on the physical mechanisms that govern the system under study. In other words, any difference between the actual (forward) time series and its time-reversed (backward) counterpart is a symptom of the presence of nonlinearities in the process that generated the observed time series. Despite the fact that several methods have been introduced to quantify the degree of time irreversibility in practice, there is not an a priori optimal approach~\cite[and references therein]{lacasa2012}. Motivated by this issue, we propose here that the amount of time irreversibility can be measured as the normalized distance (in a distributional sense) between the ordinal probability distributions estimated from the forward and backward series, \textit{i.e.} $[D_{JS}(P_{for},P_{back})/\ln 2]^{1/2}$ with $P_{for}$ and $P_{back}$ the time forward and time reversed ordinal pattern probability distributions, respectively. It is important to emphasize that the idea of implementing ordinal related tools as a way of assessing the irreversibility of a time series is not new and has been recently explored by Zanin~\textit{et al.}~\cite{zanin2018} and Mart\'inez~\textit{et al.}~\cite{martinez2018}, and that we are simply trying to give a step forward in this direction.

\begin{figure}[!ht]
\centering
\includegraphics[width=\linewidth,trim={1cm .25cm 1.4cm .35cm},clip=true]{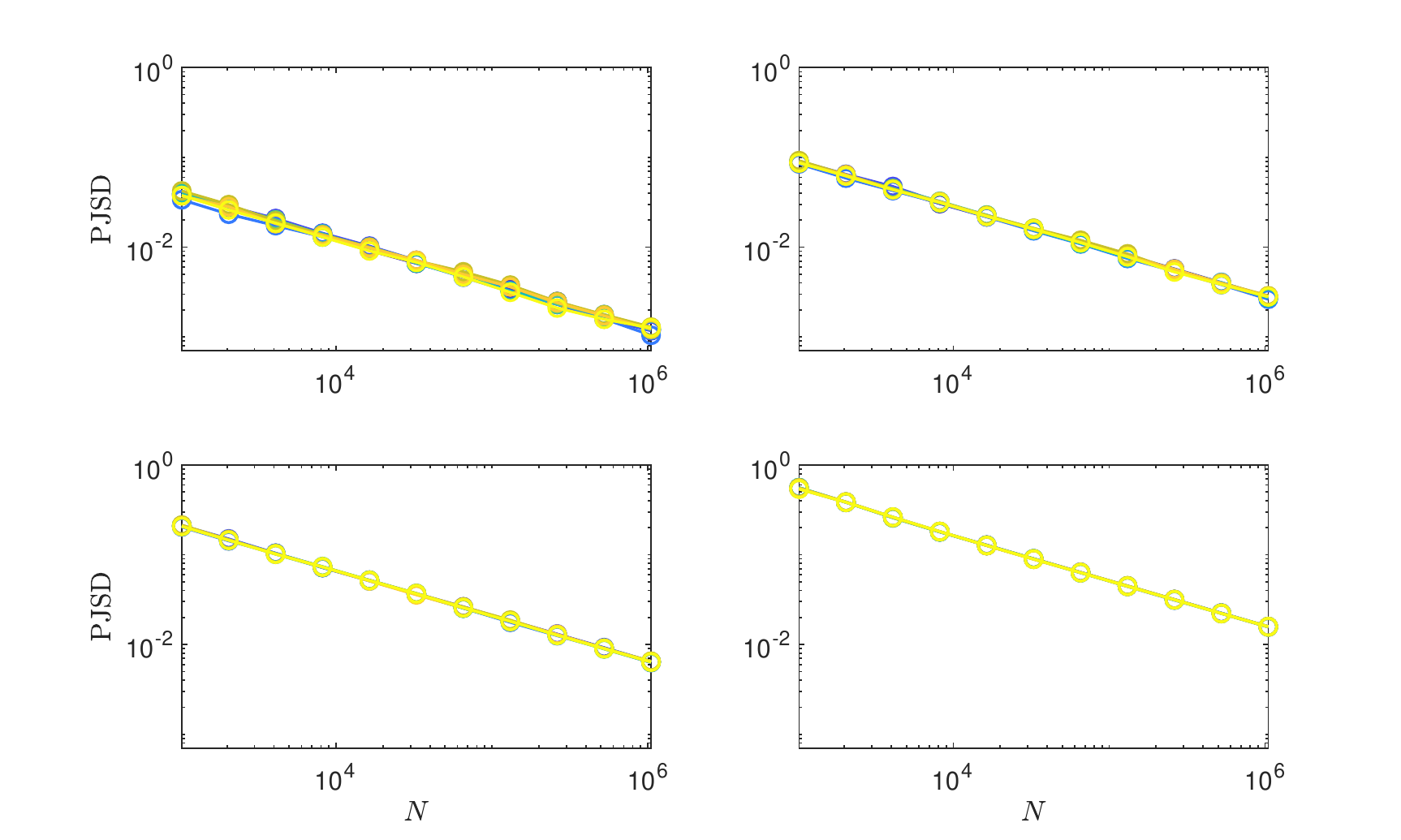}
\caption{Irreversibility analysis for fGns with several orders $D \in \{3,4,5,6\}$ (increasing from top left to bottom right) and $\tau=1$. Log-log plots of the average PJSD between the forward and backward series from one hundred independent realizations for each Hurst exponent $H \in \{0.1,0.2,\dots,0.9\}$ (increasing from blue to yellow) as a function of the time series length $N$. It should be noted that as $D$ increases the dispersion between the nine curves reduces and they are totally overlapped for $D=5$ and $D=6$.}
\label{fig:Irreversibility_fGn}
\end{figure}

In the following, several numerical analyses on theoretically validated reversible and irreversible stochastic and chaotic dynamics have been developed to characterize the performance of the proposed approach. In Fig.~\ref{fig:Irreversibility_fGn}, we have plotted, in log-log scale, the PJSD between the forward and backward series for simulated fGns with Hurst exponents $H \in \{0.1,0.2,\dots,0.9\}$ as a function of the realization length $N$ ($N \in \{2^{10},2^{11},\dots,2^{20}\}$). Average PJSD estimated with orders $D \in \{3,4,5,6\}$ and lag $\tau=1$ from an ensemble of one hundred independent numerical simulations for each value of $H$ are displayed. It is concluded that the PJSD converges asymptotically to zero when $N$ increases with the same power-law behavior $\text{PJSD} \propto N^{-1/2}$ observed in Sec.~\ref{subsec-zero-convergence}. This result is independent of the Hurst exponent $H$ (that increases from the blue to the yellow curves) and with a proportionality factor that depends on the order $D$. Consequently, the observed deviations from zero are purely related to finite-size effects, and a reversible dynamics is concluded, as expected, for all these linearly correlated stochastic noises. We have found qualitatively equivalent findings for Gaussian and uniform white noises~\cite{SM5}.

\begin{figure}[!ht]
\centering
\includegraphics[width=\linewidth,trim={.8cm .15cm 1.4cm .3cm},clip=true]{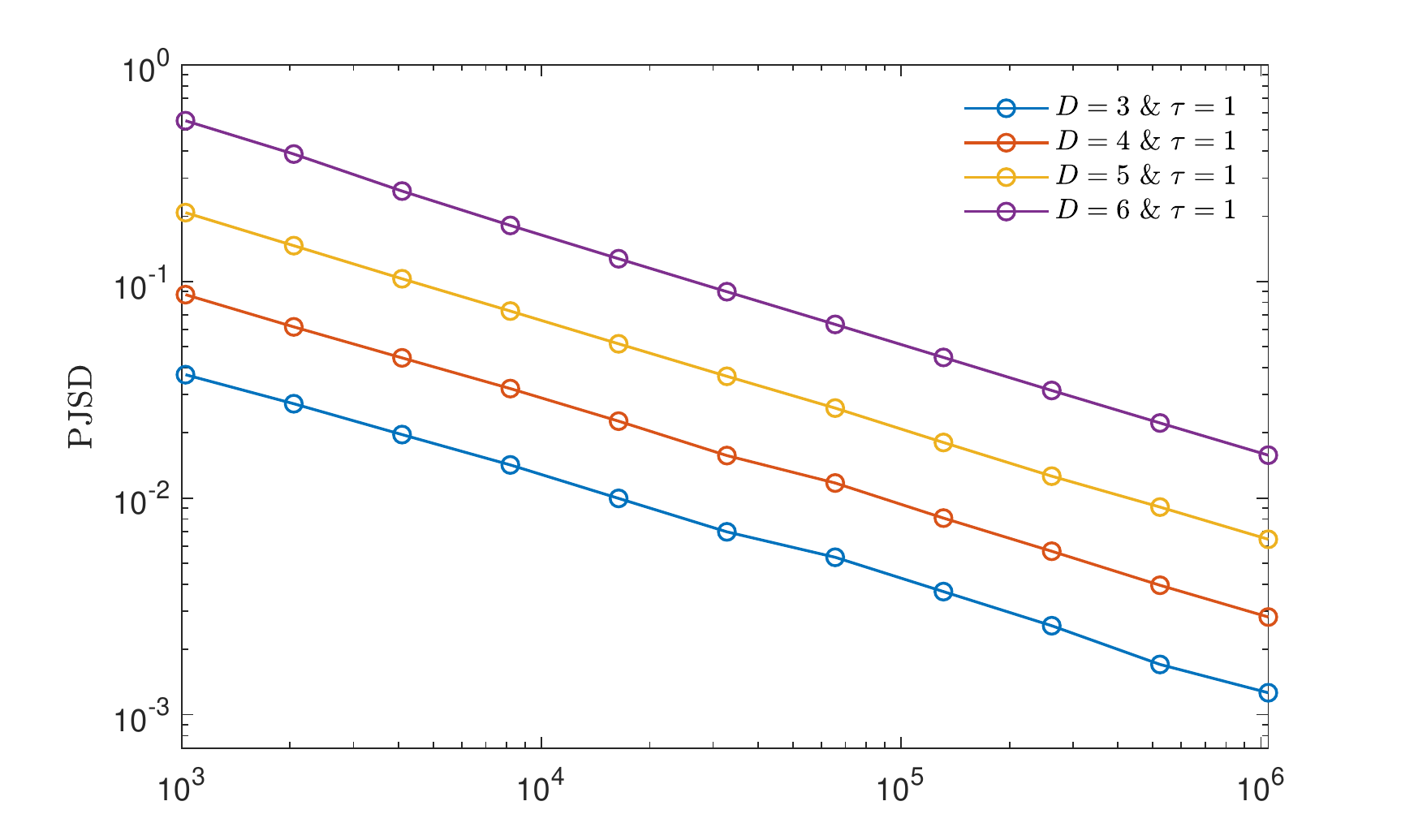}
\includegraphics[width=\linewidth,trim={.8cm .15cm 1.4cm .3cm},clip=true]{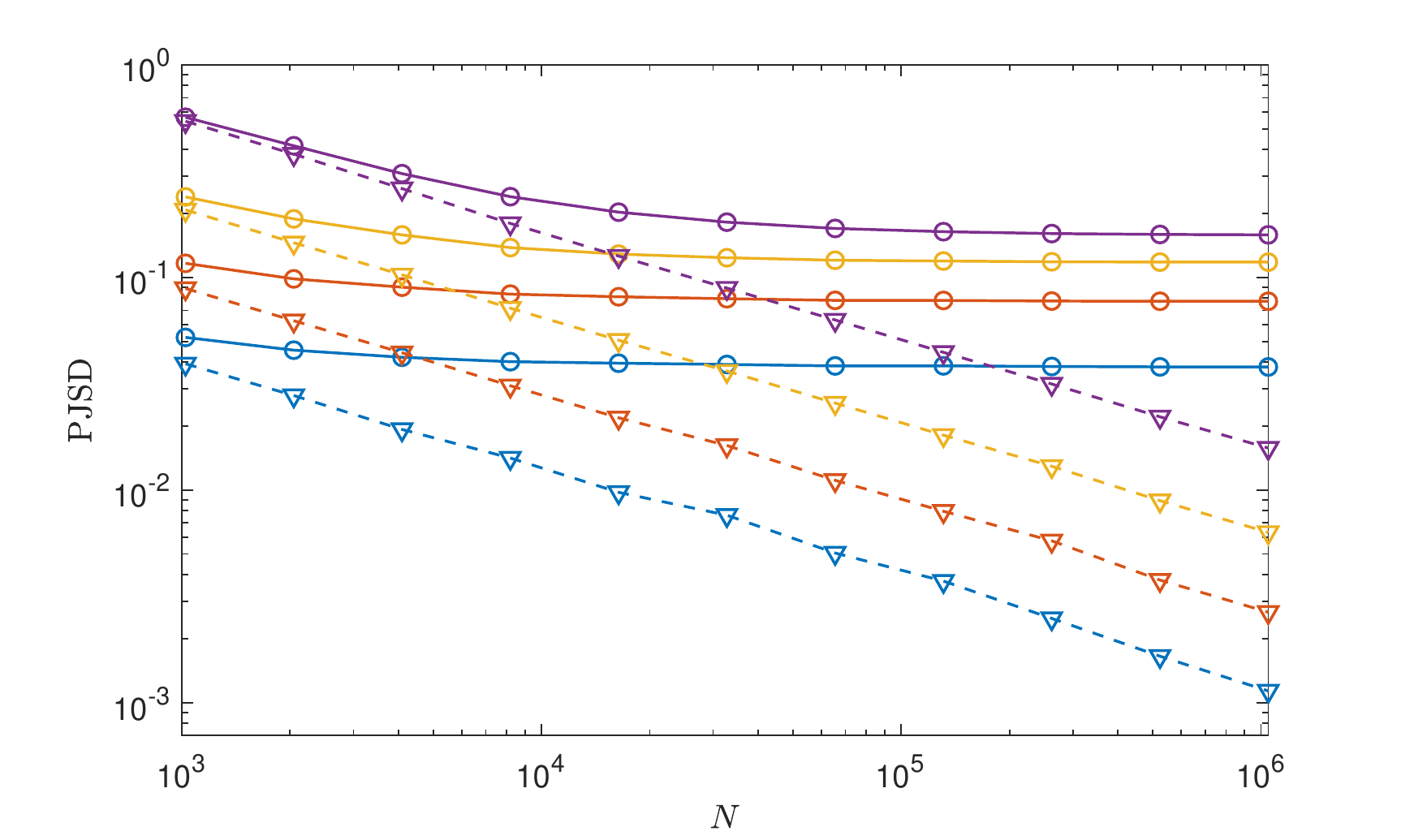}
\caption{Irreversibility analysis for the STAR (top plot) and NGRP (bottom plot) models. The PJSD between the forward and backward series estimated with several orders $D$ and $\tau=1$ is plotted (in log-log scale) as a function of the time series length $N$. Average values from one hundred independent realizations are displayed. Behaviors displayed by the standard Gaussian AR(1) process (dashed lines with triangle markers) have also been included for comparison purposes in the bottom plot.}
\label{fig:Irreversibility_stochastic}
\end{figure}

Next, we consider the irreversibility analysis of two stochastic systems: a static nonlinear transformation of a Gaussian process (commonly known with the acronym STAR) and a linear autoregressive model driven by a non-Gaussian noise (NGRP). More precisely, the STAR is given by $x_{t}=\tanh^2{(y_{t})}$ where $y_{t}$ is the first-order autoregressive AR(1) process $y_{t}=0.6y_{t-1}+\epsilon_t$ with $\epsilon_t$ pseudorandom values from the standard normal distribution while the NGRP is defined by $x_{t}=0.3x_{t-1}+\xi_t$ where $\xi_t$ are pseudorandom numbers uniformly distributed in the interval $(-0.5,0.5)$~\cite{diks1995}. These are examples of stochastic systems with static and dynamic nonlinearities and, consequently, they are time reversible and irreversible, respectively~\cite{diks1995,martinez2018}. We have developed an analysis similar to that previously described for the fGns. The average PJSD between the forward and backward series from an ensemble of one hundred independent realizations is depicted in Fig.~\ref{fig:Irreversibility_stochastic} as a function of the time series length $N$ in log-log scale. Ordinal distances are estimated with $D \in \{3,4,5,6\}$ and lag $\tau=1$. Results for the STAR model are displayed in the top plot. The PJSD tends asymptotically to zero allowing to conclude a reversible dynamics, as it was expected, for this case. Behaviors of the ordinal approach for the NGRP model are detailed in the bottom plot. We have also included the results for the standard Gaussian AR(1) counterpart as a reference (please see the dashed lines with triangle markers). An irreversible dynamics is concluded for the NGRP model since the PJSD converges to non-zero values for the different order values. Qualitatively similar results have also been found for another linear system excited by a non-Gaussian noise. It is the third-order autoregressive process $x_{t}=0.4x_{t-3}-0.3x_{t-2}+0.2x_{t-1}+\zeta_t$ with $\zeta_t$ the squaring of a white process with uniform amplitude distribution between -0.5 and 0.5~\cite{diks1995}. For a matter of space, the results for this irreversible system are detailed in the Supplemental Material~\cite{SM6}. We can conclude that the robustness of the proposed ordinal approach to distinguish between reversible and irreversible time series is confirmed in these more challenging stochastic models. Moreover, we have found that static and dynamic nonlinearities can be successfully dealt with. We have found that other dissimilarity measures have problems to unveil the irreversible nature of the NGRP model. Please see Appendix for more details.

\begin{figure}[!ht]
\centering
\includegraphics[width=\linewidth,trim={.8cm .15cm 1.4cm .3cm},clip=true]{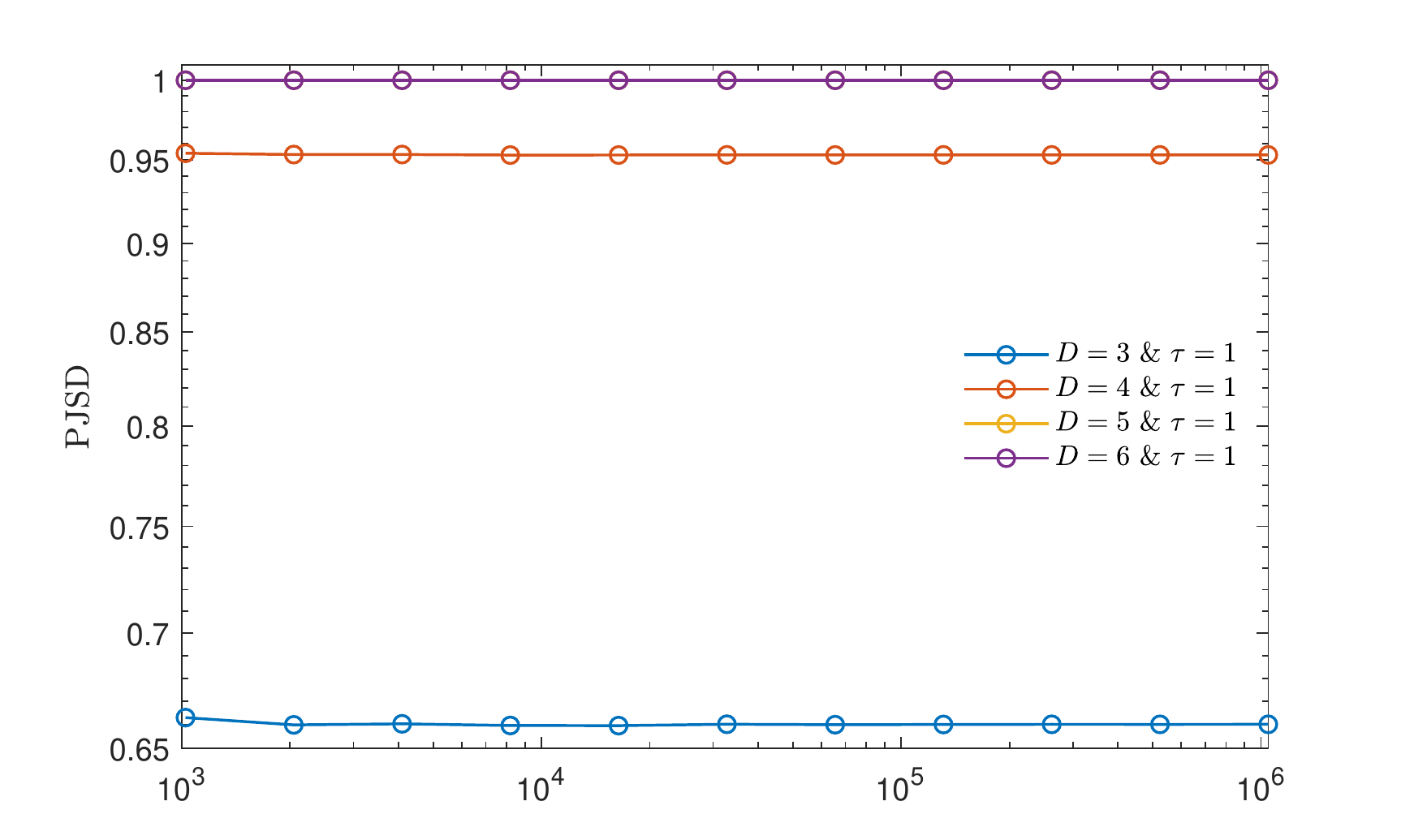}
\includegraphics[width=\linewidth,trim={.8cm .15cm 1.4cm .3cm},clip=true]{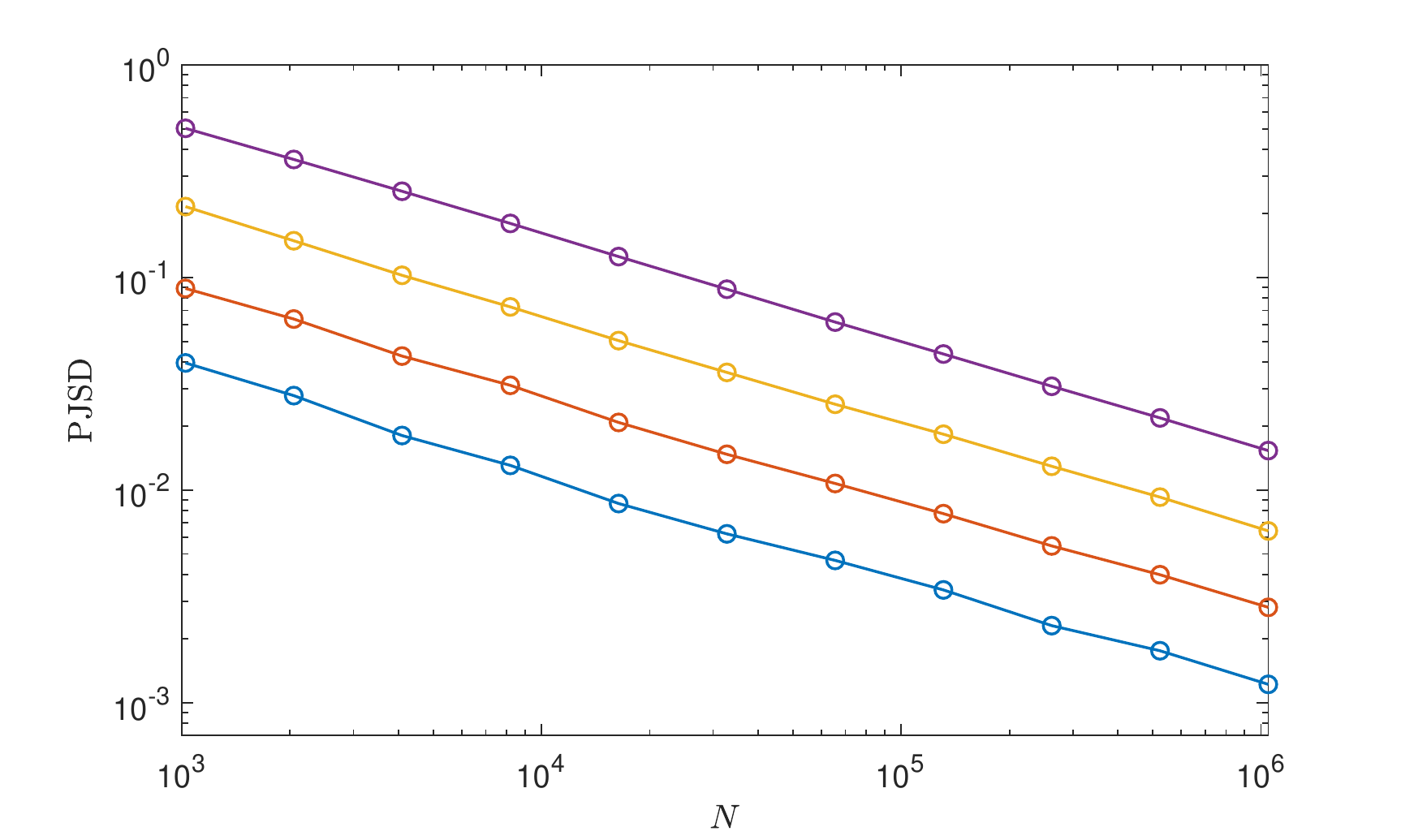}
\caption{Same as Fig.~\ref{fig:Irreversibility_stochastic} but for the logistic map in the fully chaotic regime ($r=4$) (top plot) and the $x$-component of the Arnold's cat map with $k=2$ (bottom plot). The maximal ordinal distance is obtained for the logistic map with $D=5$ and $D=6$.}
\label{fig:Irreversibility_chaotic_maps}
\end{figure}

We have also tested the performance of the proposed ordinal approach to discriminate the reversibility/irreversibility of several chaotic maps. More precisely, the logistic map in the fully chaotic regime ($r=4$), the H\'enon map ($x_{t+1}=1+y_t-ax_t^2$, $y_{t+1}=bx_t$ with $a=1.4$ and $b=0.3$) and the Arnold's cat map ($x_{t+1}=x_t+y_t$ mod(1), $y_{t+1}=x_t+ky_t$ mod(1) with $k=2$) have been analyzed. The temporal evolution of the $x$-component is considered in the two-dimensional maps. Figure~\ref{fig:Irreversibility_chaotic_maps} shows the results obtained for the logistic (top plot) and the Arnold's cat map (bottom plot). Results for the H\'enon map are qualitatively similar to that observed for the logistic map~\cite{SM7}. On the one hand, and taking into account that dissipative chaotic systems are irreversible~\cite{lacasa2012,zanin2018}, the (practically constant) non-zero estimated values reached by the PJSD for the different orders $D$ in the case of the logistic and H\'enon maps confirm irreversibility for them in full agreement with their dissipative nature. On the other hand, since the PJSD asymptotically tends to zero with $N$ following the same power-law behavior previously described for linear systems, a reversible dynamics is concluded for the Arnold's cat map as it was expected for a conservative chaotic system~\cite{lacasa2012}. We have found that the ordinal distance achieves its maximum possible value when the irreversibility of the logistic map is analyzed with the larger orders ($D=5$ and $D=6$). This is because the two ordinal pattern probability distributions under comparison have disjoint supports. This means that observed ordinal patterns of the forward time series are unobserved ordinal patterns of the reversed time series and vice versa.

\begin{figure}[!ht]
\centering
\includegraphics[width=\linewidth,trim={.8cm .15cm 1.4cm .3cm},clip=true]{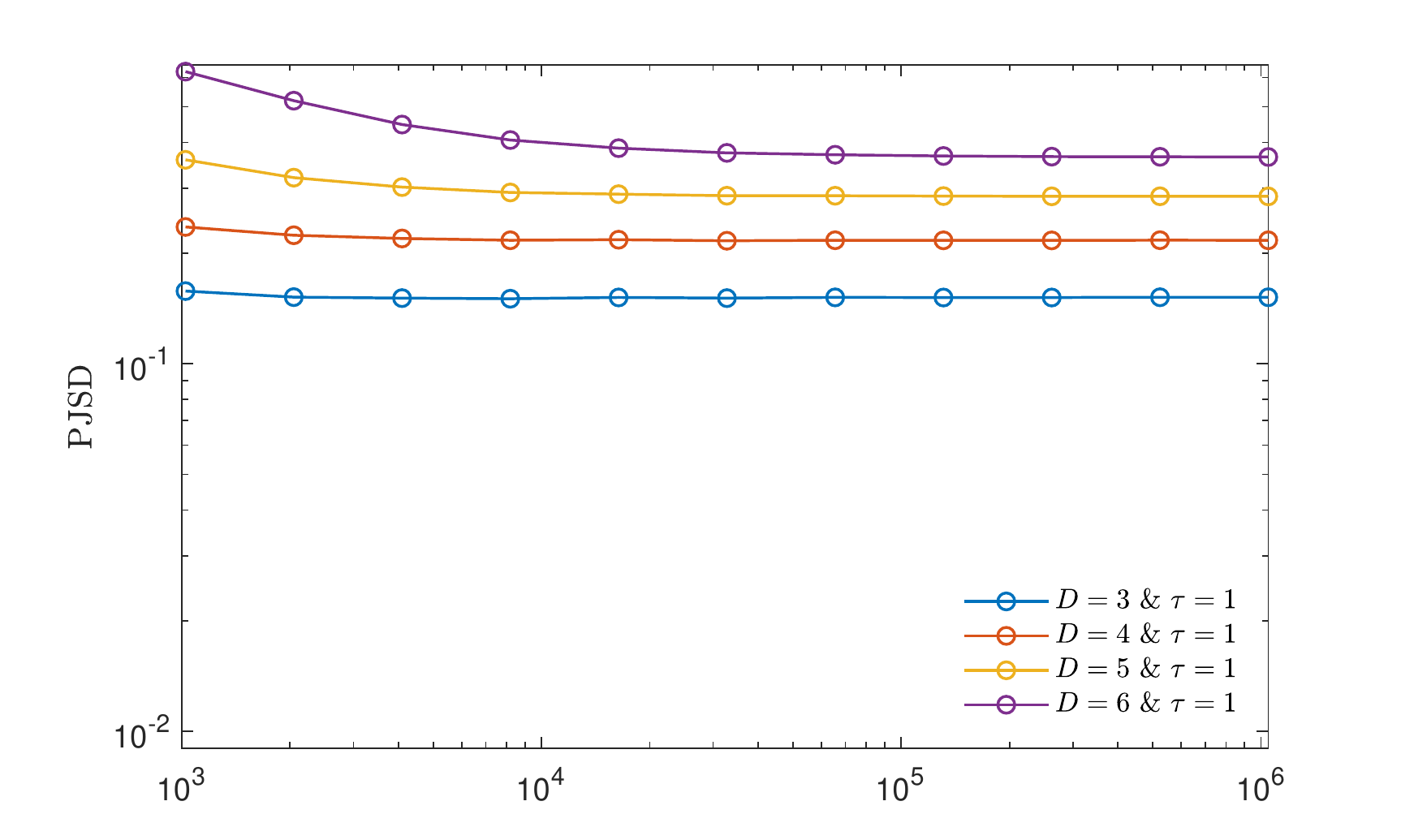}
\includegraphics[width=\linewidth,trim={.8cm .15cm 1.4cm .3cm},clip=true]{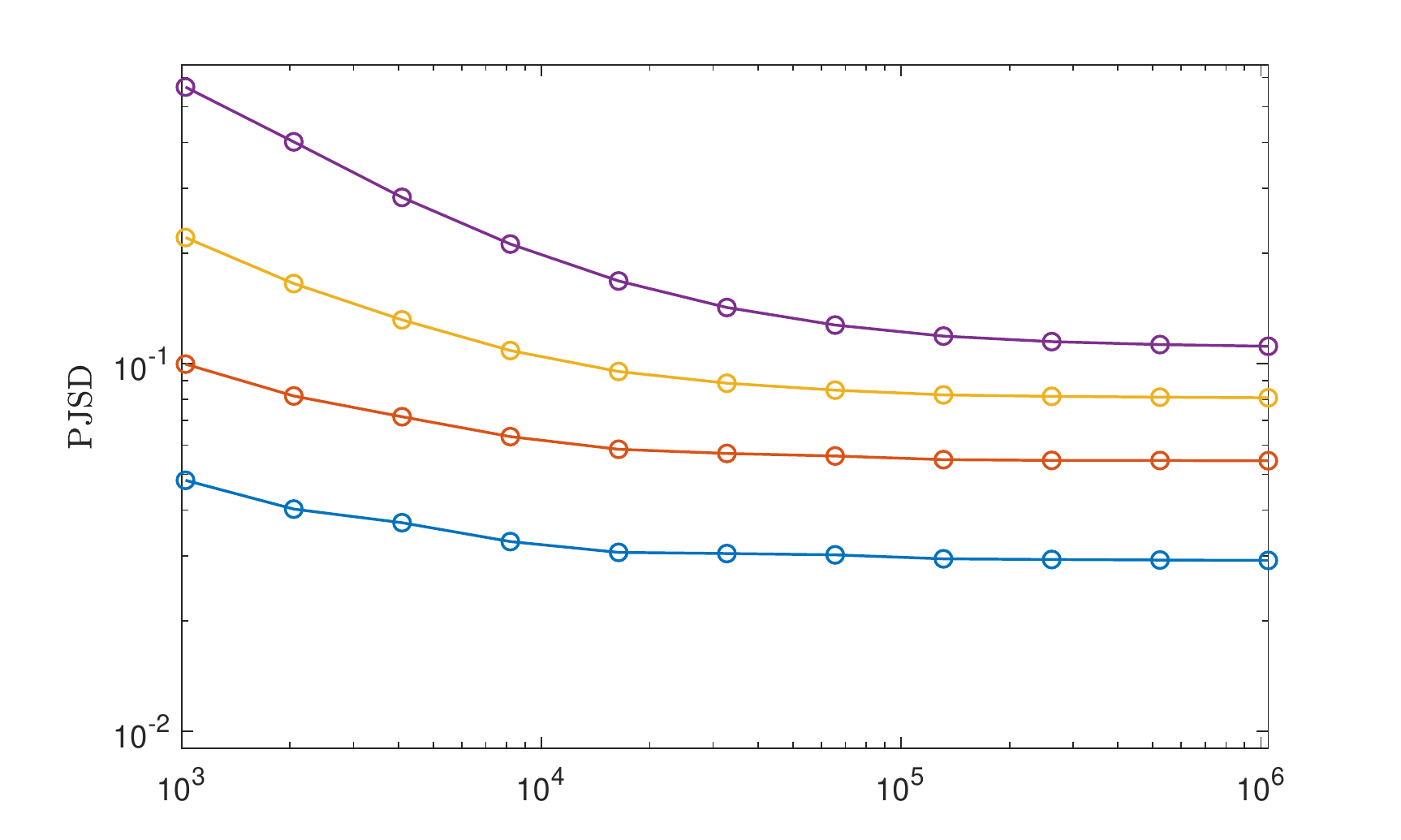}
\includegraphics[width=\linewidth,trim={.8cm .15cm 1.4cm .3cm},clip=true]{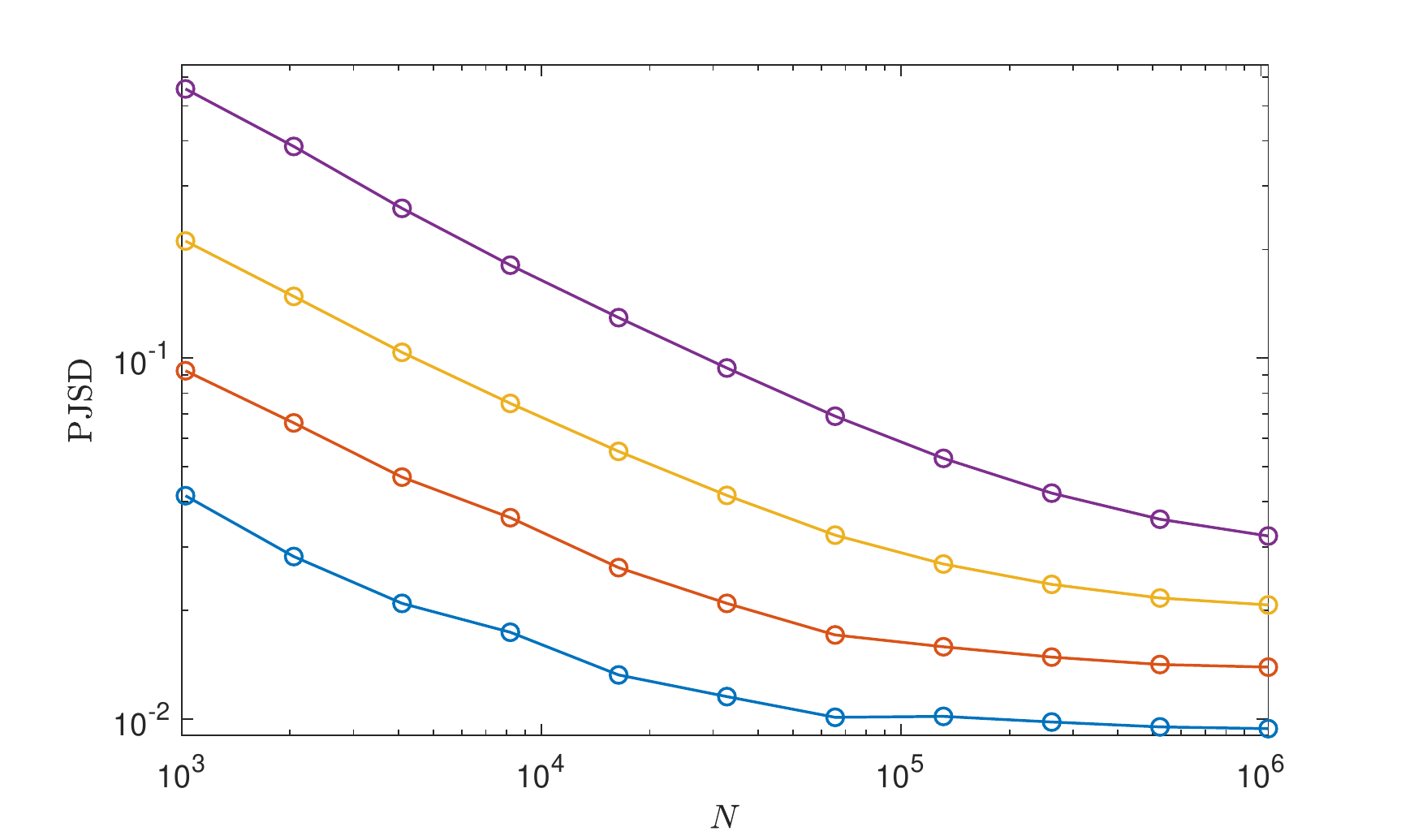}
\caption{Same as Fig.~\ref{fig:Irreversibility_stochastic} but for the $\beta$-transform with $\beta=\sqrt{20}$ (top plot), $\beta=\sqrt{200}$ (center plot) and $\beta=\sqrt{2000}$ (bottom plot). The same vertical scale is used in the three plots for easier comparison.}
\label{fig:Irreversibility_beta}
\end{figure}

Finally, we have investigated the $\beta$-transform, defined by the iterated map $x_{t+1}=\beta x_t$ mod (1), with increasing values of the parameter $\beta$. The detection of the irreversibility associated with this map becomes more challenging for larger values of the parameter since the dynamical complexity of this system increases monotonically with $\beta$~\cite{diks1995}. More precisely, a Kolmogorov-Sinai entropy equal to $\ln \beta$ is theoretically obtained~\cite{keller2014}. Figure~\ref{fig:Irreversibility_beta} illustrates the ability of the PJSD approach to unveil the irreversibility of the $\beta$-transform for three different values of $\beta \in \{\sqrt{20},\sqrt{200},\sqrt{2000}\}$. In the three cases it is observed that the PJSD between the forward and backward series converges to a non-zero value but longer time series are necessary to realize the underlying irreversible nature for larger values of $\beta$. Moreover, the ordinal distance tends to lower values as long as $\beta$ increases. Thus, it is possible to conjecture that the ordinal approach can quantify the irreversibility degree associated with this complex system. Even though further analyses are required to have a rigorous statistical approach, the current results suggest that this strategy can be useful for testing the temporal irreversibility in time series.

\section{Empirical applications}
\label{sec-exp-app}
Some examples of real data applications are detailed below. They are mainly included to demonstrate the significance and applicability of the PJSD into specific practical settings.

\subsection{Ordinal self-dissimilarity for a financial time series}
\label{subsec-osd-financial} 
\begin{figure}[!ht]
\centering
\includegraphics[width=\linewidth,trim={.8cm .1cm 1.4cm .3cm},clip=true]{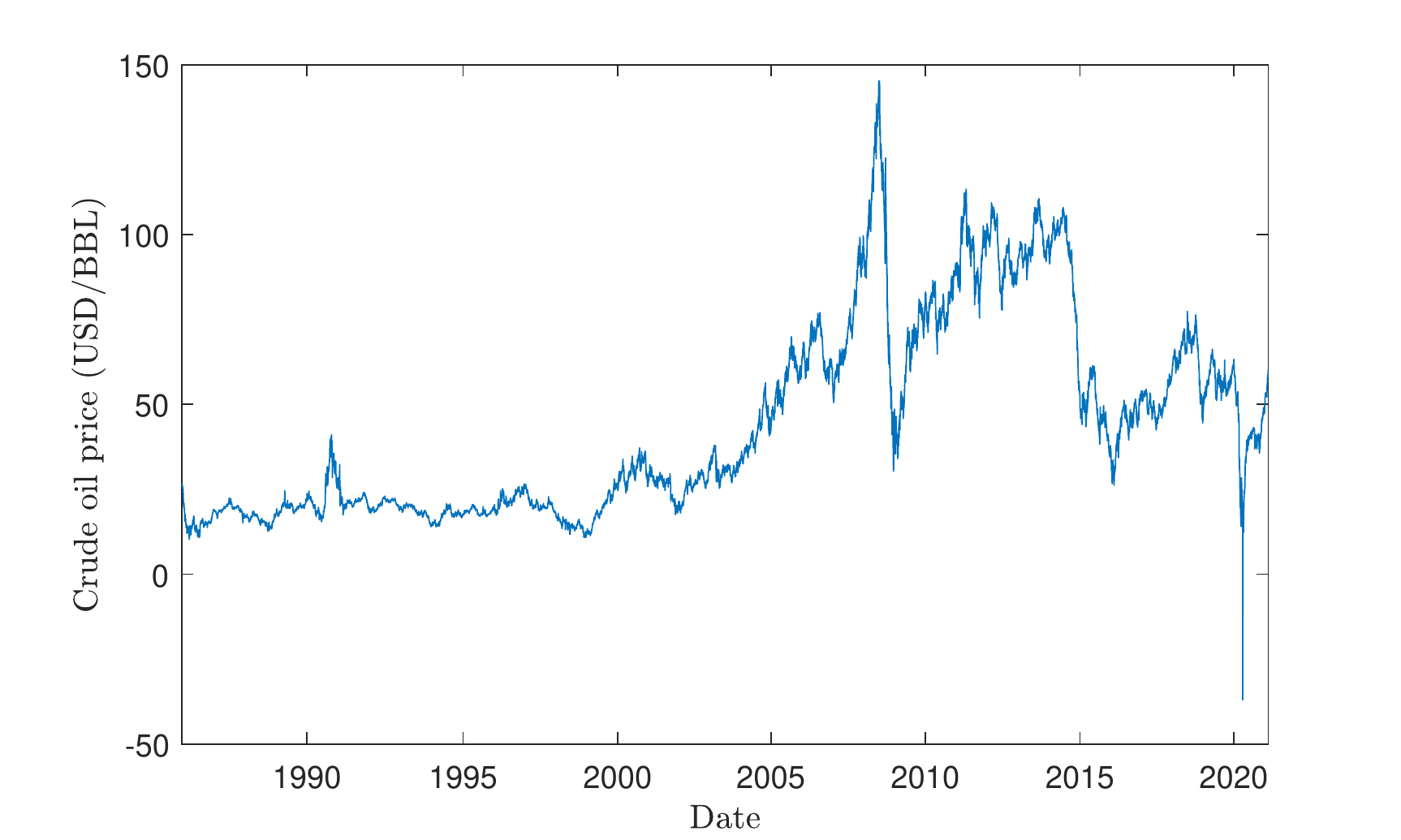}
\includegraphics[width=\linewidth,trim={.8cm .1cm 1.4cm .3cm},clip=true]{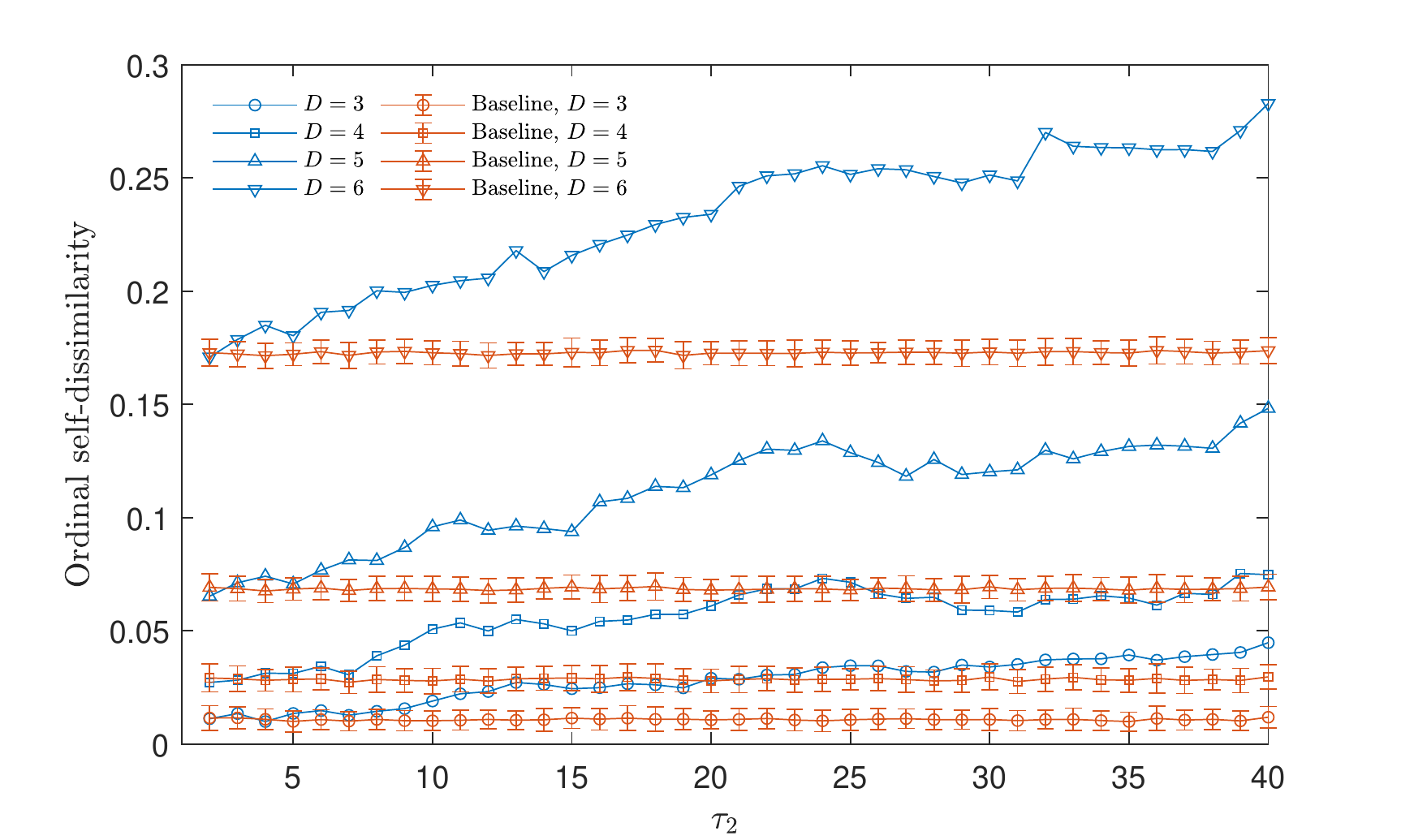}
\caption{Time series of the daily closing spot prices of the West Texas Intermediate (top plot) and its estimated ordinal self-dissimilarity as a function of the observation scale $\tau_2$ with $\tau_2^{max}=40$ for several orders $D \in \{3,4,5,6\}$ (bottom plot). Baseline references are calculated from shuffled surrogates of the crude oil prices (mean and standard deviation from estimations of an ensemble of one hundred independent shuffled realizations are displayed).}
\label{fig:SS_WTI}
\end{figure}

We have estimated the ordinal self-dissimilarity for the historical temporal price evolution of crude oil, one of the most influential commodities in the present world economy. Considering that the ordinary Brownian motion is a widely accepted first approach to model financial time series, we conjecture that the behavior for the oil prices should be analogous to that obtained for the numerical simulations analyzed in Sec.~\ref{subsec-osd}. Our main objective is to find some evidence to either support or reject this hypothesis. Daily closing spot prices of the West Texas Intermediate (WTI), often used as a benchmark in oil pricing, from January 2nd, 1986 to February 16th, 2021 have been tested. This time series of $N=8851$ oil prices, quoted in U.S. dollars per barrel and freely available to download at the U.S. Energy Information Administration (EIA) website, is displayed in Fig.~\ref{fig:SS_WTI} (top plot). At the bottom of this figure, the estimated ordinal self-dissimilarity with different orders $D$ is plotted as a function of the observation scale $\tau_2$ with $\tau_2^{max}=40$. Since observed trends are similar to those obtained for the ordinary Brownian motion (please compare with the middle plot of Fig.~\ref{fig:SS_fBm} taking into account that these results correspond to numerical realizations of length $N=10^4$ data), we can conclude that, from an ordinal self-dissimilarity perspective, this stochastic process seems to be suitable for modeling the financial data under analysis. It is also worth mentioning that our analysis is in accord with recent results obtained by Bandt~\cite{bandt2020}, which showed that the assumption of equality of frequencies of ordinal patterns in the case of the same financial time series is only justified for small values of the lag parameter ($\tau \leq 6$).

\subsection{Estimating the Hurst exponent from crude oil prices}
\label{subsec-Hurst-gait}
\begin{figure}[!ht]
\centering
\includegraphics[width=\linewidth,trim={.8cm .05cm 1.4cm .3cm},clip=true]{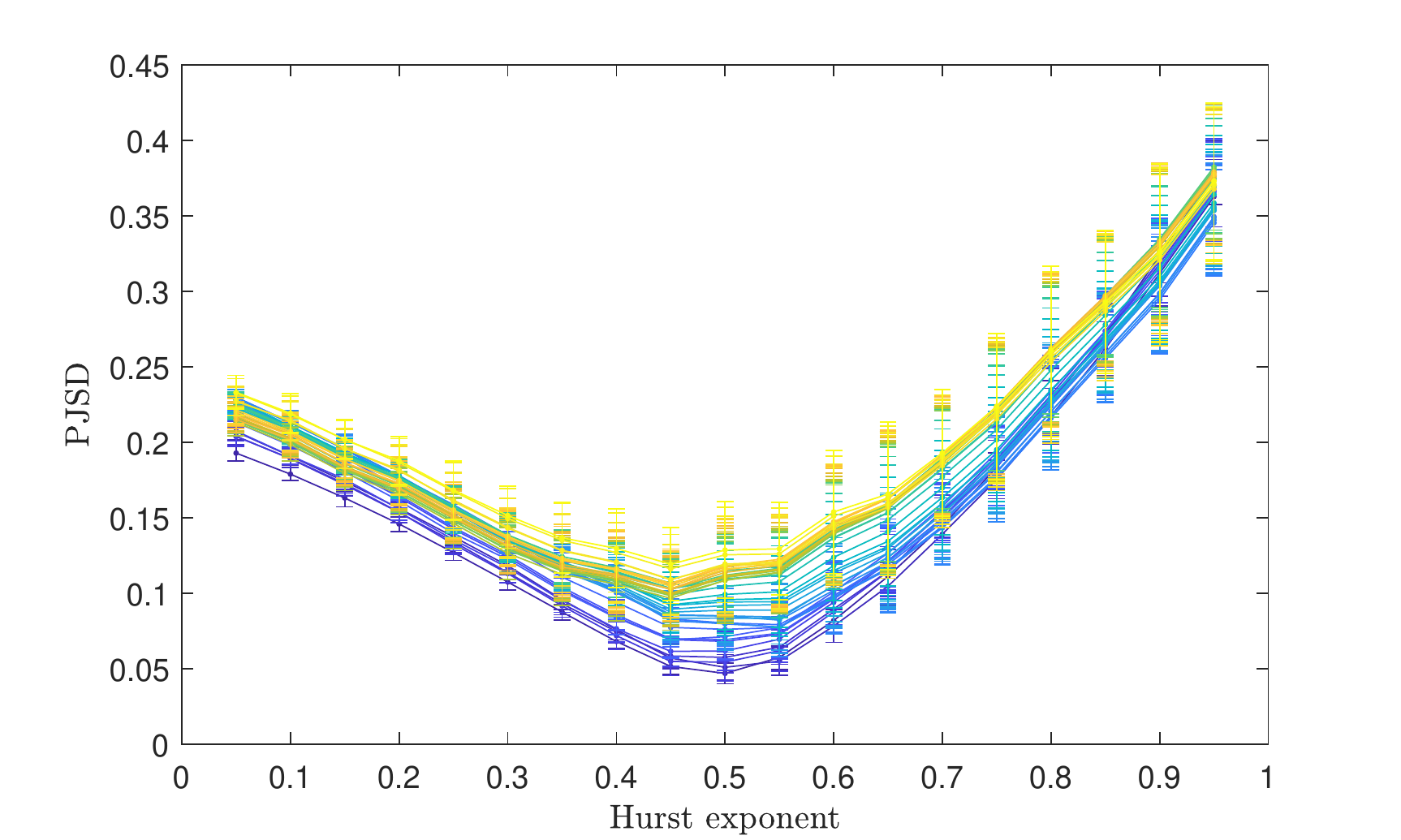}
\caption{PJSD between the WTI daily data and fBm numerical simulations of the same length ($N=8851$ data) with different Hurst exponents ($H \in \{0.05,0.1,\dots,0.95\}$). Mean and standard deviation (as error bar) from estimations with $D=4$ and different lags $1 \leq \tau \leq 40$ (increasing from blue to yellow) of an ensemble of one hundred independent numerical realizations for each Hurst exponent are shown.}
\label{fig:WTI_Hurst_D4}
\end{figure}

Although our previous application demonstrates that the ordinal self-dissimilarity obtained for time series of crude oil prices agrees with that displayed for simulations of an ordinary Brownian motion model, we seek to find further evidence in favor of this random walk behavior. If this was the case, a memoryless \textit{efficient} dynamics immune to speculative strategies would be associated with the financial time series. With this goal in mind, we propose to quantify the ordinal distance between the WTI daily data and numerical realizations of fBm with the same length ($N=8851$ data) and different Hurst exponents ($H \in \{0.05,0.1,\dots,0.95\}$). More precisely, the PJSD between the financial data and one hundred independent fBm numerical simulations for each of these Hurst exponents is estimated. A minimum of the proposed quantifier is expected for the Hurst exponent which best fits the data within an ordinal framework. Results obtained for order $D=4$ and different lags $1 \leq \tau \leq 40$ are shown in Fig.~\ref{fig:WTI_Hurst_D4}. Qualitative similar behaviors are obtained for other orders~\cite{SM8}. Even when certain deviations with the lag $\tau$ are observed, we can conclude that within the fBm family, the ordinary Brownian motion, \textit{i.e.} the fBm with Hurst exponent $H=1/2$, is the more appropriate model for the ordinal patterns of the WTI daily oil prices. This allows us to suggest the \textit{ordinal efficiency} of the oil prices. We have confirmed that a Hurst exponent $H \approx 1/2$ is also estimated with the widely accepted Detrended Fluctuation Analysis (DFA) methodology~\cite{kantelhardt2001,dfa}. It should be stressed that the success of the methodology described in this section is strongly dependent on the quality of the fBm generator and that spurious results could potentially be achieved if numerical simulations have significant bias. Beyond this limitation, we consider that the proposed ordinal approach can serve as a complementary and validation tool for more traditional Hurst exponent estimators.

\subsection{Parameter adjustment for modeling sunspot number fluctuations}
\label{subsec-model-sunspot}
Sunspots are identified as dark spots on the visible surface of the sun, with a mean lifetime of the order of a few days~\cite{blanter2006}. The number of these spots during a given period of time is a widely accepted proxy for the solar activity and, consequently, modelling its dynamics is of great interest. The best known low-frequency component of the sunspot number time series is its quasi-periodic 11-year cycle. Yet, several attempts have been made to model the interplay between this long-term variation and the presence of high-frequency fluctuations~\cite[and references therein]{shapoval2017}. Particularly, Blanter~\textit{et al.}~\cite{blanter2005} have demonstrated that the evolution of the sunspot number can be modelled as a modulated noise, whose long-term component is the well-known 11-year cycle. In a nutshell, the model consists of an autoregressive process of first order, as the high-frequency component, multiplied by a low-frequency function; $x(t)=\eta(t)[\sin(\omega t)+c]$, with $\eta(t+1)=\alpha \eta(t) + \xi(t)$, where $\alpha$ quantifies the autocorrelation of the noise $\eta(t)$ and $\xi(t)$ is an independent randomly variable uniformly distributed in $[0,1]$. Non-negative values are guaranteed by setting $c \geq 1$. It is worth mentioning that the autocorrelation parameter $\alpha$ can be related to the mean lifetime $\tau$ of sunspots ($\tau=1/(1-\alpha)$)~\cite{blanter2005}.

\begin{figure}[!ht]
\centering
\includegraphics[width=\linewidth,trim={1.25cm .3cm 1.75cm .35cm},clip=true]{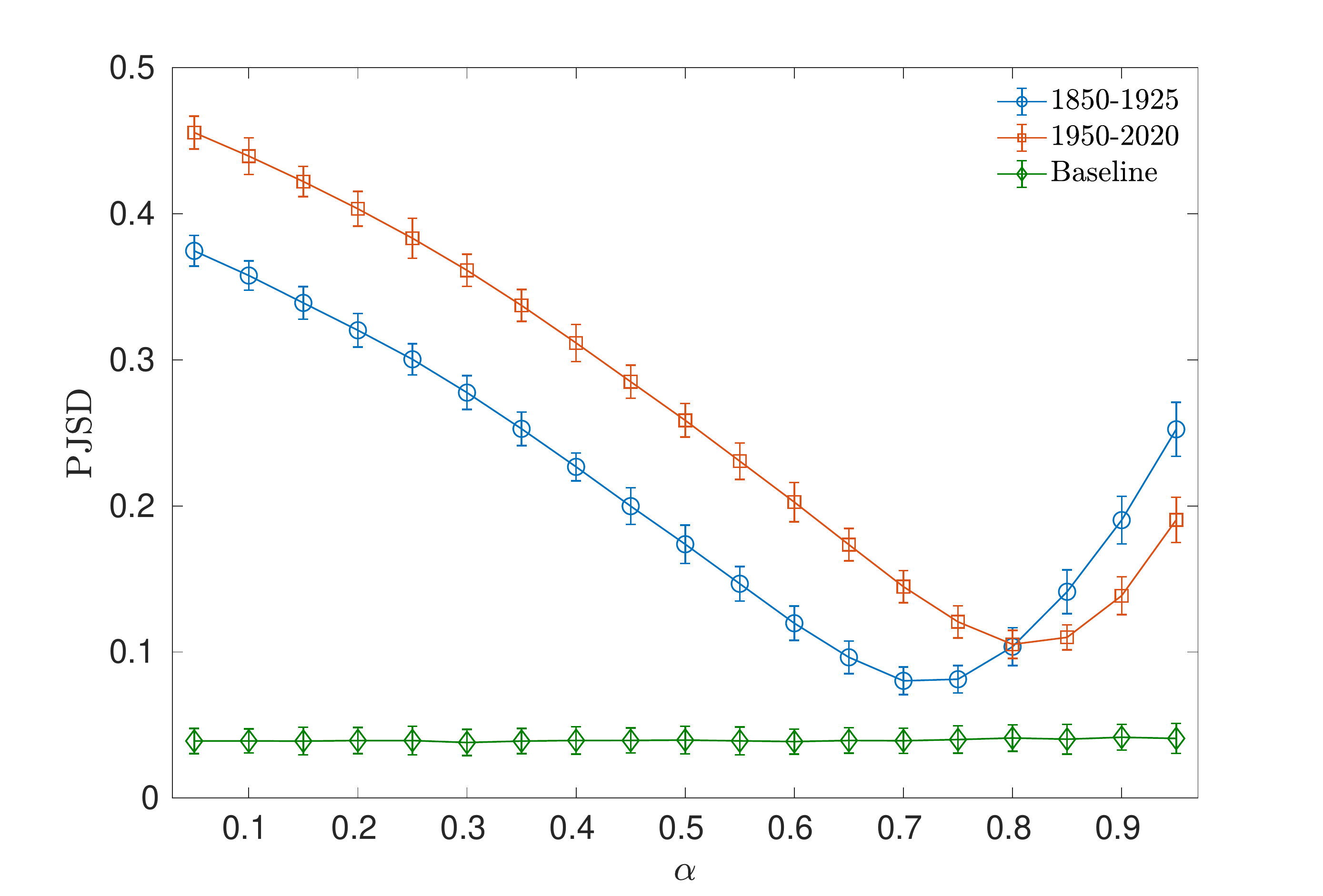}
\caption{PJSD with $D=5$ and $\tau=1$ between signals from the modulated-noise model (with $c=1.5$ and $\omega = 2\pi/11$) and the daily sunspot time series, both previously smoothed by a 11 years moving average sliding window, as a function of the autocorrelation parameter $\alpha$. Mean and three standard deviations (as error bar) calculated from the distance between the empirical sequence and one hundred independent realizations of the model are plotted.}
\label{fig:sunspot}
\end{figure}

Here, we aim to estimate the autocorrelation parameter $\alpha$ for which the model more accurately reproduces the daily evolution of the sunspot numbers. With this goal in mind, we propose to look for the $\alpha$ value that minimizes the PJSD between the modulated-noise model and the daily sampled sunspot number. Following Ref.~\cite{blanter2005}, we consider two different periods of the daily international sunspot number data from the World Data Center SILSO, Royal Observatory of Belgium, Brussels~\cite{SILSO}, spanned between 1850 to 1925 and 1950 to 2020, respectively. We have estimated the PJSD with $D=5$ and $\tau=1$ between the sunspot numbers for these two periods and one hundred independent realizations of the modulated-noise model with $c=1.5$ and $\omega = 2\pi/11$ for different values of $\alpha \in \{0.05, 0.1,..., 0.95\}$. Additionally, a baseline was established by estimating the PJSD between two independent realizations of the model. These results are depicted in Fig.~\ref{fig:sunspot}. A minimum of the PJSD is found for $\alpha=0.7$ and 0.8 for the first and second period, respectively. For these values, we can say that the modulated-noise model better fits the real observations within an ordinal framework. Similar results are found for other orders~\cite{SM9}. The change in the autocorrelation of the daily solar activity can be interpreted as an increment on the average lifetime of the sunspots in the period 1925-1950. Indeed, our results are in accordance with those obtained in Ref.~\cite{blanter2005}, where $\alpha$ was estimated from the daily sunspot signal by calculating its Markov radius of correlation for the same two time periods approximately.

\subsection{Multiscale analysis of electroencephalograms from healthy and epileptic patients}
\label{subsec-EEG-multiscale-analysis}
Trying to demonstrate the potential of the PJSD in the analysis of neurophysiological data, we have implemented it for discriminating between different brain electrical activity time series. More precisely, five sets of electroencephalogram (EEG) for different groups and recording regions have been considered: surface (scalp) EEG recordings from five healthy volunteers in an awake state with eyes open (Set A) and closed (Set B), intracranial EEG recordings from five epilepsy patients during the seizure free interval from outside (Set C) and from within (Set D) the seizure generating area, and intracranial EEG recordings of epileptic seizures (Set E). One hundred representative EEG segments of length $N=4097$ data, free of artifacts and satisfying a weak stationarity criterion, were appropriately selected for each of these sets. Please see the original paper by Andrzejak~\textit{et al.}~\cite{andrzejak2001} for further details about this database. 

\begin{figure}[!ht]
\centering
\includegraphics[width=\linewidth,trim={.8cm .15cm 1.4cm .3cm},clip=true]{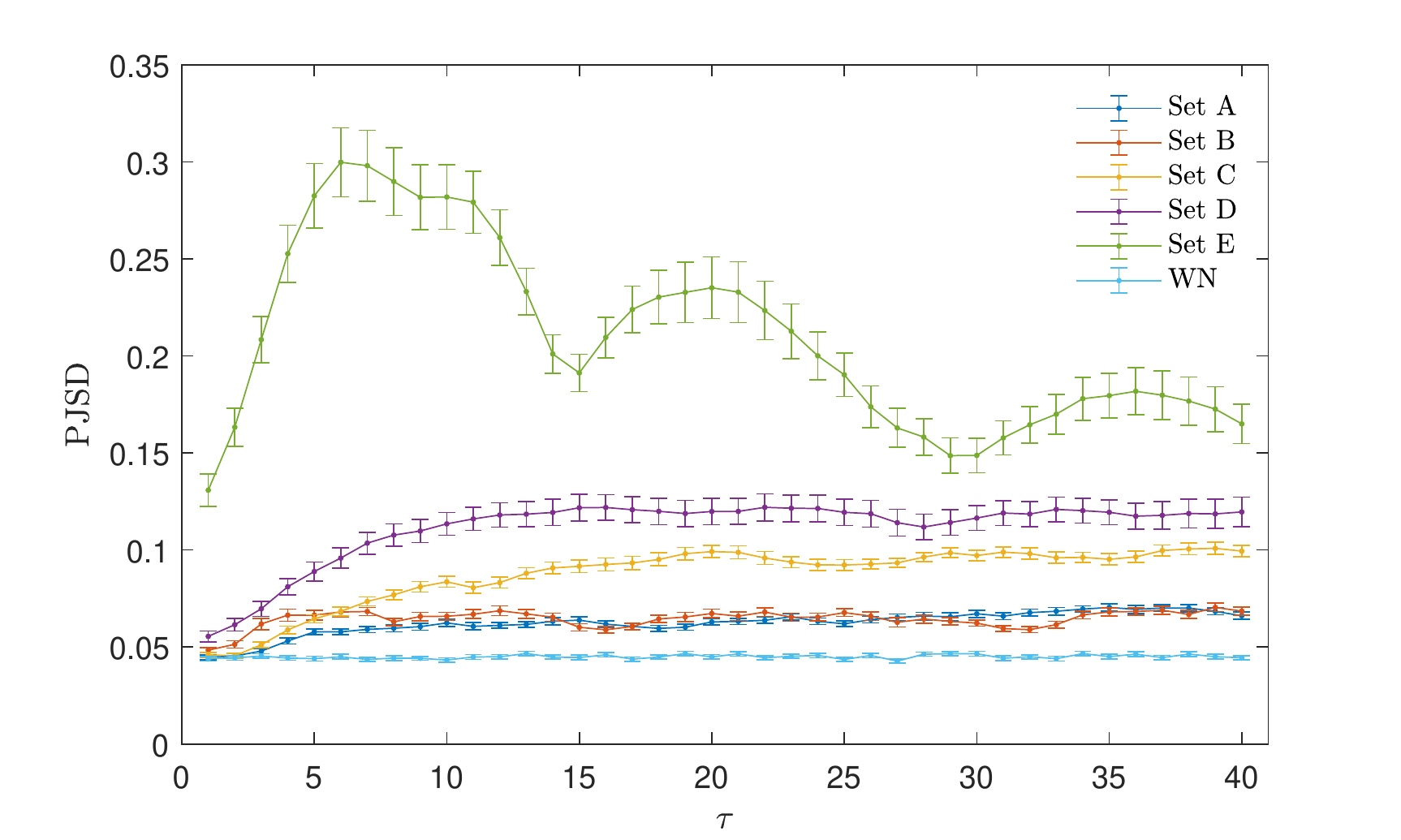}
\caption{PJSD with $D=4$ between the forward and backward EEG records as a function of the lag $\tau$. Mean and standard error (as error bar) calculated from the one hundred representative time series for each set are plotted. Results obtained for a set of one hundred independent Gaussian white noise realizations of the same length (WN) as linear reference are also displayed. The irreversibility degree and, consequently, the nonlinearity changes as a function of the considered temporal scale for the different EEG sets while the linear reference shows a constant behavior. Actually, it can be visually concluded that the discriminative power between the different datasets increases for larger lag $\tau$.}
\label{fig:EEG_discrimination_D4}
\end{figure}

It has been previously shown that the degree of nonlinearity in these sets is different~\cite{andrzejak2001,donges2013,zunino2017a,kulp2017}. Consequently, tools or statistical tests that quantify the presence of nonlinear structures can be particularly suited for the discrimination task. Considering this and also the fact that time irreversibility is a signature of nonlinearity, we propose the PJSD between the forward and backward EEG recordings as discriminatory tool. In this case, due to the limited length of the EEG recordings, we cannot study the behavior of the PJSD with the time series length $N$ as done in Sec.~\ref{subsec-reversibility}. Instead, we simply compare the estimated ordinal distances for the five EEG groups. Since the discriminative power could depend on the time scale and the optimal temporal scale is a priori unknown, a multiscale analysis, \textit{i.e.} PJSD versus $\tau$, is developed. Moreover, in order to have the reference of a linear process to compare, results obtained for an additional set of one hundred independent Gaussian white noise (WN) realizations of the same length ($N=4097$ data) have also been included. PJSD estimations with $D=4$ and $1 \leq \tau \leq 40$ for the six considered sets are displayed in Fig.~\ref{fig:EEG_discrimination_D4}. Qualitatively similar results are obtained for other orders~\cite{SMlast}.

As expected, coexistence of time scales with different degrees of time irreversibility are concluded for the complex fluctuations of the physiological time series. On the one hand, it is clearly observed that the segments of ictal activity (Set E) have the highest degree of nonlinearity for any lag $\tau$, allowing them to be clearly distinguished from the other brain electrical activities. On the other hand, larger time scales improve the discrimination between the other groups (Sets A-D and WN). Indeed, for $\tau=1$, the irreversibility degree estimated from records of Sets A-C and those from the linear WN reference are highly overlapped. The nonlinearity nature of Sets A-D seems to be better unveiled at larger time scales and, hence, a multiscale analysis is crucial to differentiate these EEG time series from the white noise linear realizations. Moreover, higher degrees of irreversibility/nonlinearity are observed for the pathological cases (Sets C \& D). As a matter of fact, this increment of time irreversibility with pathology is in perfect agreement with the results previously obtained by implementing different techniques on the same~\cite{andrzejak2001,donges2013,zunino2017a,kulp2017} and other~\cite{vanderheyden1996,schindler2016} epileptic EEG databases. It should also be noted that this finding is totally opposite to the loss of time irreversibility with disease over multiple time scales observed for human heartbeat time series and interpreted as evidence of loss of functionality and adaptability~\cite{costa2005}. Contrarily, for the epileptic brain electrical activities analyzed in this section, the time reversal asymmetry is lower in the healthy condition. It is worth adding that, even when the surface EEG recordings of healthy subjects with eyes open and closed (Sets A \& B) achieve the lower indications of nonlinear dynamics, they are clearly distinguished from the WN reference, especially for larger values of the lag $\tau$. Last but not least, EEG data from these two healthy groups show similar time irreversibility behaviors as a function of $\tau$ and can be hardly distinguished.

\subsection{Differentiating between real and simulated heartbeat time series}
\label{subsec-RR-multiscale-analysis}
In this last experimental application, we investigated the database from a challenge to generate realistic sequences of interbeat (RR) intervals~\cite{CinC2002}. More precisely, a collection of 50 (26 physiologic and 24 synthetic) RR interval time series, each between 20 and 24 hours in length, are tested. The real series were obtained from long-term ECG recordings of adults between the ages of 20 and 50 with no evidence of cardiac abnormalities, while each of twelve different generators was implemented to simulate synthetic RR interval time series (two series had been simulated by each generator). The length $N$ of these sequences is variable and it goes from a minimum of 71953 up to a maximum of 128507 data. It should also be noted that two of the generators violated the required rule that portions of real (physiologic) RR interval sequences cannot be incorporated in
the output. One of them, that generated the series designated as rr36 and rr42, used smoothed averages of real time series while the other one simply time-reversed entire (real) 24-hour series (datasets labeled as rr14 and rr16 correspond to this last not allowed strategy). Please see Ref.~\cite{moody2002} for further information about this open access database.

Our ultimate goal in this analysis is to identify a property of the underlying processes generating the time series that allows us to discriminate between the real and artificial heartbeat signals. According to some previous results, looking for signatures of nonlinear dynamical behavior seems to be an appropriate strategy for achieving robust discrimination of these data~\cite{costa2005,costa2008}. Following this hypothesis, once again the degree of time irreversibility is proposed as a potential discriminatory tool. It has also been previously demonstrated the importance of considering the multiple time scales inherent in the healthy cardiac interbeat variability for a more proper characterization of its intrinsic time reversal asymmetry~\cite{costa2005,costa2008}. Consequently, as in the analysis of the EEG database, a multiscale analysis is recommended, \textit{i.e.} PJSD between the forward and the backward time heartbeat time series are estimated and contrasted at multiple levels of temporal resolution. After several preliminary analyses, we have further found that the summation of the estimated irreversibility degree, \textit{i.e.} $\sum_{\tau=1}^{\tau=\tau_{max}}[D_{JS}(P_{for},P_{back})/\ln 2]^{1/2}$, up to a suitable maximum temporal scale, $\tau_{max}$, for a value of the order $D$ is better for the discrimination purpose. This global quantifier for characterizing the time reversal asymmetry, that considers a pre-defined range of scales simultaneously, is in line with the \textit{multiscale asymmetry index} proposed by Costa~\textit{et al.}~\cite{costa2005,costa2008}.

\begin{figure}[!ht]
\centering
\includegraphics[width=\linewidth,trim={1.25cm .3cm 1.5cm .35cm},clip=true]{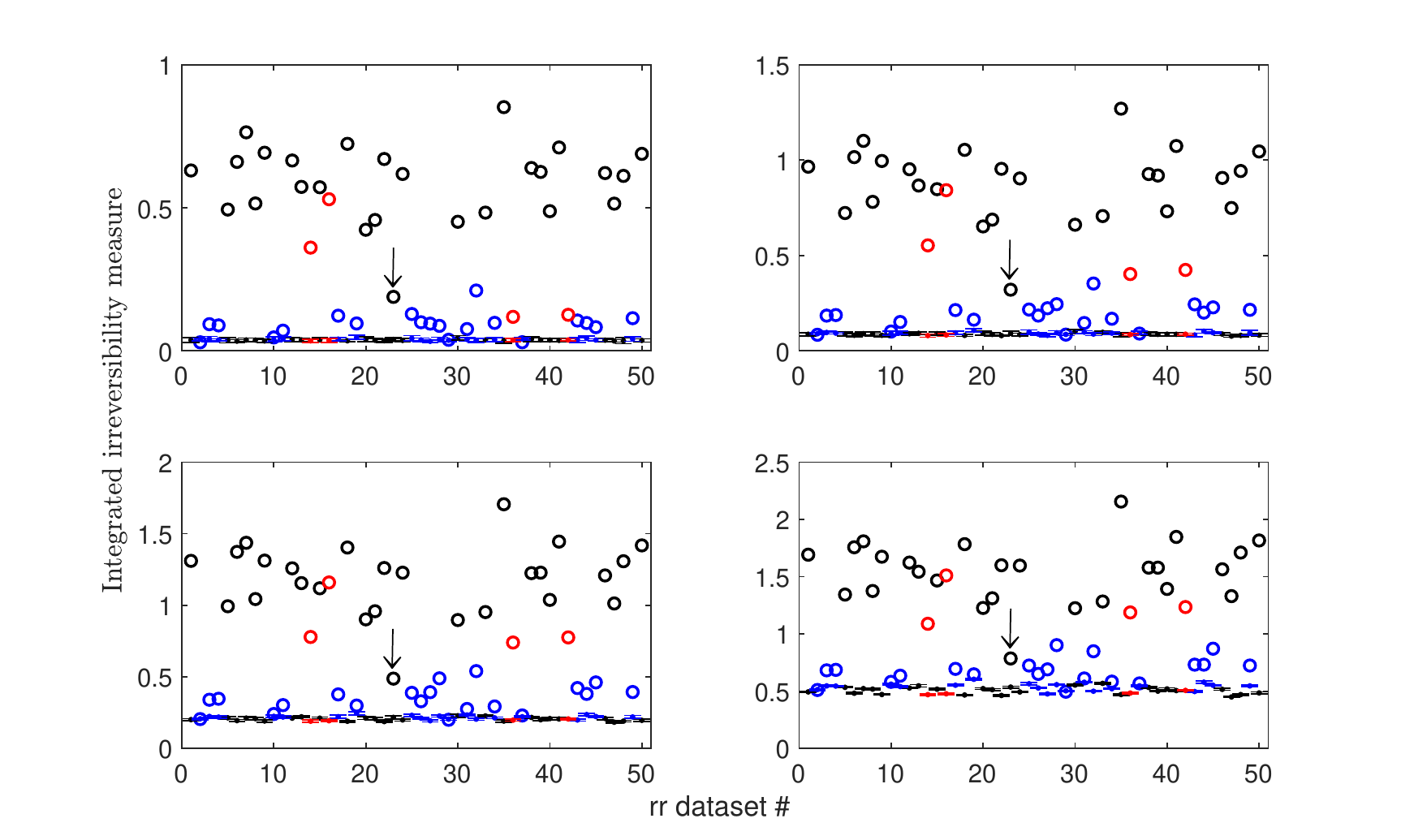}
\caption{Integrated irreversibility measure with different orders $D \in \{3,4,5,6\}$ (increasing from top left to bottom right) and $\tau_{max}=10$ for the different sequences of the RR challenge database. Results obtained for the physiologic (black circles), synthetic (blue circles) and the four problematic artificial sequences (red circles) are shown. Mean and standard deviation (as error bar) of estimations for an ensemble of ten independent shuffled realizations for each RR interval time series are also displayed in order to get a baseline for linear/reversible dynamics with the same length. The arrow indicates the true physiologic dataset rr23 with a low integrated irreversibility degree, similar to those estimated for simulated datasets.}
\label{fig:Integrated_irreversibility_measure}
\end{figure}

Results obtained for the RR challenge database by estimating this integrated irreversibility measure with several orders $D$ and $\tau_{max}=10$ are shown in Fig.~\ref{fig:Integrated_irreversibility_measure}. In order to have a linear reversible reference to compare with, ensembles of ten independent shuffled realizations for each RR interval time series have also been analyzed. Mean and standard deviation of the ten shuffled estimations of the proposed global irreversibility quantifier for each tested sequence are also displayed in Fig.~\ref{fig:Integrated_irreversibility_measure}. The value for $\tau_{max}$ has been chosen through visual comparison and it might not be the optimal one in a quantitative context. However, we consider that our qualitative analysis is enough for the illustration of the technique, and that the identification of the optimal value for $\tau_{max}$ is beyond the scope of the present work. A clear discrimination between real and simulated sequences is concluded from Fig.~\ref{fig:Integrated_irreversibility_measure}. True physiological records (black circles) have larger estimated values of the global quantifier, confirming the existence of nonlinear temporal structures in their underlying dynamics. Estimations for the synthetic datasets (blue circles) are significantly lower. Essentially, algorithms included in the challenge cannot reproduce the nonlinearity intrinsically present in the true physiological records. Only the true physiologic dataset rr23 (indicated with an arrow) shows a low irreversibility degree that makes us erroneously assume that it is an artificial record. It is worth noting here that curiously the same misclassification has been observed in the analysis developed in Ref.~\cite[please see Fig. 4]{costa2008}. This common misclassification reinforces the hypothesis suggested by the authors of the aforementioned article that unknown factors such as age, level of physical activity and /or drug effects could be potential reasons for the low level of the integrated reversal asymmetry related to this particular physiologic record. Furthermore, the results obtained for the four problematic synthetic sequences deserve special attention. On the one hand, the two records that are simply time reversal of real data, rr14 and rr16 (the two red circles on the left), are classified as physiologic as it was obviously expected. On the other hand, the other two records, rr36 and rr42 (the two red circles on the right), that are smoothed averages of real time series, show an interesting behavior as a function of the order $D$. Their integrated irreversibility degree reaches a low value for $D=3$, similar to that obtained for synthetic records. However, estimated values for the global quantifier increase as $D$ increases, achieving values similar to that obtained for true RR sequences when $D=6$. We conjecture that their inherent physiological (irreversible/nonlinear) nature remains partially hidden at the lowest orders but is successfully revealed at the highest one. We consider that all these findings confirm that the proposed multiscale time irreversibility analysis is particularly valuable for testing the quality of algorithms designed to simulate heart rate time series. We are also optimistic about its usefulness for characterizing the potential limitations of generators of other physiological signals.

\section{Conclusions}
\label{sec-con}
We have introduced the PJSD, an ordinal metric able to quantify the degree of similarity between two (or more) time series. Through several numerical and experimental analyses, we have confirmed its usefulness for characterizing and classifying time series. Comparisons against other approaches have allowed us to verify the enhanced performance of the proposed tool for some relevant purposes. Taking into account the robustness to noise effects and the invariance under scaling of the data associated with the ordinal symbolization approach, the PJSD seems to be a quantitative metric especially suited for the analysis of real-world signals. We also consider this approach to be versatile because the hypothesis to be tested can be easily set up by conveniently choosing the time series taken as reference. Actually, the diversity of applications included in this paper is explicit evidence of this. Consequently, a wide range of complex phenomena can be scrutinized with this ordinal symbolic quantitative metric. In particular, we conjecture that the PJSD can be potentially useful for quantifying the functional connectivity in complex network analysis; \textit{e.g.}, to reveal specific disease effects from brain networks analysis. Moreover, it can be easily and fastly implemented, paving the way to monitor the behavior of big data series in real time. It is also possible to hypothesize that multiple scientific fields can be benefited through its implementation. For all these reasons, we consider that the proposed ordinal distance is a useful addition to the repertoire of existing methods for complex signals analysis. Further analytical investigations that support the heuristic results detailed in this work will be undertaken in future research. Exploring possible generalizations using other divergences, such as the Jensen-Tsallis one that allows a more complete characterization in the case of heavy-tailed distributions~\cite{gerlach2016}, could be another potential avenue of study. Interested readers are welcome to contact the authors for an implemented code of the PJSD in MATLAB. This method will also be available in ordpy, a Python package for data analysis with ordinal methods~\cite{pessa2021}.

\begin{acknowledgments}
L.Z. acknowledges the financial support from Consejo Nacional de Investigaciones Cient\'ificas y T\'ecnicas (CONICET), Argentina. H.V.R. thanks for the financial support of the CNPq (Grants 407690/2018-2 and 303121/2018-1). F.O. acknowledges the Spanish State Research Agency, through the Severo Ochoa and Mar\'ia de Maeztu Program for Centers and Units of Excellence in R$\&$D (MDM-2017-0711).
\end{acknowledgments}

\appendix*
\section{Performance comparison against other approaches}
\label{Appendix}
With the aim of achieving a better characterization regarding the significance of the results obtained with the PJSD, we have developed some comparative analysis with other two quantifiers of similarity/dissimilarity between signals; namely, the information-based similarity index (IBSI)~\cite{yang2003,peng2007} and the alphabetic Jensen-Shannon divergence (aJSD)~\cite{mateos2017}. A binary coarse graining depending on the relative amplitudes of successive values is first implemented in both of them and, then, binary sequences of length $m$ ($m$-bit words) are obtained. This procedure is known as \textit{alphabetic mapping}. The relative frequencies of the $2^{m}$ possible $m$-bit words are calculated and then used to quantify the dissimilarity between two arbitrary signals. On the one hand, in the case of IBSI, the \hbox{$m$-bit} words are first sorted and ranked according to their frequencies of occurrence for each signal. Then, an average deviation of the ranks for the common words, with a suitable weighting factor that takes into account the importance (relative frequency) of the word, is proposed for characterizing the dissimilarity between the two symbolic sequences. On the other hand, for the aJSD, the Jensen-Shannon divergence between the probability distributions of the two signals under analysis, estimated following the alphabetic mapping scheme, is proposed as measure of dissimilarity. For further details please see Refs.~\cite{yang2003,peng2007} for the former tool, and Ref.~\cite{mateos2017} for the latter one.

\begin{figure}[!ht]
\centering
\includegraphics[width=\linewidth,trim={1.25cm .25cm 1.25cm .35cm},clip=true]{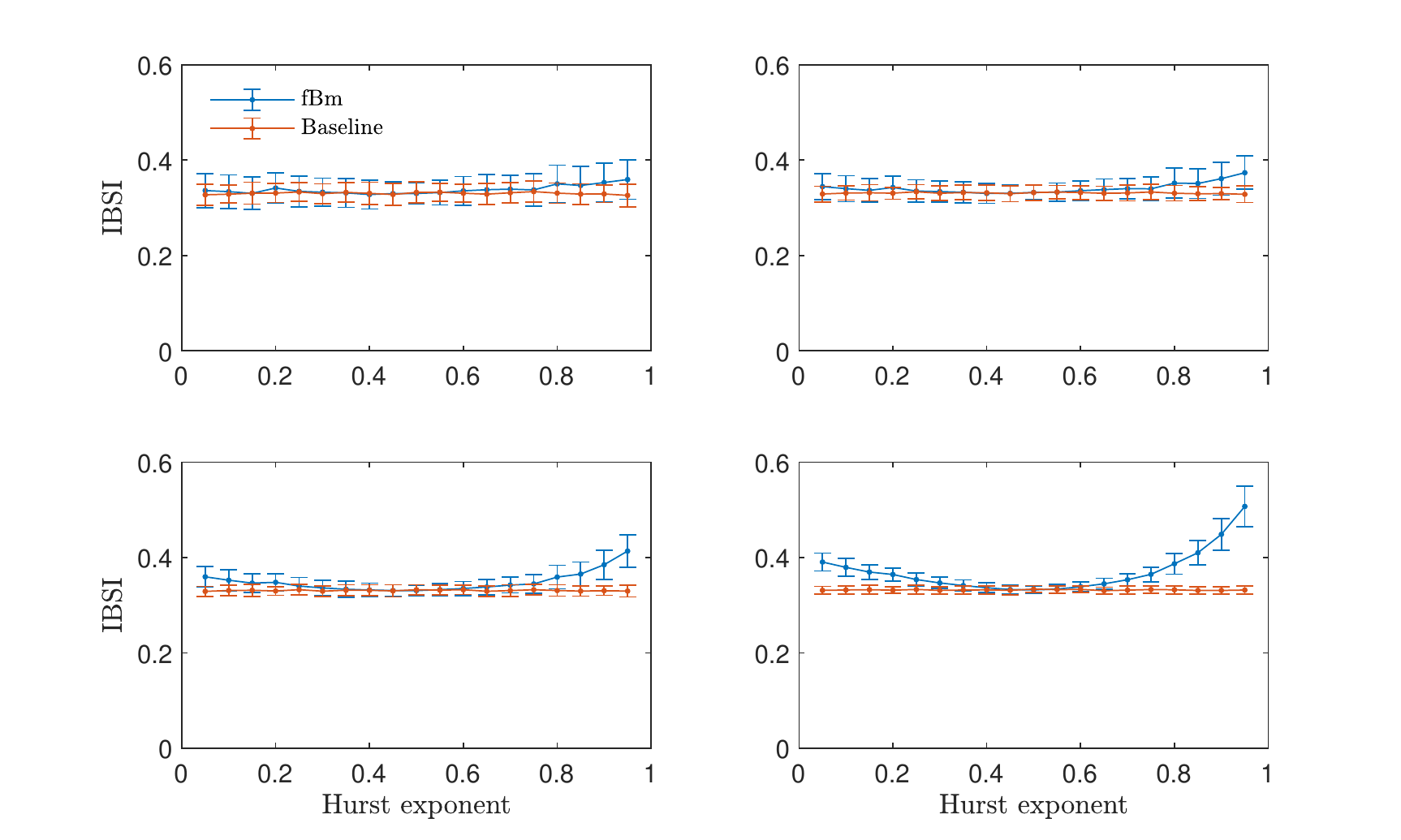}
\caption{Estimations of IBSI with several lengths $m \in \{8,9,10,11\}$ (increasing from top left to bottom right) are plotted as a function of the Hurst exponent for fBms. Mean and standard deviation (as error bar) from estimations of an ensemble of one hundred independent realizations of length $N=10^4$ data are depicted. Baseline references resulting from the analysis of a pair of shuffled surrogate realizations from each simulation are also included.}
\label{fig:IBSI_fBm}
\end{figure}

\begin{figure}[!ht]
\centering
\includegraphics[width=\linewidth,trim={1.25cm .25cm 1.25cm .35cm},clip=true]{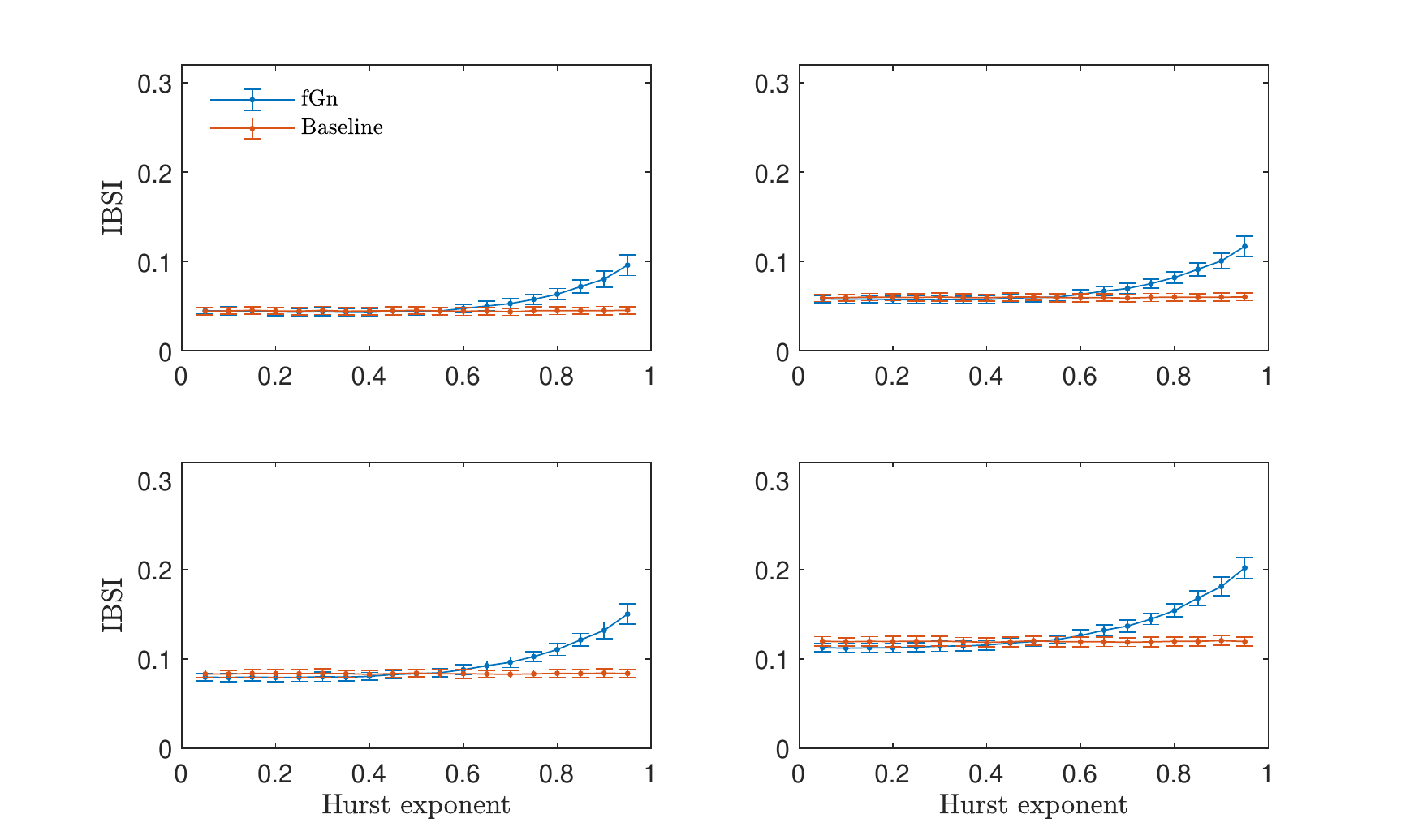}
\caption{Same as Fig.~\ref{fig:IBSI_fBm} but for fGns.}
\label{fig:IBSI_fGn}
\end{figure}

\begin{figure}[!ht]
\centering
\includegraphics[width=\linewidth,trim={1.25cm .25cm 1.25cm .35cm},clip=true]{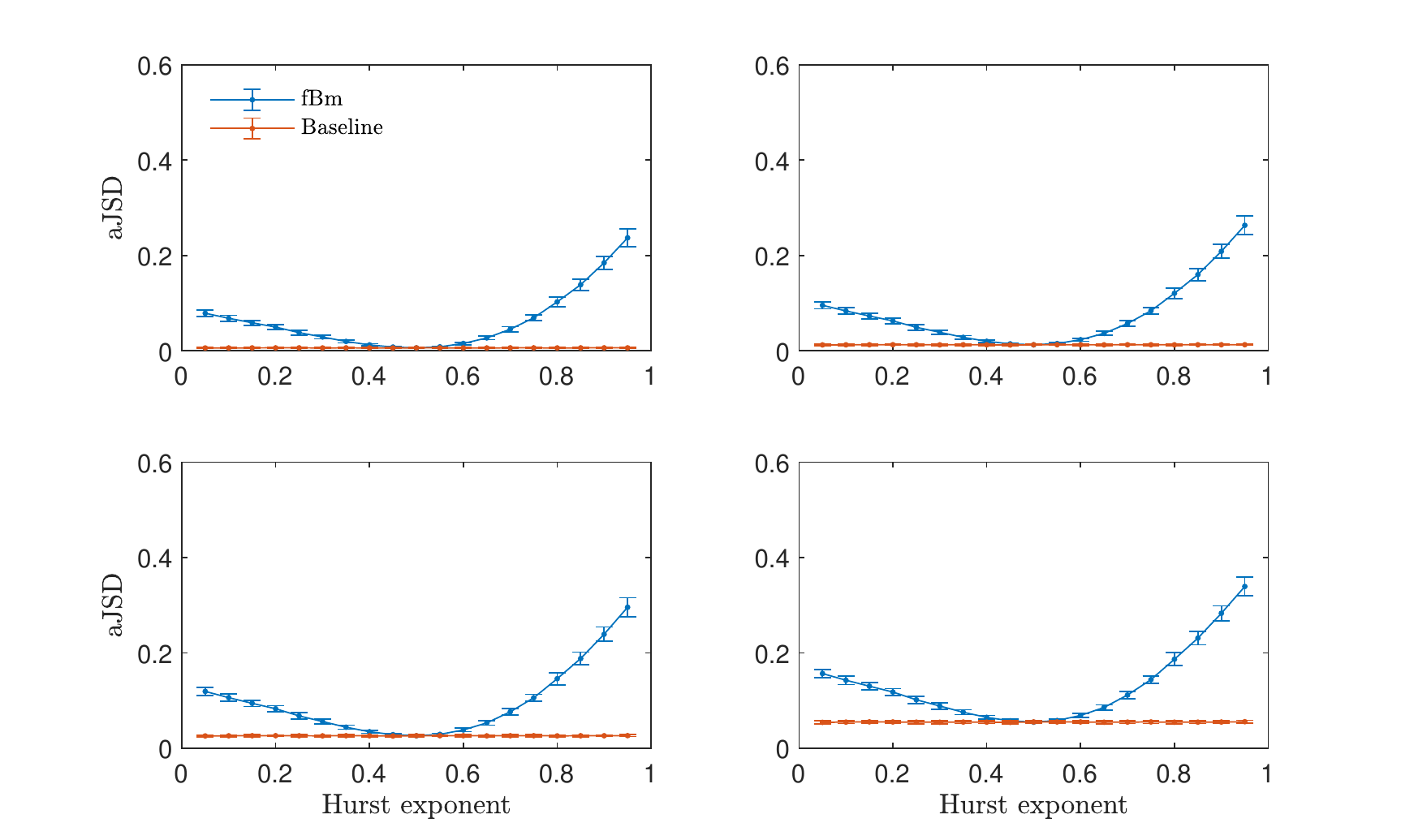}
\caption{Estimations of aJSD with several lengths $m \in \{8,9,10,11\}$ (increasing from top left to bottom right) are plotted as a function of the Hurst exponent for fBms. Mean and standard deviation (as error bar) from estimations of an ensemble of one hundred independent realizations of length $N=10^4$ data are depicted. Baseline references resulting from the analysis of a pair of shuffled surrogate realizations from each simulation are also included.}
\label{fig:aJSD_fBm}
\end{figure}

\begin{figure}[!ht]
\centering
\includegraphics[width=\linewidth,trim={1.25cm .25cm 1.25cm .35cm},clip=true]{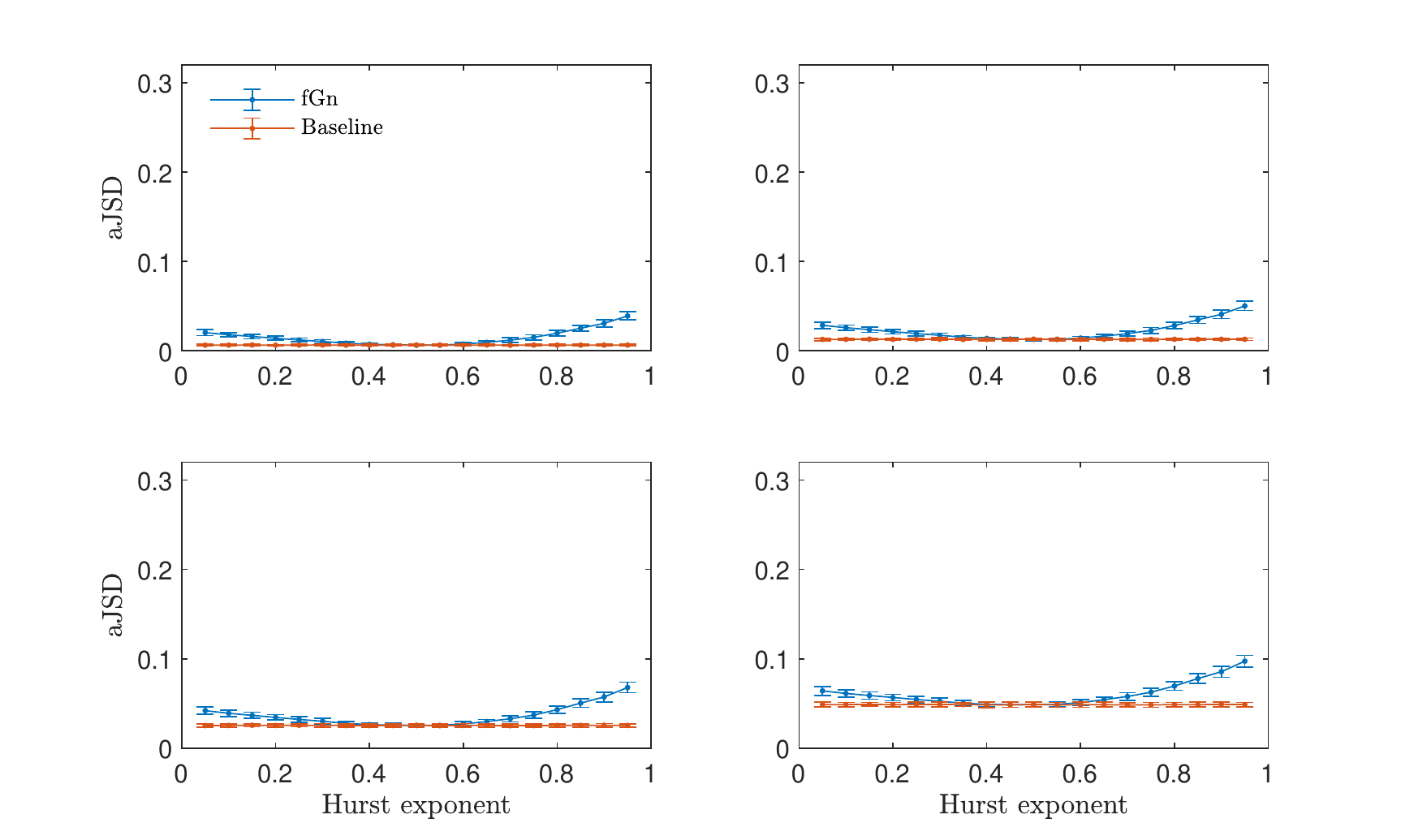}
\caption{Same as Fig.~\ref{fig:aJSD_fBm} but for fGns.}
\label{fig:aJSD_fGn}
\end{figure}

As a first numerical test, we compare performances of IBSI, aJSD and PJSD for discriminating the presence of temporal correlations in long-range correlated time series. The analysis developed is similar to that described in Sec.~\ref{subsec-colored-noises}, and, for this reason, the information is not repeated here. Both fractal stochastic models, fBm (nonstationary process) and fGn (stationary noise), with different Hurst exponent $H$ have been analyzed. Word lengths $m$ between 8 and 11 have been chosen since these values were proposed in the original papers. Results obtained for IBSI are detailed in Figs.~\ref{fig:IBSI_fBm}~and~\ref{fig:IBSI_fGn} while those for aJSD are displayed in Figs.~\ref{fig:aJSD_fBm}~and~\ref{fig:aJSD_fGn}. We have accordingly adjusted the vertical scale of these plots in order to make the comparison easier with the results obtained for PJSD (Figs.~\ref{fig:PJSD_fBm}~and~\ref{fig:PJSD_fGn}). By comparing the behavior and range of the three quantifiers, we conclude that PJSD offers the best characterization for both stochastic models. Performance differences observed in favor of PJSD are clear and evident when comparing against IBSI. Although much less noticeable, PJSD also outperforms aJSD with a larger range of estimated values that points out a higher ability for characterizing long-range memory effects.

\begin{figure}[!ht]
\centering
\includegraphics[width=\linewidth,trim={1.25cm .3cm 1.25cm .3cm},clip=true]{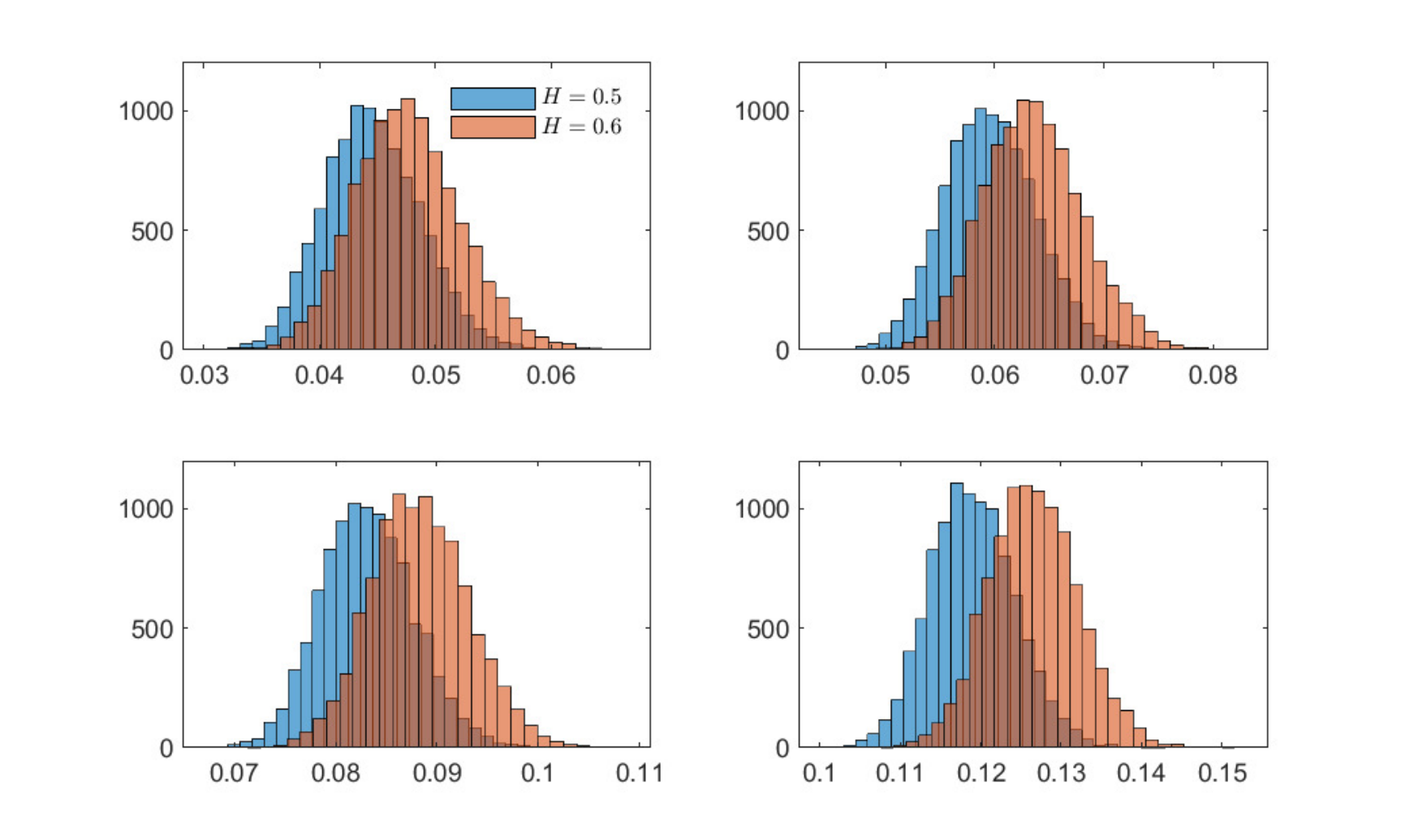}
\caption{Distributions of IBSI estimated values for ensembles of 10,000 independent numerical realizations of fGns with $H=0.5$ (fully uncorrelated dynamics) and $H=0.6$ (slightly correlated dynamics) with respect to their shuffled counterparts. Results obtained for simulations of length $N=10^4$ data with word lengths $m \in \{8,9,10,11\}$ (increasing from top left to bottom right) are shown.}
\label{fig:IBSI_fGn_hist}
\end{figure}

\begin{figure}[!ht]
\centering
\includegraphics[width=\linewidth,trim={1.25cm .3cm 1.25cm .3cm},clip=true]{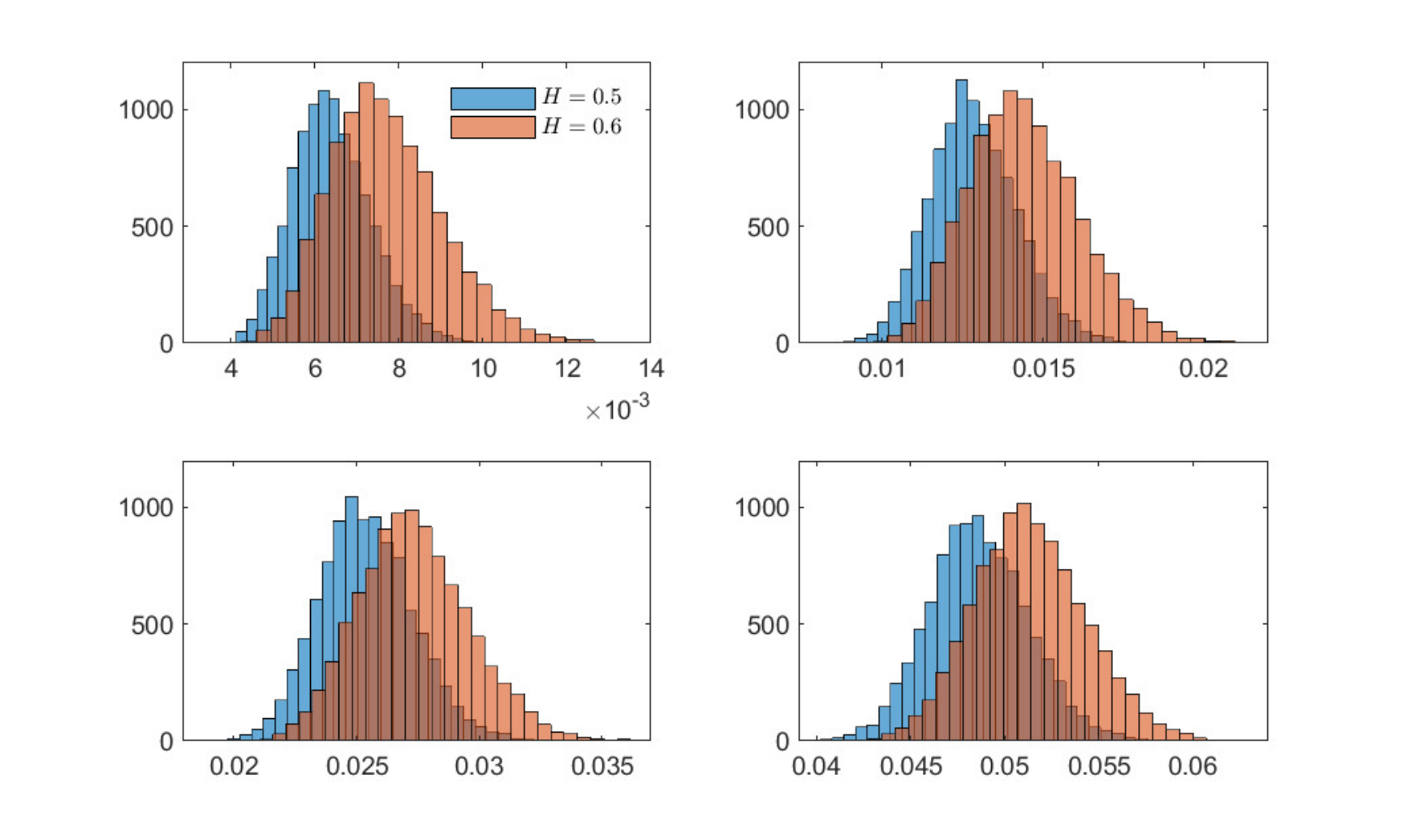}
\caption{Same as Fig.~\ref{fig:IBSI_fGn_hist} but for aJSD.}
\label{fig:aJSD_fGn_hist}
\end{figure}

\begin{figure}[!ht]
\centering
\includegraphics[width=\linewidth,trim={1.25cm .3cm 1.25cm .3cm},clip=true]{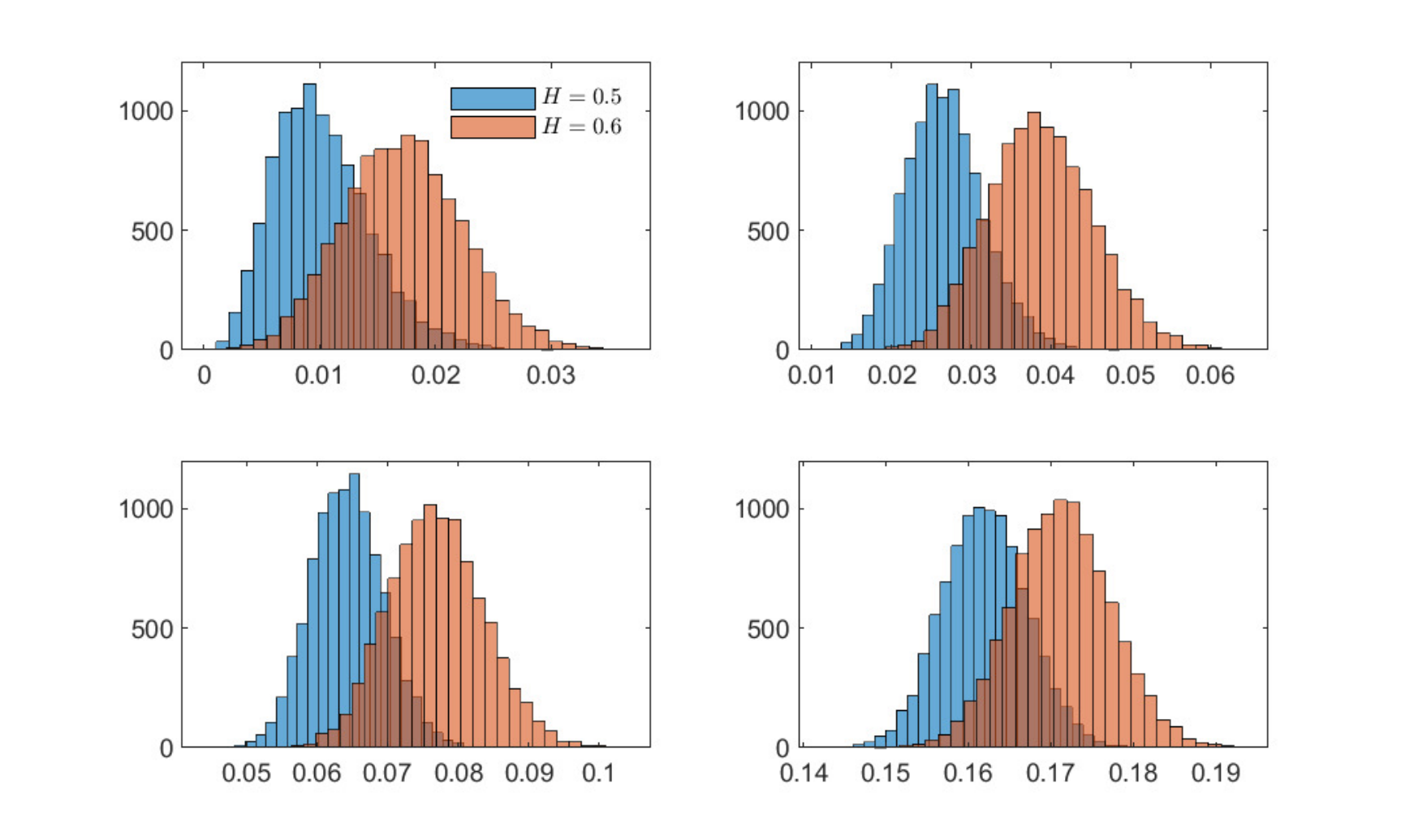}
\caption{Same as Fig.~\ref{fig:IBSI_fGn_hist} but for PJSD. In this case, results for orders $D \in \{3,4,5,6\}$ (increasing from top left to bottom right) and lag $\tau=1$ are displayed.}
\label{fig:PJSD_fGn_hist}
\end{figure}

Next, we analyze how useful the three measures result for unveiling the presence of slightly correlated dynamics. This is a relevant and well-known issue in several fields, where a fully uncorrelated dynamic is taken as the null hypothesis. Trying to provide a more rigorous statistical answer to this issue, we have estimated the three quantifiers for ensembles of 10,000 independent realizations of fGns with $H=0.5$ (fully random dynamics) and $H=0.6$ (slightly persistent dynamics). Simulations of length $N=10^4$ data were generated for this numerical test. Taking into consideration that the main objective is to identify the presence of memory effects in the time series, dissimilarities between the original and their shuffled counterparts were calculated. Histograms of estimated values for IBSI, aJSD and PJSD are depicted in Figs.~\ref{fig:IBSI_fGn_hist}, \ref{fig:aJSD_fGn_hist} and \ref{fig:PJSD_fGn_hist}, respectively. The two distributions of estimated values are better separated for the PJSD while the overlap increases in the case of IBSI and aJSD. This finding allows us to confirm a higher performance of the PJSD for uncovering the presence of slightly correlated dynamics.

\begin{figure}[!ht]
\centering
\includegraphics[width=\linewidth,trim={.8cm .15cm 1.4cm .3cm},clip=true]{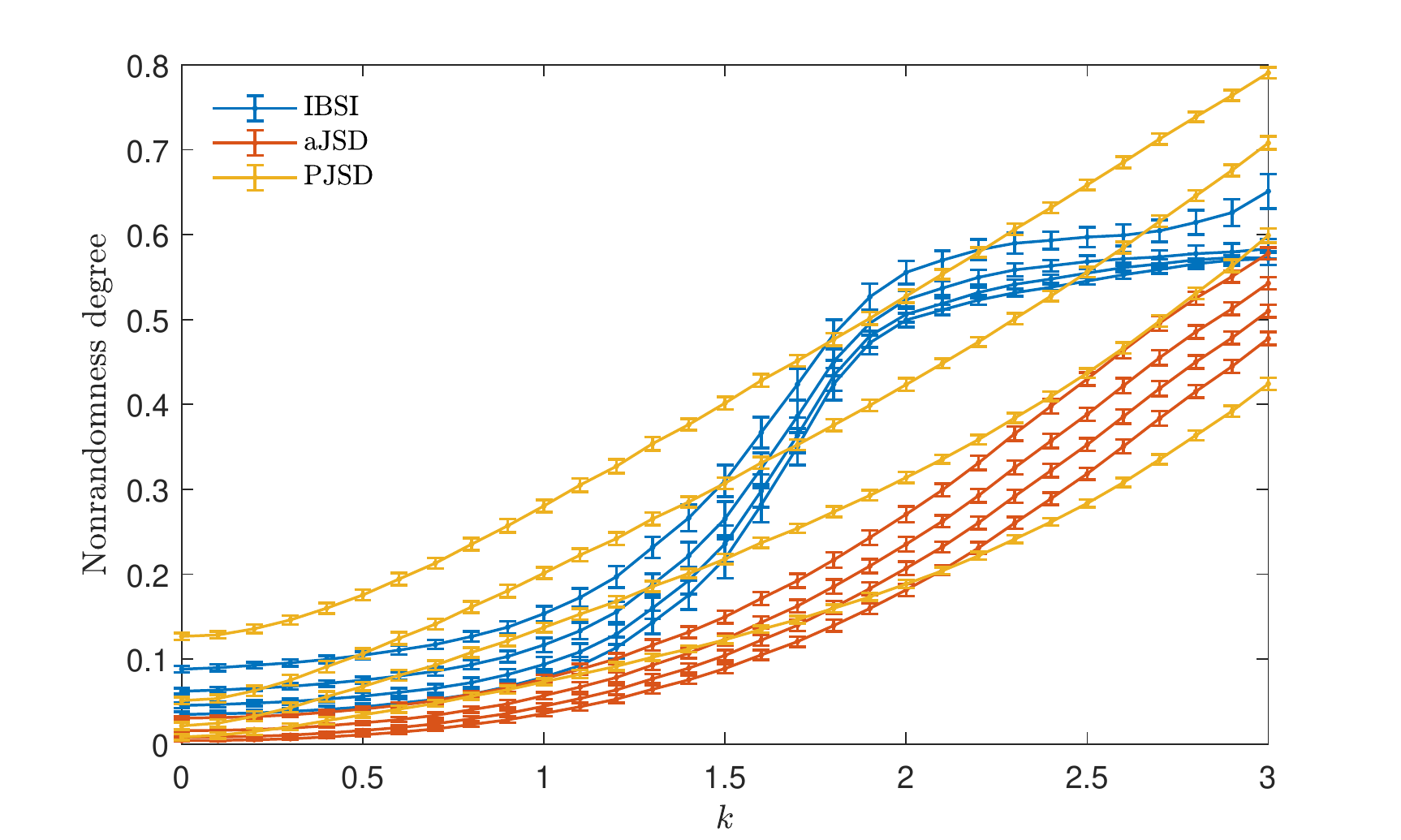}
\caption{Nonrandomness degree for colored noises $1/f^{k}$ with $k$ ranging from 0 to 3 with steps of size 0.1. Mean and standard deviation (as error bars) from estimations of an ensemble of one hundred independent realizations of length $N=2^{14}$ data are displayed. Word lengths $m \in \{8,9,10,11\}$ are considered for the IBSI and aJSD estimations while orders $D \in \{3,4,5,6\}$ with lag $\tau=1$ are used for the PJSD estimations. The value of $m$ and $D$ increases in the upward direction for each color group.}
\label{fig:Colored_noises_comparison}
\end{figure}

Dissimilarity between the original signal and its shuffled counterpart has been previously proposed as nonrandomness index~\cite{yang2003}. Indeed, this is analogous to what we have done when fGns are analyzed. Following this idea, we have found that the three dissimilarity measures show monotonic increasing behaviors when they are implemented to quantify the nonrandomness degree of colored noises. Figure~\ref{fig:Colored_noises_comparison} shows these behaviors when $1/f^{k}$ noises with $k$ between 0 and 3 and step $\Delta k=0.1$ are analyzed. The Fourier Filtering Method (FFM) has been implemented for generating the colored noises. We address the reader to Ref.~\cite{makse1996} for more details about this algorithm. Mean and standard deviation (displayed as error bars) of the three dissimilarity measures for one hundred independent realizations of length $N=2^{14}$ data for each value of $k$ have been plotted. Results for other values of $N$ ($N \in\{2^{10},2^{11},\dots,2^{17}\}$) are qualitatively similar. It is visually clear that IBSI has less accuracy for discriminating colored noises in both extremes of the power-law exponent range ($k \in [0,1]$ and $k \in [2,3]$) while aJSD suffers from a similar lack of sensitivity but only when $k \in [0,1]$. PJSD shows the best discrimination, especially for larger $D$. Moreover, an ideal quasi-linear behavior is observed for $D=6$ when $k > 0.5$.

\begin{figure}[!ht]
\centering
\includegraphics[width=\linewidth,trim={.8cm .15cm 1.4cm .3cm},clip=true]{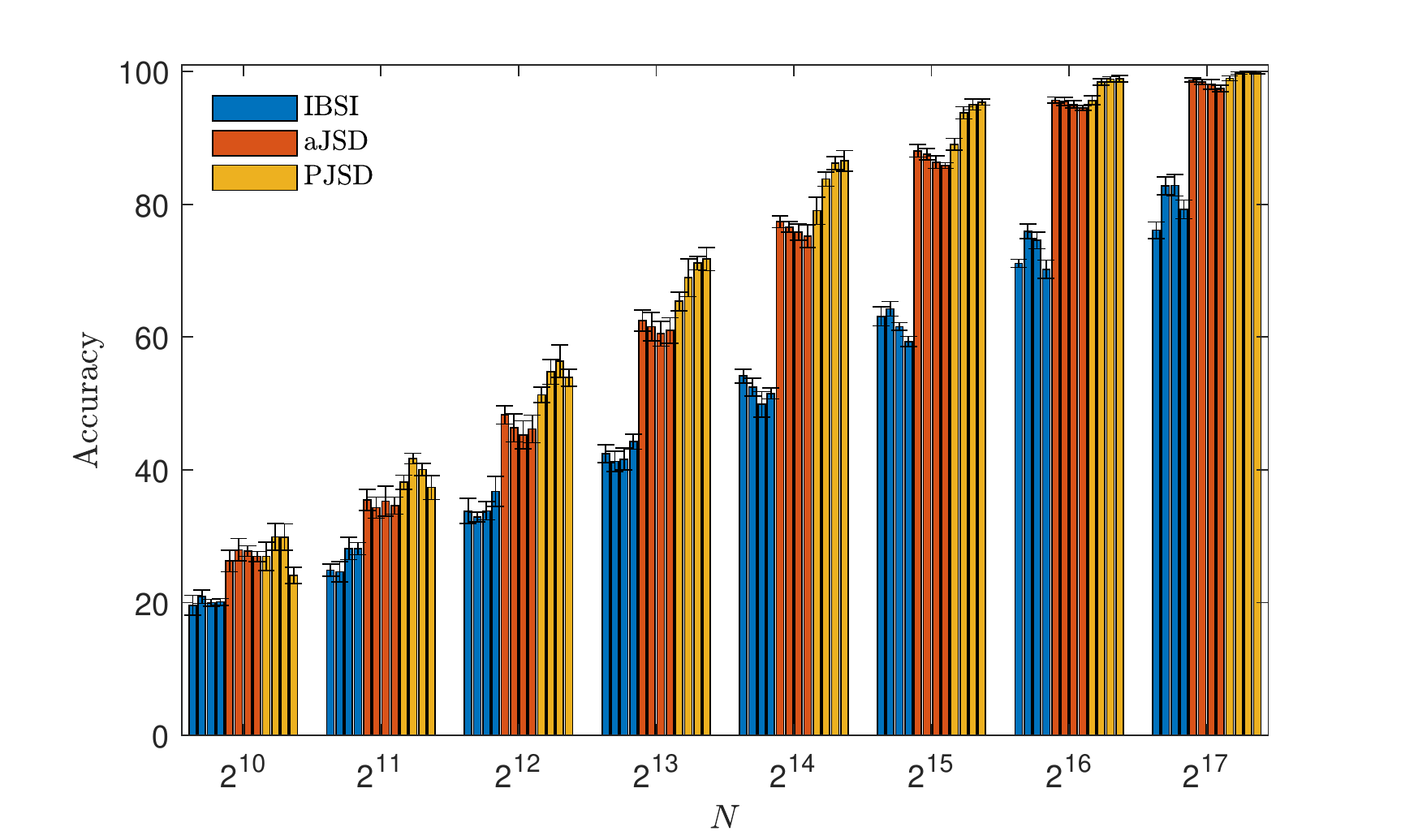}
\caption{Overall accuracy as a function of the time series length $N$ for classifying colored noises. A linear support vector machine algorithm with a 5-fold cross validation has been implemented as machine learning classifier. Word lengths $m \in \{8,9,10,11\}$ are considered for the IBSI and aJSD estimations while orders $D \in \{3,4,5,6\}$ with lag $\tau=1$ are used for the PJSD estimations. The value of $m$ and $D$ increases from left to right for each color group.}
\label{fig:Colored_noises_accuracy}
\end{figure}

Next, the performance of these dissimilarity measures for classifying different colored noises is quantified. For such a purpose, we train a linear support vector machine algorithm with a 5-fold cross validation considering the 31 values of $k$ ($k \in \{0,0.1,\dots,2.9,3\}$) as possible classes for the algorithm and the estimated nonrandomness index as single feature. We generate ensembles with one hundred simulations for each value of the power-law exponent $k$ and for each length $N$. Overall accuracies (fractions of correctly classified values of $k$) as a function of the time series length $N$ for the three dissimilarity measures are compared in Fig.~\ref{fig:Colored_noises_accuracy}. As it was expected, the overall accuracy for PJSD is enhanced regardless of the time series length $N$. Differences observed with IBSI are larger. When comparing with aJSD, differences are lower but always in favor of PJSD. The only exception is when PJSD is estimated with $D=6$ for $N=2^{10}$. This can be attributed to finite-size effects since the condition $N \gg D!$ is not satisfied in this case.

\begin{figure}[!ht]
\centering
\includegraphics[width=\linewidth,trim={.8cm .15cm 1.4cm .3cm},clip=true]{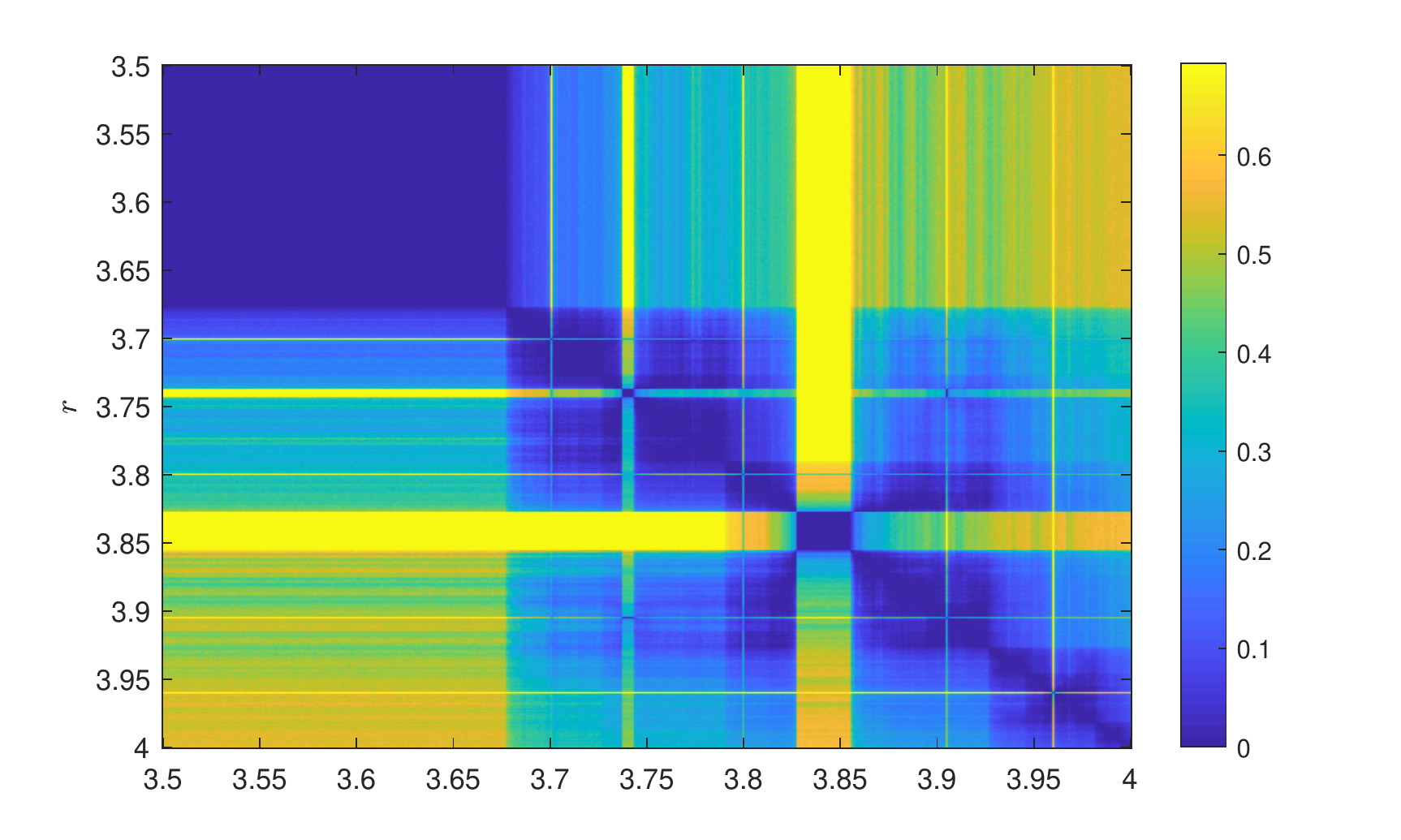}
\includegraphics[width=\linewidth,trim={.8cm .15cm 1.4cm .3cm},clip=true]{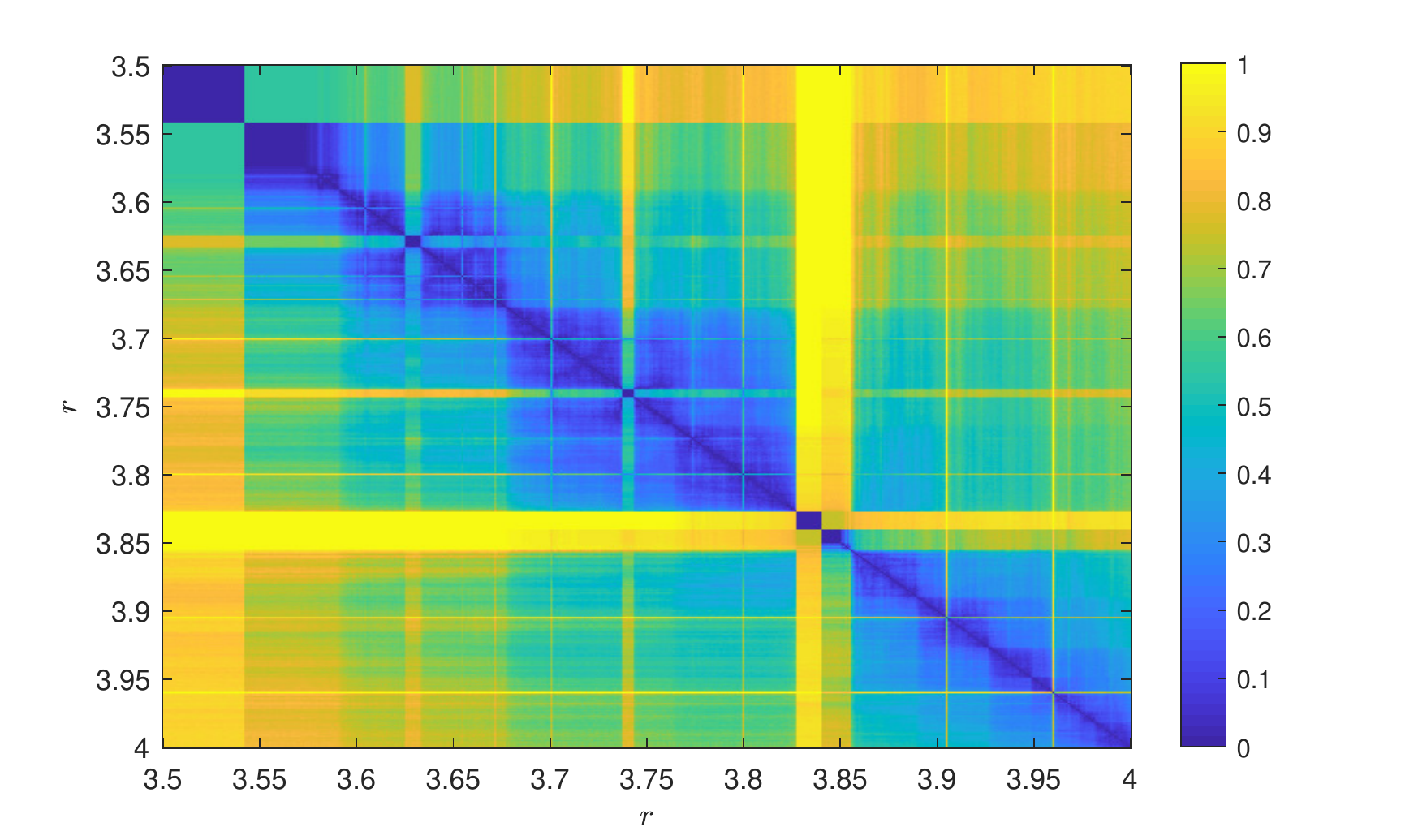}
\caption{Matrices distances obtained for the aJSD with \hbox{$m=8$} (top plot) and PJSD with $D=5$ and $\tau=1$ (bottom plot) for the logistic map with parameter $r$ between 3.5 and 4 and step $\Delta r=0.001$. Numerical realizations of length $N=10^4$ data are analyzed.}
\label{fig:Logistic_matrices_comparison}
\end{figure}

The ability to distinguish between different complex deterministic dynamics, reproducing the analysis detailed in Sec.~\ref{subsec-different-dynamics}, has also been contrasted. We have first observed that IBSI has problems to be estimated in several instances in which the dissimilarity between pure periodic and low-dimensional chaotic dynamics is considered. This is essentially due to the fact that the signals associated with these two regimes have very few common $m$-bit words, making statistically unreliable the estimation of an average deviation between their associated ranks. Actually, we have found several couples of control parameters of the logistic map for which only one $m$-bit word is shared between the simulated time series. Consequently, IBSI has been discarded for this particular analysis. The estimated matrices for the aJSD and PJSD in the case of logistic map for 501 equidistant values of the control parameter $r \in [3.5,4]$, are compared in Fig.~\ref{fig:Logistic_matrices_comparison}. For a fairer comparison, we have fixed the values of $m=8$ and $D=5$ in order to have a similar number of elements in the probability distributions ($2^8=256$ and $5!=120$). By comparing the two matrices, it is concluded that PJSD achieves a more accurate discrimination of the different deterministic dynamics, especially in the periodic windows. Moreover, expected minima along the diagonal are also better defined in the PJSD matrix distance, evidencing a more robust distinction between nearby values of the control parameters. It is worth mentioning here that we have confirmed that results obtained for aJSD with larger $m$ ($m \in \{9,10,11\}$) are very similar to those obtained for $m=8$ and that they do not show significant improvements. This behavior is different to that observed for the PJSD (Fig.~\ref{fig:Logistic_analysis}), for which improved discrimination between dynamics is reached for larger orders $D$.

\begin{figure}[!ht]
\centering
\includegraphics[width=\linewidth,trim={.8cm .15cm 1.4cm .3cm},clip=true]{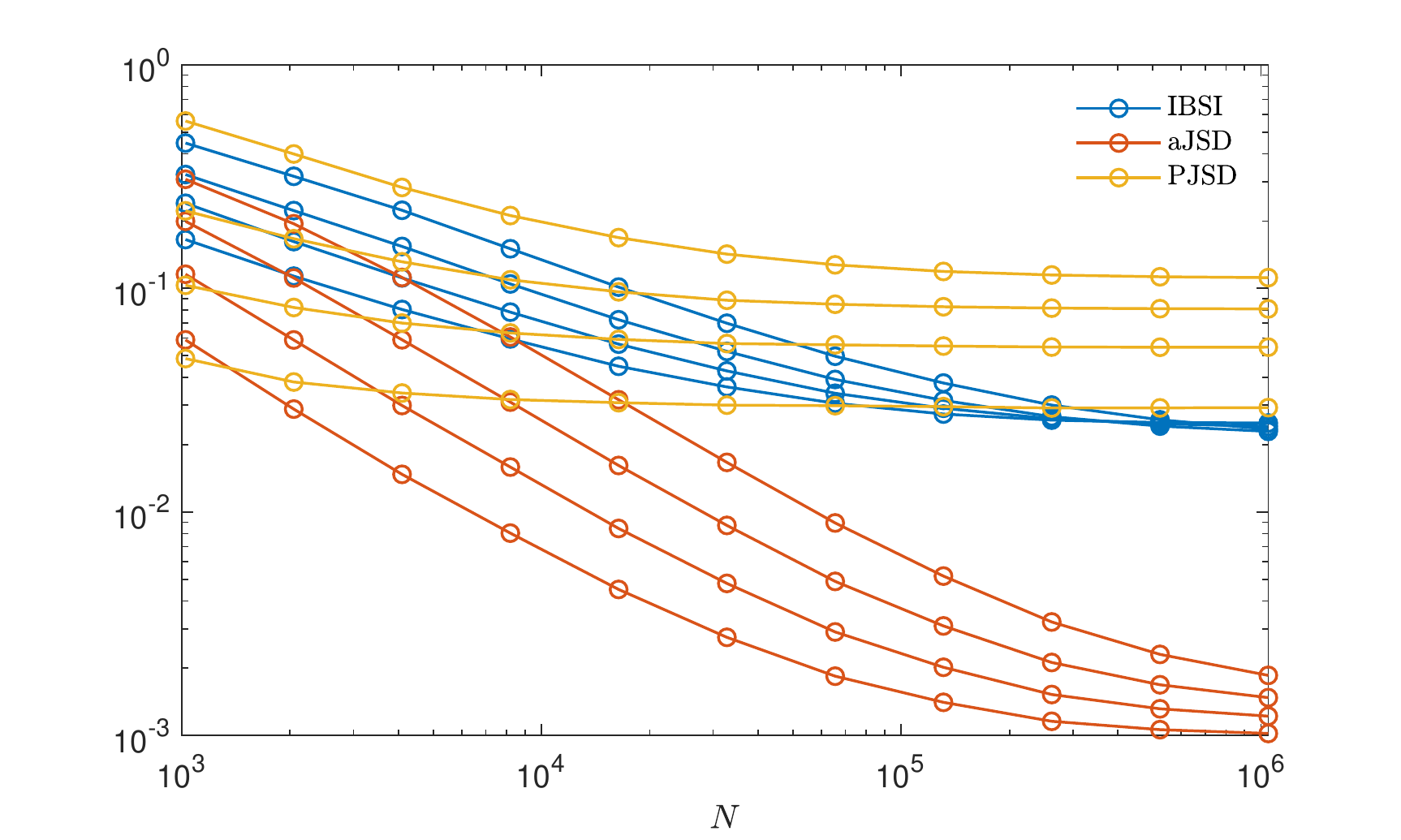}
\caption{Irreversibility analysis for the $\beta$-transform with $\beta=\sqrt{200}$. Dissimilarities between the forward and backward series are plotted (in log-log scale) as a function of the time series length $N$. Average values from one hundred independent realizations are displayed. Word lengths $m \in \{8,9,10,11\}$ are considered for the IBSI and aJSD estimations while orders $D \in \{3,4,5,6\}$ with lag $\tau=1$ are used for the PJSD estimations. The value of $m$ and $D$ increases in the upward direction for each color group.}
\label{fig:Irreversibility_beta_comparison}
\end{figure}

\begin{figure}[!ht]
\centering
\includegraphics[width=\linewidth,trim={.8cm .15cm 1.4cm .3cm},clip=true]{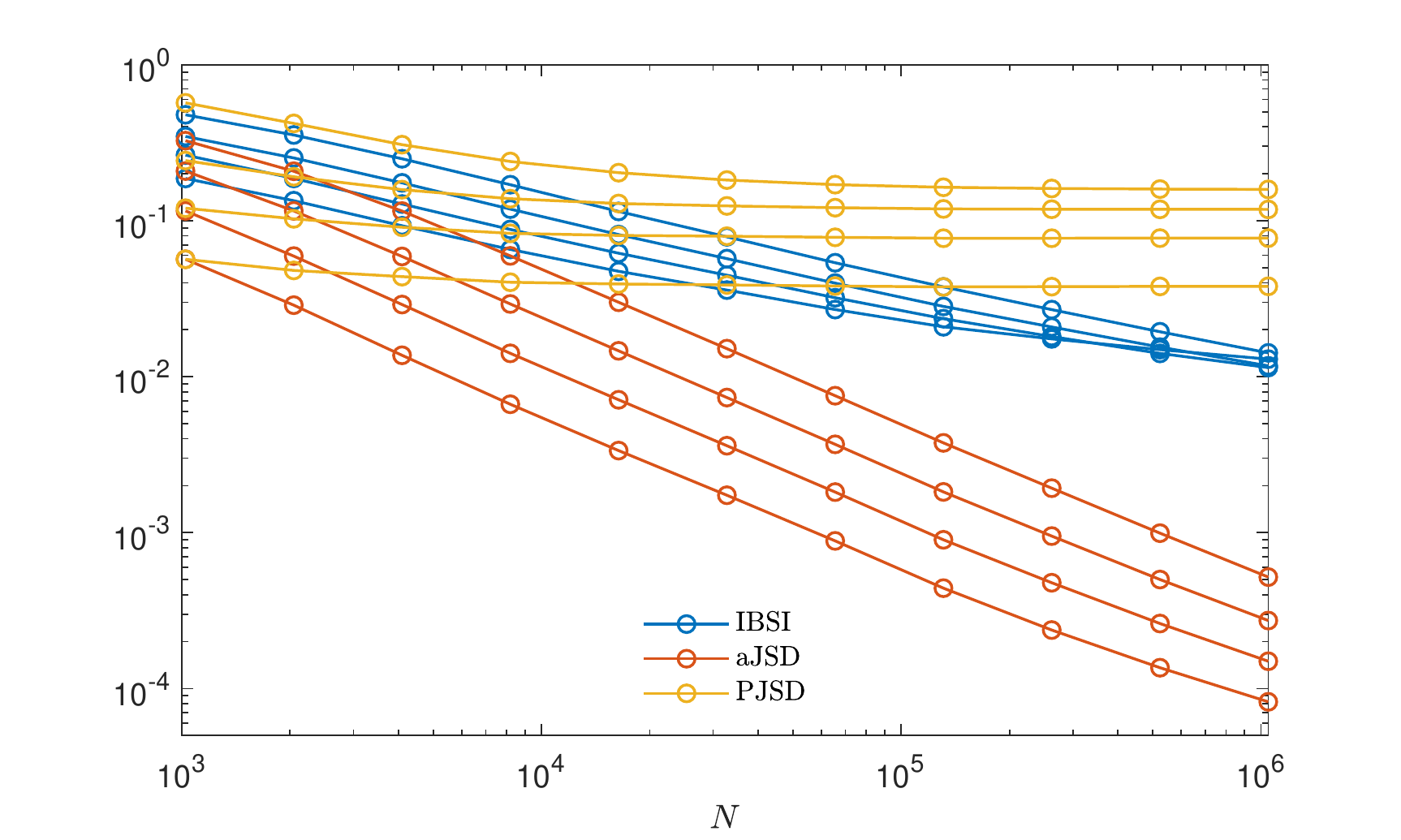}
\caption{Same as Fig.~\ref{fig:Irreversibility_beta_comparison} but for the NGRP model.}
\label{fig:Irreversibility_NGRP1_comparison}
\end{figure}

Finally, we have confirmed that IBSI and aJSD are also useful to discriminate between reversible and irreversible time series following a procedure analogous to the one proposed for the PJSD. That is, the behavior of the dissimilarity between the forward and backward series as a function of the time series length is analyzed to conclude in favor of a reversible (asymptotic convergence to zero) or irreversible (convergence to a non-zero value) dynamics. We have identified, however, a couple of cases in which both IBSI and aJSD have clear difficulties within this framework. The first one is the $\beta$-transform with a large value of the parameter $\beta$. Figure~\ref{fig:Irreversibility_beta_comparison} illustrates the irreversibility analysis for IBSI, aJSD and PJSD when $\beta=\sqrt{200}$. It is concluded that the convergence to a non-zero value is realized for IBSI and aJSD but for larger values of $N$. Consequently, these two quantifiers need longer time series to identify the irreversible character of the dynamics. With short time series both of them may be unable to conclude the irreversible nature of the original dynamics. The second case that deserves special consideration is the NGRP model. According to the results displayed in Fig.~\ref{fig:Irreversibility_NGRP1_comparison}, IBSI and aJSD are not able to unveil the underlying irreversible nature. In fact, on the contrary, reversibility could be erroneously concluded from the aJSD analysis. Undoubtedly, PJSD offers an enhanced characterization of this stochastic system with dynamic nonlinearities.

\bibliography{PJSD}

\begin{thebibliography}{102}
\bibitem [{RWn()}]{RWnullmodel}%
  \BibitemOpen
  \href@noop {} {}\bibinfo {note} {It should be noted that the random walk
  references for the fBm processes are generated by integrating the randomly
  resampled normalized fluctuations associated with each
  simulation.}\BibitemShut {Stop}%
\end{thebibliography}

\clearpage
\includepdf[pages=1-17,pagecommand={\thispagestyle{empty}, \clearpage}]{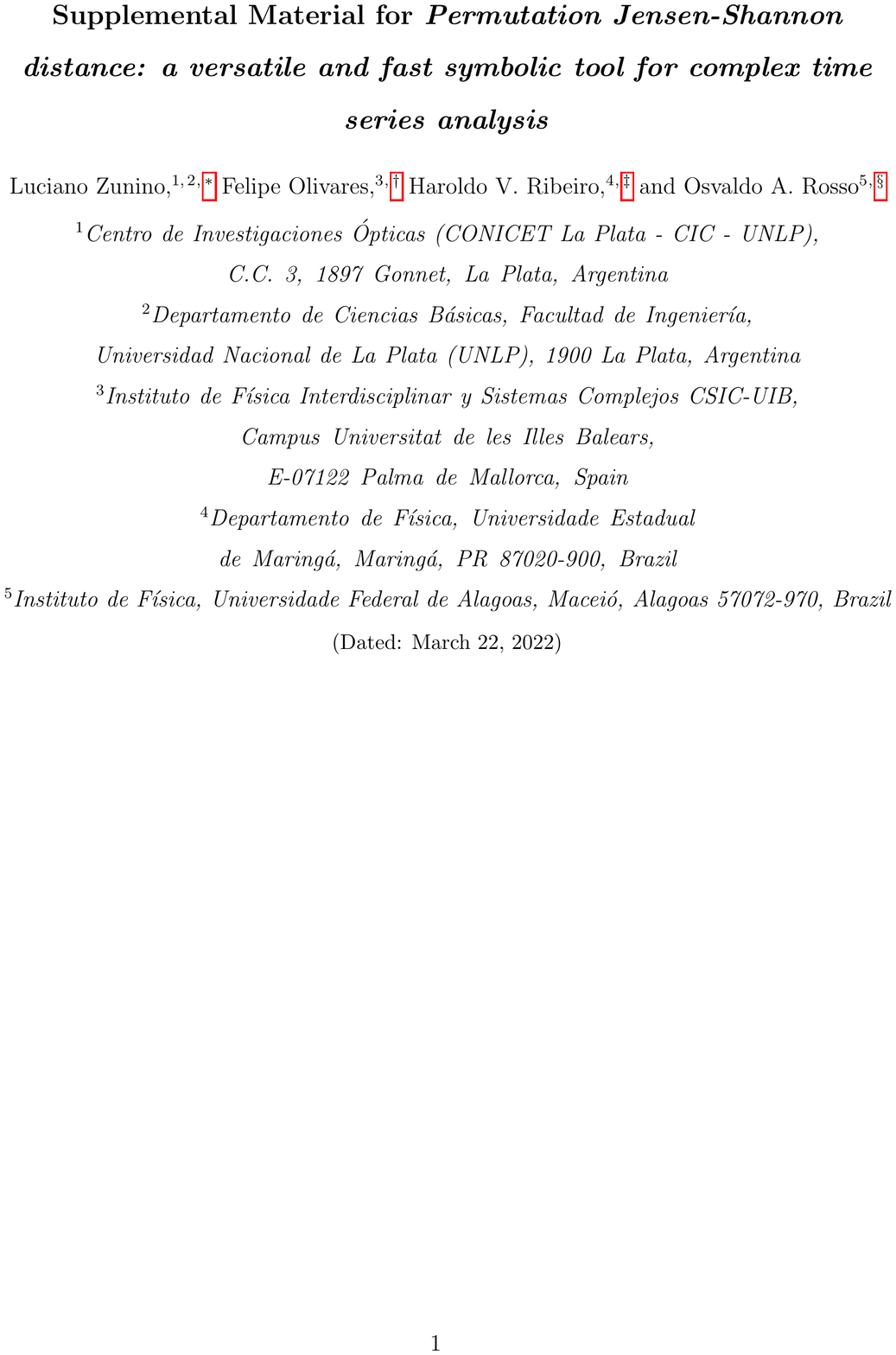}
\includepdf[pages=18,pagecommand={\thispagestyle{empty}}]{PJSD_Supp_Material.pdf}

\end{document}